\documentclass[pra,twocolumn,showpacs,superscriptaddress,floatfix,nofootinbib]{revtex4-2}

\usepackage{color}
\usepackage{hyperref}  
\usepackage{graphicx}
\usepackage{amsfonts}
\usepackage{footmisc}

\usepackage{fancyhdr}
\let\a=\alpha \let\b=\beta  \let\g=\gamma  \let\d=\delta \let\e=\varepsilon
\let\z=\zeta  \let\h=\eta   \let\th=\theta \let\k=\kappa \let\l=\lambda
\let\m=\mu    \let\n=\nu    \let\x=\xi     \let\p=\pi    \let\r=\rho
\let\s=\sigma \let\t=\tau   \let\f=\varphi 
\let\ch=\chi  \let\ps=\psi   
\let\G=\Gamma \let\D=\Delta  \let\Th=\Theta\let\L=\Lambda \let\X=\Xi
         
\let\O=\Omega 

\font\tenmib=cmmib10\font\sevenmib=cmmib7\font\fivemib=cmmib5%
\mathchardef\Bl   = "0515  

\def\Bb   {{\mbox{\boldmath$ \beta$}}}

\def\Bd   {{\mbox{\boldmath$ \delta$}}}

\def\Bl   {{\mbox{\boldmath$ \lambda$}}}

\def\Bs   {{\mbox{\boldmath$ \sigma$}}}

\def\BTh  {{\mbox{\boldmath$ \Theta$}}}

\def\BF   {{\mbox{\boldmath$ \Phi$}}}
\def\BPs  {{\mbox{\boldmath$ \Psi$}}}

\def\BDpr {{\mbox{\boldmath$ \partial$}}}

\def\eqalign#1{\null\,\vcenter{\openup\jot
  \ialign{\strut\hfil$\displaystyle{##}$&$\displaystyle{{}##}$\hfil
      \crcr#1\crcr}}\,}

\def\AA{{\mathcal A}}\def\CC{{\mathcal C}}
\def\EE{{\mathcal E}}\def\DD{{\mathcal D}}\def\TT{{\mathcal T}}
\def\NN{{\mathcal N}}\def\MM{{\mathcal M}}
\def\PP{{\mathcal P}}\def\LL{{\mathcal L}}

\def\uu{{\V u}}\def\kk{{\V k}}\def\xx{{\V x}}

\def\T#1{{#1_{\kern-3pt\lower7pt\hbox{$\widetilde{}$}}\kern3pt}}
\def\Tdpr{{\T{\BDpr}}}
\def\uU{{\T {\bf u}}}

\def\ie{{\it i.e.\ }}
\def\dpr{{\partial}}
\def\defi{{\buildrel def\over=}}
\def\rhs{{\it r.h.s.}}\def\lhs{{\it l.h.s.}}
\def\otto{\,{\kern-1.truept\leftarrow\kern-5.truept\to\kern-1.truept}\,}
\def\Pprod{\prod^{\kern-1mm\raise.0mm\hbox{$\leftarrow$}}}
  
\newdimen\xshift \newdimen\xwidth \newdimen\yshift \newdimen\ywidth

\def\ins#1#2#3{\vbox to0pt{\kern-#2pt\hbox{\kern#1pt #3}\vss}\nointerlineskip}

\def\eqfig#1#2#3#4#5{
\par\xwidth=#1pt \xshift=\hsize \advance\xshift
by-\xwidth \divide\xshift by 2
\yshift=#2pt \divide\yshift by 2
{\hglue\xshift \vbox to #2pt{\vfil
#3 \includegraphics{#4.eps}
}\hfill\raise\yshift\hbox{#5}}}

\def\V#1{{\bf #1}}
\def\lis#1{{\overline#1}}
\let\wt=\widetilde
\let\wh=\widehat

\def\tende#1{\,\vtop{\ialign{##\crcr\rightarrowfill\crcr
 \noalign{\kern-1pt\nointerlineskip} \hskip3.pt${\scriptstyle
   #1}$\hskip3.pt\crcr}}\,}
\def\eg{{\it e.g.\ }}

\def\0{\noindent}
\def\*{\vskip2mm}
\def\media#1{\langle #1 \rangle}
\def\Eq#1{\label{#1}}
\def\equ#1{(\ref{#1})}

\font\titolo=cmbx12%
\def\iniz{\setcounter{equation}{0}}
\def\be{\begin{equation}}\def\ee{\end{equation}}
\renewcommand{\theequation}{\arabic{section}.\arabic{equation}}

\newcounter{appendice}
\def\APPENDICE#1{
\setcounter{appendice}{#1}
\appendix
\renewcommand{\theequation}{\Alph{appendice}.\arabic{equation}}%
\renewcommand{\thesection}{\Alph{appendice}}%
}

\def\alert#1{{\color{ired}#1}}
\definecolor{iblue}{RGB}{65,105,225}
\definecolor{ired}{RGB}{220,20,60}
\definecolor{igreen}{RGB}{50,205,50}
\definecolor{ipurple}{RGB}{75,0,130}
\definecolor{iochre}{RGB}{218,165,32}
\definecolor{iteal}{RGB}{51,204,204} 
\definecolor{imauve}{RGB}{204,51,153}

\def\txt{\textstyle}

\iniz

\begin{document}

\let\titolo=\bf

\alert{\centerline{\bf \Large Nonequilibrium and}
    \centerline{\bf \Large Fluctuation Relation}}\vskip1mm
 
\centerline{\bf Giovanni Gallavotti} \centerline{\today}

{\vskip3mm}
{\halign{\small #&{\it #}\cr
    \hglue1.2cm&This review is dedicated to Joel Lebowitz to\cr
    &witness my gratitude for his constant guidance\cr
    &and teaching which influenced indelibly my work.\cr}}
  
{\vskip3mm}
\noindent {\bf Abstract}: {\it A review on the fluctuation relation,
  fluctuation theorem and related topics.
  Journal of Statistical Phys.:
https://doi.org/10.1007/s10955-019-02376-3, arxiv 1906.10069}  {\vskip3mm}

\setcounter{section}{0}
\label{sec0}
\iniz

{\tiny
\tableofcontents
}
\*\*

\def\SEC{\ Introduction}
\section{\SEC}
\label{sec1}
 
Many works investigate the Fluctuation Relation, (FR): but its
interpretation is often very far from the original one proposed in
\cite{GC995} and in subsequent works, see for instance \cite{Ga013b}.
Here I present the original  point of view referring also a few
of its interpretations, consequences and related conjectures.

In this review only the foundations of the theory are discussed, as I
cannot present the many (pertinent) developments that followed,
starting with the early ones, \cite{JP998b,Ku998,EPR999,LS999,Ma999}.

The FR arises from a simple theorem on dynamical systems, the
{\it Fluctuation Theorem}, (FT).  The FT applies, under further
suitable assumptions, to {\it Anosov systems}: which can be
considered as playing a role analogous to that played for non
chaotic systems by the harmonic oscillators, although sometimes
they are considered an abstract mathematical
notion.\footnote{\small\eg ``Whether or not speculations
concerning such hypothetical Anosov systems are an aid or a
hindrance to understanding seems to be an aesthetic question'',
\cite[p.221]{Ho999}.
\label{H1}}

In natural observations initial data are generated by a well
defined procedure, that is sometimes called a {\it protocol}, but
are always affected by unavoidable errors, no matter how
carefully one fixes the protocol.

Therefore initial data are generated with a probability distribution:
defined by the protocol and {\it unknown}. Yet it is subject to the
fundamental assumption that it is a probability distribution on the ``phase
space'' $M$ (a smooth Riemannian manifold here) which {\it admits a
  density} with respect to the volume of $M$. The probability of $x\in M$
being in a open set $dx$ around $x$ has the form $\r(x)dx$ where $\r$ is
some continuous function (or slightly more general). This is an assumption
which should not be overlooked: it {\it cannot be proved} but it is always
assumed (at least tacitly) and is, therefore, a {\it law of nature}, with
the far reaching consequence that it leads to the determination of the
probability distributions of the stationary states in equilibrium as well
as in nonequilibrium systems: see below.

The connection with Physics is established via the hypothesis
(called {\it Chaotic Hypothesis}, (CH)) stating that ``all''
systems exhibiting chaotic motions can be treated for many
purposes as {\it Anosov systems}.  Informally, in such systems
an observer co-moving, in phase space $M$, with a point $x$ sees
it as a ``saddle point'' (mathematically a ``hyperbolic fixed
point''), while it wanders invading an attracting surface $\AA$
in $M$, The notion of Anosov maps and flows,
\cite{AA966,Sm967},\cite[Ch.4]{GBG004}, is briefly recalled in
the footnote.%
\footnote{\small If $M$ is a smooth (\ie
$\infty$-differentiable)
bounded manifold and $S$ is an invertible smooth (\ie
$\infty$-differentiable together with the inverse $S^{-1}$) map
on $M$, the system $(M,S)$ is an Anosov map if
  \\ (a) at every point $x\in M$ there are two complementary
  tangent planes $T_s(x)$ and $T_u(x)$, transverse in $x$, which
  depend continuosly on $x$, are covariant in the sense that the
  Jacobian $\dpr S(x)^{\pm1}$ acts on the plane tangent to the
  attracting set so that $\dpr S(x)^{\pm1}
  T_\g(x)=T_\g(S(x)^{\pm1}), \g=u,s$,
  \\ (b) furthermore there are $C>0,\l<1$ such that
  $|\dpr S^n(x) v|< C \l^n |v|, n>0$ if $v\in T_s(x)$
  and $|\dpr S^{-n}(x) v|<  C \l^n |v|, n>0$ if $v\in T_u(x)$,
  \\
  (c) there is a point whose orbit is dense in $M$.
  \\
  The definition of Anosov flow is similar: the covariant
  mutually transversal planes are now three: $T_u(x),T_s(x)$ on
  which expansion and contraction take place under action of the
  flow $S_t$ as in the map case and a third $1$-dimensional
  tangent plane parallel to $f(x)$, if $\dot x=f(x)$ is the
  differential equation defining the flow $S_t$; this is a
  tangent vector supposed not zero, $|f(x)|>0$, and which, of
  course, neither expands nor contracts under the $\dpr S_t(x)$.
  \\
The stable and unstable planes of Anosov systems (maps or flows)
can be {\it integrated} (\ie there is a smooth surface everywhere
tangent to them) to define stable and unstable manifolds
$W_u(x),W_s(x)$ which are dense on $M$.\label{hyperbolic}}

Suppose that evolution ($\infty$-smooth, for simplicity) of a mechanical
system is defined by a map $S$ on a phase space $M$ and is attracted by an
$\infty$-smooth surface $\AA\subset M$ on which $S$ is an Anosov map
$S\AA\otto\AA$, as considered here in most cases (for simplicity). It
transforms an initial datum $x$ into the new datum $Sx$ in a single time
step; then the main property of the dynamical system $(M,S)$ is that the
evolution is chaotic and a phase space point, with exceptions forming a set
of zero volume, moves accumulating at the attracting surface $\AA$
densely.\footnote{\small This means that for all times $t_0$ the closure of
  the trajectory of $\{S^tx\}_{t>t_0}$ is $\AA$.}  Such $S$ will be
called a {\it map with an Anosov attractor}.\footnote{\small Evolutions
  defined by maps arise typically when studying evolutions in continuous
  time through observations triggered by ``timing events'', \ie
  observations that are made every time a specified event takes place: for
  instance every time that the evolving trajectory crosses a given surface
  in phase space. These are referred as observations performed on a {\it
    Poincar\'e's section}.%
\label{Poincare' section}}

The key property of maps with an Anosov attractor is that the fraction of
time asymptotically spent in any open (or just measurable and with $>0$
volume) phase space region $dx$, defines a stationary probability
distribution $\m(dx)=\m(Sdx)$ which is independent of the initial $x$,
again except for data $x$ forming a set of $0$ phase space volume,
\cite{Si977,Si994}. The probability $\m$ is concentrated on the attracting
surface $\AA$, \ie $\m(\AA)=1$; the data outside the attracting surface
$\AA$ evolve while exponentially attracted by $\AA$,
\cite{Si968a,Si968b}.

Likewise if the evolution is instead described by a flow $x\to S_tx$,
generated by a ($\infty$-smooth) differential equation on
$M$, $\dot x=f(x)$, which on $\AA\subset M$ is an Anosov flow and $\AA$ is
an attractive smooth surface, then any initial point $x\in M$, with
exceptions forming a set of zero volume, moves accumulating densely at
$\AA$ and spending a well defined fraction of time in every measurable
region of $M$ with positive volume: thus uniquely defining a stationary
probability distribution $\m(dx)=\m(S_t dx)$, concentrated on $\AA$, which
is independent of the initial $x$, again except for data forming a set of
$0$ volume in phase space $M$, \cite{Si977,Si994,BR975}: such flow $S_t$
will be called a {\it flow with an Anosov attractor}.

\0Formally: \*\0{\it Theorem: Suppose that a smooth dynamical system is
  defined on a manifold $M$ containing a globally attracting smooth surface
  $\AA$ on which the motion is an Anosov map or flow. Then all initial data
  $x$, aside from a set of zero volume, evolve visiting open sets $D$ with
  a time frequency $\m(D)$: the probability distribution $\m$ is
  independent on the protocol generating $x$. The probability $\m$ is
  ergodic and mixing at exponential rate for all smooth observables.}  \*

Phase space $M$ can be a smooth manifold or, more generally, an open set in
$R^n$ for some $n$, which is a domain of attraction of $\AA\subset M$. For
systems with a global Anosov attractor (flows or maps) the associated
unique stationary probability distribution $\m(dx)$ that defines the
statistical properties of the evolution, (\ie frequencies of visit to
regions of $M$), is called the {\it SRB distribution},
\cite{RT971,Ru995,Ru989}.

The just mentioned theorems on maps or flows become relevant for systems
evolving chaotically in the cases in which the following hypothesis holds
{\it i.e.}, as its name suggests (as intended in \cite{GC995,BG997,BGG997}, see
also the warning in \cite[endnote 18]{GC995}), always when motions are
{\it empirically chaotic}: \*

\0{\it Chaotic hypothesis (CH):} {\it A chaotic evolution takes place on a
  phase space $M$ being attracted by a bounded smooth attracting surface
  $\AA\subset M$ and on $\AA$ the map $S$ (or the flow $S_t$) is an Anosov
  map (or flow).}  \*

The SRB distribution $\m$ has support on the attracting set $\AA$: so
smoothness of $\AA$ is a strong assumption.%
\footnote{\small Which contrasts the picture of $\AA$ as a fractal set. In
  systems of $\sim10^{20}$ molecules with a fractal attractor of dimension
  $6\cdot10^{19}+3.141$ this means that it 'behaves' as a smooth surface of
  dimension $6\cdot 10^{19}$; or in a Navier-Stokes fluid (an
  $\infty$-dimensional system) at large Reynolds number $R$ an attracting
  set of dimension $R^{\frac93}+.33$ 'behaves' as a smooth surface of
  dimension integral part of $R^{\frac93}$.} Of course there is the
possibility that $\AA$ is a surface of dimension lower than that of $M$, as
specified in the CH.\footnote{\small Often, if the dynamical system depends
  on a parameter $\e$, the chaotic motion might occupy, asymptotically, an
  attracting set $\AA_\e$ with a dimension dependent on $\e$ and equal to
  that of the full phase space only for a small (if any) interval of
  variability of $\e$.}

The CH is a general and heuristic (more restrictive) interpretation of
original ideas on turbulence phenomena, \cite{Ru980}, and has been
introduced in \cite{GC995,GC995b} to interpret simulations on evolutions
with attracting set coinciding with the full phase space, and extended to
the more general case in which the attracting surface is lower dimensional,
\cite{BG997,BGG997}).

Hereafter the CH will
be supposed to hold {\it for all dynamical systems considered}, unless
stated otherwise.

The above formulation in which the Anosov system is realized on a
attracting surface $\AA\ne M$, rather than on the full phase space, already
hinted in \cite{GC995}, has become relevant as soon as attempts were
undertaken to apply CH to systems for which the strict inclusion
$\AA\subset M$ was manifest, \cite{BG997,BGG997}.

Besides smoothness of the attracting set the further strong assumption of CH
is that the evolution on $\AA$ is hyperbolic in the sense of Anosov. Both
aspects of the CH can be simultaneously weakened by supposing that the
motion has an attractor which satisfies the ``Axiom A'', \cite{BR975};
however such generality will not be envisaged here.

It will be convenient to distinguish between {\it attractor} and {\it
  attracting set} $\AA$: the latter is an invariant set approached
asymptotically by all points $x$ in its {\it basin} of attraction (which is
an open set around $\AA$): $\lim_{t\to\infty} distance (S_tx,\AA)=0$; while
an attractor $A\subset \AA$ is an invariant subset dense on $\AA$
which has full SRB measure (\ie $\m(A)=1$) and minimal Hausdorff dimension,
often smaller than the dimension of $\AA$ which, in turn, could be {\it much
smaller} than the dimension of $M$.

The SRB distributions have strong ergodic properties (see Sec.\ref{sec14}
for some details) and in particular the average value over time of a smooth
observable $O(x)$ is reached exponentially fast: however, more
generally, it is possible that in $M$ there are several attracting sets
$\AA_i$, each with its own SRB distribution, just as in equilibrium
statistical mechanics there are cases in which the Gibbs state is not
unique and the extremal ones correspond to different phases. It will appear
that the analogy is a deep one, see Sec.\ref{sec7}.

Anosov systems are chaotic systems whose properties can be studied in {\it
  great detail}: certainly they correspond to an idealization of chaos; but
it should be kept in mind that Statistical Mechanics arose from the
idealization, far more surprising, that microscopic motion could be
regarded as periodic, \cite{Bo866,Cl871,Bo868,Ma879}, see also
\cite[Sec.6\&7]{Ga016}.

\def\SEC{\ Stationary Distributions (SRB)}
\section{\SEC}
\label{sec2}

Let time evolution on $M$ be a map $S_r$, (or a flow $S_{r,t}$), that
may depend on a parameter $r$ (or on more but imagine, to simplify, that
only $r$ will be varied). Then as $r$ changes the stationary SRB distribution
$\m_r(dx)$, for the system $(M,S_r)$, changes and the {\it collection} of
such distributions will be called $\EE^{mc}$ and its elements will be
thought of as {\it ensembles of stationary states}. 

Volumes of regions in phase space $M$ {\it change}. in general, when
transformed by the discrete evolution map $S$ or by the flow $S_t$
generated by a differential equation $\dot x=f(x)$; and the rate of change
per unit volume can be measured from the Jacobian matrix $J(x)_{ij}=\dpr_i
S(x)_j$ for maps (here $\dpr_i\defi \dpr_{x_i}$) or by the matrix
$J_{ij}=\dpr_i f_j(x)$ in the case of a flow. And the {\it phase space
  contraction rate} is defined by:
\be \kern-3mm
\eqalign{\s(x)=&-\log |\det J(x)|,\qquad{\rm discrete\ evolution} \cr
  \s(x)=&-{\rm div} f(x)\equiv -{\rm Tr} J(x), \ {\rm continuous\ 
    evol.}\cr}\Eq{e2.1}\ee
Changing variables or the metric on phase space implies a change of $\s(x)$
into $\s(x)+ u(S_rx)-u(x)$ for a suitable function $u(x)$:\footnote{\small\eg a
  change in variables $y=w(x)$ leads to $u(x)=-\log |\det
  \dpr_{x_j}{w_i}|$.}  which implies that the time average
$\lim_{T\to\infty} \frac1T\sum_{j=0}^T \s(S_r^jx)$, if existing, does not
depend on the metric used on $M$. Likewise and with the same implication on
the averages, in the continuous evolution systems, $\s(x)$ changes into
$\s(x)+\dot u(x)$ (where $\dot u(x)\defi\sum_j \dpr_{x_j} U(x) \,f(x)_j$
for a suitable function $U(x)$).\footnote{\small\eg changing $x$ into $y=w(x)$
  leads to $U(x)= -\log |\dpr_{x_j}w_j|$.}

The time average $\s_+$ of $\s$ \ie of $n\to\s(S^nx)$ or $t\to\s(S_tx)$
coincides, except for a set of $x$'s with $0$ volume, with the average
$\s_+=\int \s(x)\m(dx)$ with respect to the SRB distribution $\m$.

Remark that $\s_+$ is a quantity which has the dimension of an inverse time
in the case of continuous systems while it is dimensionless for maps, (as
time is an integer in the case of maps). It will play an important role in
the following, particularly when $\s_+\ne0$, as it sets a time
scale\footnote{\small Since the CH-evolutions that we consider proceed towards a
  bounded attracting set it is $\s_+\ge0$, \cite{Ru996}.} that will be
called the {\it dissipation time scale}.

\def\SEC{\ Symmetries, (time reversal)}
\section{\SEC}
\label{sec3}
\iniz

Trying to evince information from the statistical properties of the stationary
distributions of a time evolution, discrete or continuous, it is
important to take into account symmetries of the underlying equations of
motion, \cite{DGM984}.

There are not many such symmetries and a key role is played by the
fundamental symmetries, like translation and rotation invariance or time
reversal, often enjoyed by the molecular constituents of the systems of
interest and perhaps reflected by their macroscopic properties.

Of particular interest will be systems in which the evolution is $S$,
discrete in time, or is a continuous time evolution $S_t$, which satisfies a
``{\it time reversal\,}'' symmetry, \ie such that there is an
$\infty$-smooth map $x\to Ix$ on the phase space $M$, {\it independent or
  smoothly dependent} on any parameter that might affect the dynamics, with
the property

\be\eqalign{
 &(a)\  I S^{-1}=S I,\qquad I^2=1, \qquad {\rm discrete \ evol.}\cr
  &(a')\ IS_{-t}=S_t I,\qquad I^2=1, \qquad {\rm cont.\ evol.}\cr
  &(b)\ I \ {\rm is\ isometric}\cr
}\Eq{e3.1}\ee
For isolated particle systems $I$ is just the reversal of all velocities
and it is a basic law of nature in Newtonian physics.

A typical situation is described in the following section presenting a
simple, but quite general, model of a nonequilibrium system.

The model will also illustrate the notion of {\it phase space contraction}
and its relation with the thermodynamic notion of {\it entropy
  generation}. It will appear that although there is a relation between
entropy creation rate and phase space contraction, still the two notions
are quite different. {\it Nevertheless} their difference can be expressed
as a variation of a suitable phase space observable evaluated at successive
map iterations or, in the cases of flows, as a {\it time derivative} of a
suitable observable: therefore it has no influence, or a controlled one,
on the {\it average phase space contraction}, \cite[Ch2.5]{Ga013b}.

\def\SEC{\ Example: reversible dissipation}
\section{\SEC}
\label{sec4}
\iniz

The system consists in $N\equiv N_0$ particles in a container
$\CC_0$ and of $ N_a$ particles in $n$ containers $\CC_a$ which
play the role of {\it thermostats}: their positions will be
denoted $\V X_a,\,a=0,1,\ldots,n$, and $\V X\defi(\V X_0,\V
X_1,\ldots,\V X_n)$.  Interactions will be described by a
potential energy

\be
W(\V X)=\sum_{a=0}^{n} U_a(\V X_a) +\sum_{a=1}^n W_a(\V X_0,\V X_a)
\Eq{e4.1}\ee
{\it i.e.} particles in different thermostats only interact indirectly, via
the system. All masses will be $m=1$, for simplicity.

\eqfig{110}{80}{}{fig2}{}

\0{\small Fig.1:\label{Fig1}\it The reservoirs occupy finite regions outside
  $\CC_0$, \eg sectors $\CC_a\subset R^3$, $a=1,2\ldots$. Their particles are
  constrained to have a {\it total} kinetic energy $K_a$ constant, by
  suitable forces $\V F_a$, so that the reservoirs ``temperatures'' $T_a$,
  are well defined, by $K_a=\sum_{j=1}^{N_a} \frac12\, (\dot{\V
    X}_{a,j})^2\defi \frac32 N_a k_B T_a\defi \frac32 N_a\b_a^{-1}$.  The
  set-up, classical and quantum, is introduced in \cite{FV963}.}

\vskip2mm
Particles in $\CC_0$ may also be subject to external, possibly
non conservative, forces $\V F(\V X_0,{\V E})$ depending on a few strength
parameters ${\V E}=(E_1,E_2,\ldots)$. It is convenient to imagine that
the forces due to the confining potentials determining the geometrical shape
of the region $\CC_0$ are included in $\V F$, so that one of the
parameters is the volume $V=|\CC_0|$. See Fig.1.

Following Sec.\ref{sec1} the statistical properties of the stationary
states of the system should be described, assuming the CH, by the SRB
distributions $\m_{\V E}$ on phase space.

The equations of motion are:
\be\eqalign{
\ddot{\V X}_{0i}=&-\dpr_i U_0(\V X_0)-\sum_{a}
\dpr_i W_a(\V X_0,\V X_i)+\V F_i\cr
\ddot{\V X}_{ai}=&
-\dpr_i U_a(\V X_a)-
\dpr_i W_a(\V X_0,\V X_i)-\a_a \dot{\V X}_a
\cr}\Eq{e4.2}\ee
where the last term $-\a_a \dot{\V X}_a$ is a {\it phenomenological} force
that implies that thermostats particles keep constant total kinetic energies
$K_a=\frac32 N_a k_B T_a$: $\a_a$ is therefore (as checked by
direct computation of the time derivative of $K_a=\frac12\, \dot{\V
  X}_{a,j}^2$ defined by
\be\a_a \,\defi\,\frac{L_a-\dot U_a} {3N_a k_B
    T_a}\Eq{e4.3}
\ee
where $L_a=-\partial_{\V X_a} W_a(\V X_0,\V X_a)\cdot \dot{\V X}_a$ is
the work done per unit time by the forces that the particles in
$\CC_0$ exert on the particles in $\CC_a, a>0$; here $k_B$ denotes
Boltzmann's constant.

The exact form of the forces that have to be added in order to insure the
kinetic energies constancy should not really matter, within wide
limits. But this is a property that is not obvious and which is much
debated.\footnote{\small The above thermostatting forces choice
  can be seen to coincide with the ones obtained via Gauss' {\it
    least effort} principle for ideal anholonomic constraints applied to
  the constraints $K_a=const$, see
\cite[Ch.2]{Ga013b}: this is a criterion that 
has been adopted in several simulations,
\cite[Sec.5.2,p103]{EM990}. {\it Independently} of Gauss' principle it is
immediate to check that if $\a_a$ is defined by Eq.\equ{e4.3} then the
kinetic energies $K_a$ are, strictly, constants of motion.}

The work $L_a$ in Eq.\equ{e4.3} will be interpreted as {\it heat} $\dot
Q_a$ ceded, per unit time, by particles in $\CC_0$ to the \hbox{$a$-th}
thermostat.  The
{\it entropy production rate} due to heat exchanges between the system and
the thermostats can, therefore, be naturally defined by
\be\s^0(\dot{\V X},\V X)\defi\sum_{a=1}^{N_a} \frac{\dot Q_a}{k_B
T_a}\Eq{e4.4}\ee
because the ``temperature'' of $\CC_a$ remains constant, and at
stationarity the thermostats can be regarded in thermal equilibrium.

It should be stressed that here {\it no entropy notion} is introduced for
the stationary state: only {\it variation} of the thermostats entropy is
considered and it should not be regarded as a new quantity because, in the
stationary states, the thermostats should be considered in equilibrium at a
fixed temperature.

A question is whether there is any relation between $\s^0$ and the phase
space contraction $\s$ of Eq.\equ{e2.1} for the equations of motion in
Eq.\equ{e4.2} (\ie minus the divergence of the equations
Eq.\equ{e4.2}). 

The latter, in the recent literature, has been identified
with the entropy production: and in the present case can be immediately
computed by the appropriate differentiation of Eq.\equ{e4.2},\equ{e4.3} and
is (neglecting $O(\min_{a>0} N_a^{-1})$) 
\be\txt\s(\dot{\V X},\V X)={\mathop\sum\limits_{a>0}}
{\txt\frac{3N_a-1}{3 N_a}} \frac{\dot Q_a-\dot U_a}{k_B T_a} =
{\mathop\sum\limits_{a>0}} \frac{\dot Q_a}{k_B T_a}-\dot U
\Eq{e4.5}\ee
where $U=\sum_{a>0} \frac{3N_a-1}{3 N_a} \frac{U_a}{k_B T_a}$.  Hence in
this example, physically interesting, in which the thermostats are
``external'' to the system volume (unlike to what happens in several
examples in which they act inside the volume of the system), the phase
space contraction is not the entropy production rate, \cite[Ch.2]{Ga013b}.  {\it
  However it differs from the entropy production rate by a total time
  derivative}.

Consequence: entropy creation rate $\s^0$ and phase space contraction
$\s$ differ, {\it but their time averages coincide}.

This is relevant because the definition Eq.(\equ{e4.4}) {\it has meaning
independently of the equations of motions} and can, therefore, be
suitable for experimental tests, \cite[Ch.5-2.Ch.3,Ch.4]{Ga013b}. 

It should be stressed that the numbers $N_a$ of particles in the reservoirs,
$a>0$, enters through $\frac{3 N_a-1}{3N_a}$, hence is is essentially
independent on the thermostat sizes (provided large).

Finally I mention that the identification, up to a total time derivative,
of phase space contraction with entropy production rate can be shown, as
discussed in Sec.\ref{sec7}\,,\,\ref{sec15} below, to cover the entropy
production rate in systems whose evolution can be approximated by
macroscopic continua equations, like fluids described by Navier-Stokes
equations. In the sense that, again, phase space contraction of the
interacting particles systems that underlay the macroscopic equations is
related, in the stationary states, to the entropy production rate
independently defined in classical nonequilibrium Thermodynamics,
\cite{DGM984}: and, at most, it differs from it by a total time derivative,
\cite[Ch.4]{Ga013b}.

\def\SEC{\ Hamiltonian dissipation?}
\section{\SEC}
\label{sec5}
\iniz

No entropy production is possible in a stationary state of a
Hamiltonian system (\ie an isolated system). At least not if it is
finite. However things are different when the system is in contact
with infinite systems.

As an example consider a Hamiltonian version of the model of
Sec.\ref{sec4}, Fig.1 above.  If the constraints on the kinetic energy in
the containers $\CC_a, a=1,\ldots,n$, are removed and the containers are
extended to infinity, interesting stationary states can be obtained from
initial configurations which, in each $\CC_a$, are naturally chosen from
the {\it canonical equilibrium ensemble} with density $\r_a$ and
temperature $T_a$ and from any distribution\footnote{\small With density on
  the phase space $\CC_0\times R^{3N_0}$.} for the particles in
$\CC_0$. The initial distribution will be called $\m_0$ and, although not
invariant in time, it may evolve towards an invariant one, $\m$ (as
reasonable as this looks, however, a mathematical proof of this is {\it far
  from known}). The existence of $\m$ will be assumed in this example.

Since the containers are infinite the stationary state that will be reached
can be expected to keep an average kinetic energy per particle remaining
$\frac32 k_B T_a$, identically equal to the initial value. For rather
general models of microscopic interaction between particles, it can be
shown, see for instance \cite{GP010b}, 
that the time evolution of the particles in $\CC_a$ that are far from the
boundary of $\CC_0$, are little affected by the interactions with the
particles in $\CC_0$ and the average kinetic energy per particle in each
$\CC_a$ will be an exact constant of motion, equal to the initial
$\frac32k_B T_a$, for all finite times (although it is still possible that
in the limit of infinite time this might change).\footnote{\small The
  physical picture is that the energy generated by work performed by the
  active forces on the particles in $\CC_0$ and by the interactions between
  the particles in $\CC_0$ and those in the thermostats is ceded to the
  thermostats creating in them heat currents $J_a$ which
  decrease as the inverse of the square distance to $\CC_0$: so the
  thermostats remain asymptotically, as the distance from $\CC_0$ tends to
  $\infty$, in equilibrium.}

Consider as initial distribution $\m_0(dx)$ which is a product of
independent canonical distributions in each container $\CC_a,a\ge0$, with
given densities and temperatures $\r_a,T_a\equiv \frac1{k_B
  \b_a}$.\footnote{\small But the choice of $T_0$ has no particular physical
  meaning and the distribution in $\CC_0$ could be replaced by ``any''
  distribution, with some density on the $X_0,\dot X_0$ variables.}

Although {\it now purely Hamiltonian} the system is infinite
and the phase space volume {\it measured by the evolving distribution}
$\m_t(dx)=\m_0(S_{-t} dx)$, changes per unit time by $\s(x)$ with:

\be \s(x)=\frac{d}{dt} \log \frac{\m_0(S_{-t}dx)}{\m_0(dx)}\equiv
\sum_{a=1}^n \b_a \dot Q_a+\b_0 \dot Q_0\Eq{e5.1}\ee
where $\s(x)$ is computed from the equations of motion as:

\be \kern-3mm\eqalign{
  \dot Q_a=& L_a=-\dpr_{\V X_a} W_a(\V X_0,\V X_a)\cdot\dot {\V X}_a,
  \qquad a\ge1\cr
  \dot Q_0=&\dot K_0+\dot U_0\cr}\Eq{e5.2}\ee
and $\b_0 \dot Q_0=\b_0\,(\dot K_0+\dot
  U_0)=\b_0\,\V F\cdot \dot{\V X}_0-\sum_{j>0}(\dot U_{0j}-\dot Q_j))$ 
is a time derivative so that it {\it does not contribute} to the time average 
$\s_+=\lim _{T\to\infty}\frac1T\int \s(S_t x) dt$, \cite[Ch.4]{Ga013b}.
\*

\0{\it Remark:} If $\m_0$ is defined, as proposed above, as a product of
independent canonical distributions it might be surprising that the system
in $\CC_0$ plays a special role: could one write the same formulae with
$\CC_1$ playing the role of $\CC_0$ and find that $\dot Q_1$ is a total
derivative of $U_1+K_1$?
\\ However $U_1+K_1$ is infinite unlike $U_0+K_0$, so nothing can be
concluded about its time derivative. If the regions $\CC_j$ were finite
then the system would evolve and all $\dot U_a+\dot K_a$ would
uninterestingly average to $0$: so $\CC_0$ plays a special role and is the
only container for which $U_0+K_0$ and its time derivative are meaningful.
\*

The expression Eq.\equ{e5.1}, and the irrelevance of the contribution from
$\dot Q_0$ to the average of $\s$ (see Eq.\equ{e5.2}) is the key to the
interpretation of fluctuations on $\s$ in systems modeled by particles:
even in cases (essentially in all experimental settings) in which the
evolution is not describable in terms of equations of motion, in the sense
that the equations of motion are not analytically known.

In such cases the average of $\s(x)$ (equal to that of
$\s^0(x)=\sum_{a\ge1}\b_a \dot Q_a$) {\it is still accessible} via measurement
of the heats exchanged and the temperatures of the reservoirs with which
heat is exchanged. Of course it is a delicate and difficult task to measure
{\it all} such quantities, \ie the full entropy production.

It is remarkable, and it will be discussed in Sec.\ref{sec6}, that quite
generally, under the Chaotic Hypothesis, that it becomes possible to obtain
a general, universal, property of the (rare) fluctuations of the entropy
production.

\def\SEC{\ Fluctuation Relation (FR)}
\section{\SEC}
\label{sec6}
\iniz

The {\it Fluctuation Relation} deals with the entropy production
$\s^0(x)=\sum_a\frac{\dot Q_a}{k_B T_a}$, see \equ{e5.1},\equ{e5.2}, or
more generally, with the phase space volume contraction rate
$\s(x)=-\sum_i\dpr_{x_i} f_i(x)$ and it applies to {\it finite} systems
whose evolution takes place on a phase space $M$, via an equation $\dot
x=f(x)$, and is {\it time reversal symmetric} in the sense of
Sec.\ref{sec3} and hyperbolic on $M$ (\ie it is a Anosov
system).\footnote{\small A positive $\s(x)$ means that the volume {\it
    contracts} near $x$: hence in stationary states the average $\s_+$ of
  $\s$ must be $\ge0$, \cite{Ru996}.} It equally deals with evolutions
which are time reversible Anosov maps: attention will be mostly
concentrated on the continuous time case, to avoid repetitions.

The $\s(x)$ is defined in terms of the metric on $M$ and via appropriate
covariant derivatives $\dpr_{x_i}$: but it is simpler to imagine that $M$
is a Euclidean space with coordinates measured in prefixed units, so that
$\dpr_{x_i}$ are the usual partial derivatives.

In general $\s(x)$ {\it depends} on the metric and changing metric on $M$ the
expression of $\s(x)$ changes; however the variation can be, in general,
expressed by the time derivative of a suitable observable so that the long
time averages $\s_+$ of $\s(x)$ {\it do not depend on the metric}, like most
physically interesting observables (\eg the Lyapunov exponents). See
comments following Eq.\equ{e2.1} and Eq.\equ{e5.2}

Given an attracting set $\AA$ on which the evolution is an Anosov system,
\ie motion on $\AA$ is a smooth continuous hyperbolic
flow\footref{hyperbolic} $x\to S_tx$, or a smooth discrete hyperbolic
map\footref{hyperbolic} $x\to Sx$; let $\s_\AA(x)$ be the {\it surface
  contraction rate} on the surface $\AA$ and consider the
quantity

\vglue-6mm
\be \eqalign{
  p=&\frac1\t \int_0^\t \frac{\s_\AA(S_t)}{\s_+} dt \qquad {\rm continuous\ 
    time}, \ \t>0\cr
  p=&\frac1\t\sum_{k=0}^\t \frac{\s_\AA(S^kx)}{\s_+}\qquad{\rm discrete\ time},
  \ \t \ {\rm integer}
\cr}\Eq{e6.1}\ee
where $\s_+$ is the infinite time average of $\s_\AA(x)$, which is
$x$-independent, aside exceptional $x$'s in a $0$-volume set, and
coincides with the average of $\s$ with respect to the SRB distribution
$\m_{srb}$ on $\AA$, \ie the probability distribution generated by the motion of
any point $x$ chosen randomly with some density with respect to the volume,
\ref{sec1}.

Then consider the stationary  $\m_{srb}$-probability of the above variable $p$,
Eq.\equ{e6.1}, and define its {\it large deviation rate} as a function $\z(p)$
such that:

\be P_{srb,\t}(D)= e^{ \t\,\max_{p\in D}\z(p) +o(\t)}\Eq{e6.2}\ee
for all domains $D$ (\ie for all regions $D$ which are closures of their
interiors).

If the dynamical system is an Anosov system (or, more generally, if is satisfies
the ``axiom A'') then
\*
\0{(a)} the rate $\z(p)$ exists and is defined in the interior of an
interval $[p',p'']$ containing $p=1$,
\\(b) it is analytic in $p$ if $p'<p''$ while it is $-\infty$ for
$p\not\in [p',p'']$
\\(c) $\s_+=0$ if and only if the SRB distribution $\m$ admits a density
over the attracting surface, \cite{Ru996}. 

\* A simple universal result
for reversible Anosov systems, is the following {\it Fluctuation Theorem},
\cite{GC995},\cite{Si977,Ga995b}.
\*
\0FT: {\it If the evolution is time reversal symmetric then the rate
  function\footnote{\small The rate $\z(p)$ is defined so that the
    probability of finding $\frac1\t\int_0^\t \frac{\s(S_tx)}{\s_+}dt dt\in
    [p,p+\d p]$ is $\exp \t \max_{[p,p+\d p]}\z(p)$ for $p\in (-p^*,p^*)$
    where $p^*\ge1$; it exists and is analytic if $\s(x)$ is the phase
    space contraction of an Anosov evolution.}  verifies the symmetry
  property:
    \be \z(-p)=\z(p)-p \s_+\Eq{e6.3}\ee
for $p\in [-p^*,p^*]$ with $p^*\ge1$.
}
\*

An immediate consequence is that if the Chaotic Hypothesis is considered
valid, time reversibility holds and the attracting set can be supposed to
coincide with the full phase space (so that $\AA=M$ and $\s_\AA\equiv\s$),
then the entropy generation rate or more generally the phase space
contraction rate are expected to satisfy the large deviation property which
is, in this case, called {\it Fluctuation Relation}, FR, and it is {\it
  informally} written as

\be \log\frac{P_{srb}(p)}{P_{srb}(-p)}=\t\,1\, p\, \s_+\, +\, o(\t)
\Eq{e6.4}\ee
and more precisely formulated, as existence of a ``large deviation'' rate
$\z(p)$ satisfying Eq.\equ{e6.3}: where the $1$ is inserted for later
reference.

While the formal probability density for the events $\pm p$, \ie
$P_{srb}(\pm p)$, is a difficult quantity strongly dependent on the
dynamical system, the interest of the FR is that Eq.\equ{e6.3},\equ{e6.4}
are, under the above assumptions, an {\it exact symmetry} of $\z(p)$ and,
at least in some cases, FR {\it deals with a quantity ($\s_\AA(x)$) which
  has physical meaning} (entropy generation rate) and mathematical meaning
(phase space contraction rate): therefore a check of Eq.\equ{e6.4} can
become a test of the chaotic hypothesis.

The fluctuations relation is, for time reversible evolutions, a symmetry of
the SRB distributions. {\it However} it requires that:
\*
\0(i) the motion on the attracting surface $\AA$ has the Anosov property,
\\
(ii) and {\it at the same time} it is reversible; hence, in the frequent
cases in which $\AA$ is not the full phase space but just a smooth surface
in it, it should be $I\AA=\AA$, quite {\it unlikely} if $I$ is the usual time
reversal symmetry (\ie velocities reversal),\\
(iii) furthermore, if (i) and (ii) hold, the FR concerns the fluctuations
of the surface area of $\AA$, and {\it not of the full volume}: which is
very hard to access, as $\AA$ itself. \*
\0and the three conditions strongly limit a literal
applicability of FR and lead to the analysis of further properties of the
considered evolutions, see Sec.\ref{sec9}\,,\,\ref{sec17}.

Nevertheless it can be applied to systems that are only mildly out of
equilibrium. If the system, remaining time reversal symmetric, is set out
of equilibrium by the action of small forces and is in contact with
thermostats with small differences of the respective temperatures, call
$\e$ a parameter measuring the size of the forces and of the temperature
differences. Then, if for $\e=0$ the system has the Anosov property on the
full phase space, it will continue to have such property also for small
$\e\ne0$, because Anosov systems are structurally stable,
\cite{AA966}. Hence the attracting set $\AA$ will remain identical to the
full phase space and time reversal will be a symmetry of the motions on the
attracting set and the FR assumptions will remain verified. This was the
case of the systems to which the FR has been applied, \cite{GC995}, and
tested, \cite{BGG997}, to explain the fluctuations of the phase space
contraction observed in the simulation in \cite{ECM993}.

In attempting to test or use the FR in systems which are not very small it
is not reasonable to hope that the smallness of the above $\e$ does not
depend on the system size, although important cases (lattices of coupled
Anosov maps) are known in which $\e$ can be taken independent of the size,
\cite{BK995,GBG004}: therefore it might be thought that FR becomes
irrelevant in most interesting cases, \footref{H1}. Clearly more properties
are needed to deal with the systems that are not small perturbations of
Anosov systems.  In Sec.\ref{sec9} the {\it applicability far beyond the
  latter cases} will be discussed.  \*

\0{\it Remark:}\label{cil} Often the Eq.\equ{e4.5} raises the question
``how can it be relevant'' as the Boltzmann's constant in the denominator
is likely to give a huge value to the inverse time scale $\media{\s}$ which
determines the time scale over which the FR yields predictions? For
instance imagining to put $1 \,cm^3$ of steel (with faces of $1\,cm^2$) in
contact between two reservoirs at temperatures $T=300{}^oK$ and $T+\d
T=310{}^oK$ the average of the entropy production rate, ${\dot Q}\frac{\d
  T}{k_B T^2}$, can be expressed via the steel thermal conductivity $\ch$
as $\ch (\frac{\d T}T)^2 \frac{\D}{k_B}$: and the result is $\sim
10^{18}\,sec^{-1}$, see also \cite[p.4]{STX005}. If FR could be applied
{\it literally} there would be no way to see a heat flow from cold to warm
during $10^{-6} sec$ before ``trying'' to see it, say once every second,
for at least $\sim 10^{18}/{10^6}$ times (\ie $\sim 10^3$ billion
days). See Sec.\ref{sec9} for a possible answer to the problem.
\footnote{\small The problem is
considered, by some colleagues, a ``disaster'' for FR, making it
physically irrelevant.}  \*

The FR bears {\it formal} similarity with identities arising in the
evolution of equilibrium states, or more generally with the evolution of
initial distributions on phase space {\it which are symmetric under time
  reversal but not stationary}. The deep difference between the latter
identities and the above FR is briefly commented in Sec.\ref{sec19}) below.

Unfortunately the name ``fluctuation relation'' has been often used in all
cases, causing great confusion to loom on the subject.
\*

\def\SEC{\ Nonequilibrium ensembles. Ensembles equivalence}
\section{\SEC}
\label{sec7}
\iniz

In general given an evolution equation on a phase space $M$ depending on
one or more parameters, denoted $\V E=\{\n,E,\ldots\}$, the SRB stationary
states, \ie the distributions that are generated by all points of $M$,
excepting a subset of $M$ with $0$ volume, form a collection $\EE^c$ of
probability distributions $\m_{\V E}$ parameterized by the given parameters
and each of which can be called an ``ensemble''.

Even in the cases, considered in this section, in which there is only one
parameter $\n$ and CH holds, there might be several distinct attracting
surfaces and therefore more than a single SRB distribution $\m_\n$: if so,
further parameters will have to be added to distinguish the various
possibilities, {\it just} as done in equilibrium statistical mechanics in
presence of phase transitions to distinguish the different pure phases,
\cite{Ru969,LR969,Ga000}.

A key question is whether the same system can be described by {\it
  different} equations of motion. There are several instances in which this
is possible: for instance a fluid motion can be equally well described by,
say, a Navier-Stokes equation or by a
(far more complex) collection of molecules, in contact with a thermostat and
at given density, at least if attention is given to
observations depending on large scale properties and performed over long
time scales, \cite{Ma867-b}.

Even for Navier-Stokes (NS) fluids there might be several different
equations, simpler than the ultimate molecular models, that can describe
the class of phenomena considered relevant in given physical situations.

For instance it has been convincingly argued that macroscopic transport
coefficients can be obtained by replacing the equations of motion of
molecules by simple(r) models, suitable for simulations, obeying modified
equations of motion which can even be non-Newtonian: in the context of
molecular simulations this has been originated in the early '80s,
\cite{Ho999,EM990}. A first example of equations alternative to the NS
equations to describe a developed turbulence flow is found in \cite{SJ993}.

It is natural to consider, together with the collection of SRB
distributions $\m_\n\in\EE^c$ for a given equation depending on a parameter
$\n$, the collection $\m'_E\in\EE'$ of SRB distributions corresponding to a
different equation parameterized by a new parameter denoted $E$, which on
physical or just heuristic grounds describes equivalently the same class of
phenomena, \ie predicts the same properties for large classes of
observables.
\*

\0{\it Remarks:} (i) The equivalence should mean that it is possible to
establish a correspondence between the ensembles (\ie the distributions) in
$\EE^c$ and the ones in $\EE'$ so that for each $\n$ there is a
corresponding $E(\n)$ and the average of ``most'' observables in the
$\m_\n\in \EE^c$ and $\m'_{E(\n)}$ in $\EE'$ should
coincide (or be close).
\\
(ii) This is quite analogous to the description of equilibrium states in
Statistical Mechanics (SM): the canonical distribution $\m^V_\b$ of $N=\r V$
molecules of a gas in a container of volume $V$ depends, at fixed density
$\r$, on a parameter (inverse temperature) $\b=(k_B T)^{-1}$ and the
microcanonical distribution $\m^{'V}_E$ depends on a parameter (total energy)
$E$. If $E$ and $\b$ are so related that
\vglue-6mm
\be\m^{'V}_E(\sum_{i=1}^N\frac1{2m} p_i^2)=\frac32 N \b^{-1}\Eq{e7.1}\ee
then the average of ``many'' observables $O$, $\m^{'V}_E(O)$, is equal or
close to $\m_\b(O)$.  \\
(iii) Actually, in SM, in the limit as
$V\to\infty,\ \r=N/V\ {fixed}$, for any {\it local observable}. \ie
depending only on the configuration of the molecules located in a finite
region, the canonical and microcanonical averages are not only close but
{\it strictly equal}, at least in absence of long range forces or of phase
transitions, \cite{Ru969,Ga000}.  \\
(iv) And, still considering (SM), in presence of phase transitions at $\b$
it will be necessary to label the distributions in $\EE^c$ by further
parameters $a$: in this case the distributions in the ensemble $\EE'$ will
also have to be distinguished by an equal number of parameters $a'$ and a
correspondence between $a$ and $a'$ can be established so that, under the
condition Eq.\equ{e7.1}, it is still $\m^{'V}_{a',E}(O)=\m^V_{a,\b}(O)$, for
local observables, in the
limit $V\to\infty$.  \*

A first example is obtained by considering a system described by equations
on $x\in R^n$ which are obtained as follows\*

\0(a) let $\dot x= G(x)$ be a time
reversible equation for the time reversal $Ix=-x$ (\ie $G(x)=G(-x)$) \\
(b) add a reversible forcing $f(x)$, with $If=fI$ (\ie $f(x)=f(-x)$)
\\
(c) and a 
dissipative term, $-\n Lx$, with $L$ linear and positive $(Lx\cdot
x>0$, if $x\ne0$) depending on a parameter $\n$, whose effect is to
balance, in the average, the ``energy'' injected by the forcing
\*

The complete equation has therefore the form

\be \dot x=G(x)+f(x)-\n L x \Eq{e7.2}\ee
At fixed $f$ and for each ``friction'' $\n>0$ small enough, the evolution
will be supposed to satisfy the CH and to lead to a unique
stationary (``SRB'') distribution $\m_\n$.

The collection of the SRB distributions $\m_\n$, as $\n$ varies, will be
denoted $\EE^c$, and each of them defines a {\it nonequilibrium ensemble}.

Next consider a {\it different} equation in which the friction coefficient
$\n$ in Eq.\equ{e7.2} is replaced by a multiplier $\a(x)$ so defined that a
selected observable $\O$ is an exact constant of motion. For instance the
cases $\O(x)= x^2$ or $\O(x)=(x\cdot Lx)$ lead to new equations of motion
$\dot x=G(x)+f(x)-\a(x) L x$ with, respectively:

\be \eqalign{
  \a(x)=&\frac{x\cdot G(x)+ x\cdot f(x)} {x\cdot Lx}\qquad{\rm or}\cr
  \a(x)=&\frac{Lx\cdot G(x)+ Lx\cdot f(x)} {L x\cdot Lx}\cr}
   \Eq{e7.3} \ee
 Then the stationary states for the new equations form a collection of
 stationary states $\EE'$ with elements parameterized by the value $E$ of
 the constant of motion $\O$ introduced by the multiplier $\a$.
 
   Quite generally the motion generated by the new equations is eventually
   restricted to a bounded region, because of the action of the friction
   and of conservation laws possibly valid for the time reversible system in
   absence of forcing.

Therefore for $\n$ small it can be expected that in the stationary states
$\a(x)$ fluctuates leading to a {\it homogenization phenomenon}, \ie to the
property that in the stationary state for the new equation
\be \dot x=G(x)+f(x)-\a(x) L x\Eq{e7.4}\ee
large classes of observables have the same averages in the distribution
$\m^c_\n$ and in the distribution $\m'_{E}$ belonging to the new ensemble
$\EE'$, of stationary distributions for Eq.\equ{e7.4}, {\it provided} $\n$
and $E$ are kept related by $\n=\m'_{E}(\a)$:
\be \lim_{\n\to0} \m^c_\n(O)= \lim_{\n\to0} \m'_{E}(O),\Eq{e7.5}\ee
or, equivalently, if $E=\m_\n(\O)$.  More formally: \*

{\it If motions following Eq.\equ{e7.2} eventually develop on a ball $M$
  generating a family $\EE^c$ of stationary distributions parameterized by
  $\n$ and if the motions following the Eq.\equ{e7.4} are also eventually
  confined in a ball $M'$, generating a family $\EE'$, then Eq.\equ{e7.5}
  holds for arbitrarily fixed observables $O$, provided the correspondence
  between $\m_\n^c$ and $\m'_{E(\n)}$ is such that $\n=\m'_{E(\n)}(\a)$ or,
  equivalently, $E(\n)=\m_\n(\O)$.}  \*

As in the case of ensembles equivalence in equilibrium not all ensembles
are equivalent, not even in the thermodynamic limit, therefore the
observables $\O$ defining the ensemble have to be selected on a case by
case basis.

The above statement has been tested in a few cases: involving strongly
truncated NS equations, \cite{GRS004},\footnote{\small Doubts have been
  raised in \cite{RM007}: which might be related to the use of a rather
  large value of $\n$ in a strongly truncated NS equation in 2D: it is
  hoped that the latter results will be tested again at smaller $\n$ (in
  spite of computational difficulties).} Lorenz96 equations, \cite{GL014},
shell model for turbulence, \cite{BCDGL018}.

The conjecture will be analyzed in some detail, and considerably
strengthened, in Sec.\ref{sec17} for the stationary states of the
incompressible NS equation with periodic boundary conditions. But it is
convenient to discuss first in which sense the FR can be made relevant for
systems irreversibly evolving in presence of strong friction, and to
exhibit a few more applications of the FR to classical and new problems.

In particular a {\it key problem} is whether the FR can be of any utility
if the attracting set $\AA$ is a surface of dimension lower than that of
$M$ and, although the evolution equations remain reversible, it is
$I\AA\ne\AA$, \ie reversibility does not hold as a symmetry for motions on
the attracting set (as, instead, required for the validity of the FR), see
comments (i-iii) in Sec.\ref{sec6}\,.

\def\SEC{\ Strong Dissipation: attracting set size. Lyapunov pairs.}
\section{\SEC}
\label{sec8}
\iniz

As forcing and dissipation increase the attracting set $\AA$ may become a
small subset of phase space: and, if the CH holds, it
becomes a smooth surface of dimension {\it lower} than the full dimension
of phase space.

In this case although the motion on $\AA$ is a Anosov system
it may appear,  at first, that it does not even make sense to ask whether
a FR holds because:

\* \0(1) it is not possible to think that it could express properties of
the volume contraction on $\AA$: the main difficulty is that time
reversal symmetry, essential for the FR, is lost since, even if the
equations of motion are time reversible, the action of the time reversal $I$
will likely map the attracting set $\AA$ into a ``repelling'' set $I\AA$
disjoint from $\AA$.  The motion on $\AA$ remains hyperbolic, as assumed by
CH, but it no longer has the desired symmetry in time.  \\
(2) Furthermore the FR deals with the volume contraction on the full phase
space but the hyperbolic character (assumed by the CH) of the motion on the
attracting set could establish, {\it if} for some reason a new time
reversal $\wt I$ were spawned as a symmetry on $\AA$, a property of the
contraction of surface elements in $\AA$; however their analysis would
require a, highly unlikely, detailed understanding the geometry of the
attracting surface $\AA$.  \*

The latter two, seemingly insurmountable, difficulties are however
intertwined and tend, in several cases, to ``compensate''. We begin with
a simple case.

A remarkable property was discovered for a Hamiltonian
evolution with $n$ degrees of freedom for $x=(\V p,\V q)$ with $H(\V p,\V
q)=\frac12\V p^2+V(\V q)$ and {\it subject also} to a friction force $-\n\V
p$.  Namely, under very general conditions on the potential $V$ (typically
just boundedness of the surfaces $H=const$), the Lyapunov exponents
$\l_0\ge\l_1\ge \l_{d/2-1}\ge\ldots\l_{2n-1}$ of the motion are such that,
\cite{Dr988},
\be \frac12(\l_j+\l_{2n-1-j})=-\frac\n2\qquad j=0,\ldots,n-1\Eq{e8.1}\ee
In other words the symplectic symmetry of the Hamiltonian systems (which in
absence of friction implies $\l_j+\l_{2n-1-j}\equiv0$) leaves, in presence
of friction, Eq.\equ{e8.1} as a ``remnant'', at least {\it if the friction
  force has the simple form} $-\n\V p$. Furthermore Eq.\equ{e8.1} holds
identically for the eigenvalues $\l_j(x)$ of the Jacobian matrix $J(x)$ of
the flow at each point $x$.

{\it Remarkably} the relation Eq.\equ{e8.1} has been extended, \cite{DM996}, to
the time reversible cases in which the friction $-\n \V p$ is replaced by a
force $-\a(\V p,\V q) \V p$ with the multiplier $\a$ such that evolution
conserves the total kinetic energy $\frac12\V p^2$ exactly, as in some of
the simplest thermostat models, \cite{CELS993}, \ie $\a=-\frac{\V
  p\,\cdot\,\dpr_{\V q} V(\V q)}{\V p^2}$.

In the latter systems Eq.\equ{e8.1} not only holds with $\n$ replaced by
the time average of $\a$ but it follows from the stronger property that the
evolution $t\to S_tx$ is such that, given $t_0>0$, the matrix $W=\dpr_i
(S_{t_0} x)_j$ has the property that the logarithms of the
eigenvalues of $(W^TW)^{\frac12}$
are $t_0$ times
$\l_{t_0,j}(x)>0,\,j=0,\ldots 2n$ (depending on $t_0$), which satisfy
$\frac12(\l_{t_0,j}(x)+\l_{t_0,2n-1-j}(x))=$ $-\frac\n2$ or respectively
$-\frac12\media{\a}$, where the average
is intended over $S_tx$ for $t\in[0,t_0]$.%
\footnote{\small The proof of the pairing symmetry in the above mentioned
  cases is that the Jacobian matrix $\dpr_i (S_t x)_j|_{t=0}$ is seen to be
  the sum of the Jacobian for the Hamiltonian flow of $H(\V p,\V
  q)-\frac\n2 pq$ plus the identity times $-\frac\n2$,
  \cite{Dr988}.  In the case of $\a$ a similar property holds replacing
  $\n$ with $\a$, as is seen via a calculation.  If $J(t)=\dpr_i (S_t x)_j$
  then $J(t)PJ(t)^T=P e^{-n\n t}$ (or $P e^{-n\int_0^t\a(x(t))dt}$) where
  $P=\pmatrix{0&{\bf 1}\cr{\bf-1}&0\cr}$ (${\bf 1}$ being the $n\times n$
  identity) because $J(t)\defi J_0e^{-\frac12 \n t}$ with $J_0$ a symplectic
  matrix so that $ J_0PJ_0^T=P$ (proposition 24, Sec.3.12 in
  \cite{Ga983}). Therefore let $v$ be an eigenvector $J_0^TJ_0 v=\l v$ then
  the following chain of identities, using $P^2=-1$ shows that $\l^{-1}$ is
  an eigenvalue, with eigenvector $Pv$:
  $$\eqalign{&J_0^TJ_0 v=\l v \ \to\ 
    P J_0^TJ_0 v=\l P v\ \to\ 
    -J_0^{-1}P J_0 P P v=\l P v \cr
    &\to
  \ J_0^{-1}J_0^{-1T} Pv=\l Pv\ \to\ (J_0^T J)^{-1} Pv=\l Pv\cr}$$
  impliyng pairing to $-\frac\n2$ (respectively to $-t^{-1}\int_0^t\frac12
  \a(x(t))dt$) for the matrix $(J(t)^TJ(t))^{\frac12}$.
  \cite{Dr988,DM996}.\label{OR}}

Eq.\equ{e8.1}, called {\it pairing symmetry}, is certainly very special,
\cite{Ho999}, but it suggests, \cite{BGG997}, that the dimension of the
attracting set $\AA$ is equal to {\it twice the number of non negative
  Lyapunov} exponents: because it suggests that the pairs with two negative
exponents simply correspond to the phase space compression in the
directions that ``stick out'' of the surface $\AA$.

If so the latter directions certainly do not contribute to the contraction
of the surface of $\AA$.\footnote{\small Failure to realize the difference
 between Anosov motions on the attracting surface versus Anosov motions on
 the entire phase space in chaotic systems satisfying the CH together with
 time reversal symmetry is responsible for statements,
 \cite[p.220]{Ho999}, like: ``{\it If there {\it were} such systems then
it could be proved that they would generate relatively simple
attractors, with equal numbers of positive and negative Lyapunov
exponents. Because the simple geometric argument of Section 7.8 shows
that nonequilibrium attractors are actually generated by {\it any}
stable time reversible, steady dynamics, the applicability of the
Anosov proofs is evidently rare to vanishing}'',
\cite{GC995},\cite{BG997},\cite{Ga998}.}

An arbitrary number of negative exponents\label{Hoo} can be added to any
spectrum by adding arbitrarily many dimensions whose coordinates contract
to $0$. It is only if there is a pairing symmetry that the negative pairs
can be {\it conjectured} to be unambiguously identified: and it can be
hoped that the same remains valid if the pairing is only approximate, which
is a property that is {\it often encountered}, see Sec.\ref{sec18} for
examples.

This idea has been discussed in the analysis of
a simulation dedicated to tests of the CH and FR in a system with
pairing symmetry, \cite[Sec.6]{BGG997}, and its relevance for strongly
dissipative systems like the Navier-Stokes flows has been proposed in
\cite[Sec.5]{Ga997b}, see Sec.\ref{sec18} below.

\*
\0{\it Remarks:}
(1) Accepting the above proposal, the dimension of the attracting surface $\AA$
is determined when the system has a (possibly approximate) pairing
symmetry, and it is identified as twice the number of non negative Lyapunov
exponents.
\\
(2) It is worth stressing the general difference between the latter dimension,
that will be called {\it fluctuation dimension} of $\AA$ (or {\it
  fd-dimension}), and the Kaplan-Yorke dimension of $\AA$: the Kaplan-Yorke
dimension (or {\it ky-dimension}) is a measure of the fractal properties of
the SRB attractor contained in the attracting set $\AA$ and it is not
larger than the fluctuation dimension; with which it coincides if the
SRB distribution has a density on phase space. In general the ky-dimension
is a fraction of the fd-dimension.
\\
(3) The above discussion, {\it heuristically} proposes how to determine the
dimension of the attracting surface under the CH when the pairing symmetry
holds.
\*

However it is unclear whether the pairing symmetry, exact or approximate, can be
of any help to address the second of the above difficulties, \ie the lower
dimension of the attracting set and the accompanying breakdown of time
reversal symmetry for the motions confined to $\AA$, the only ones of
statistical interest, and the consequent apparent irrelevance of the phase
space contraction $\s(x)$, to which {\it also} contribute the contracting
directions sticking out of the attracting surface.

The second of the two difficulties mentioned is addressed in the next section,
on the basis of the proposal in \cite{BG997,Ga997b}.

\def\SEC{\ Dissipation. Time Reversal \& FR.}
\section{\SEC}
\label{sec9}
\iniz

Consider a time reversible evolution depending on a forcing parameter and,
still assuming the CH, suppose the forcing, hence the dissipation, to grow
so strong that the attracting set $\AA$ becomes a surface of dimension
smaller than that of phase space. 

Then the time reversal symmetry $I$ is
{\it spontaneously broken} in the sense that it ceases to be a symmetry for
the motions that develop on the attracting set. It remains a symmetry for
the motions in phase space, but it has little relevance for the statistical
properties (with respect to the SRB distribution) of the motions because,
asymptotically, they are attracted to $\AA$. The time reversal
image $I\AA$ of $\AA$ is quite generally a {\it repeller} and no motion
(except a set of data with $0$ volume) evolves towards it.

Therefore a natural question is whether the continuing existence of the
global time reversal symmetry $I$ can be accompanied by a map $\wt I$ of
$\AA$ to itself which is still a smooth isometry,
with $\wt I^2\equiv1$ and $S_t\wt I=\wt I S_{-t}$ in the flow case or $S\wt
I=\wt IS^{-1}$ in the case of maps.

The question has been analyzed in \cite{BG997} where a {\it geometric
  property} has been identified which, when holding, shows that a ``local
time reversal symmetry'' $\wt I$, defined as a map of the attracting set
$\AA$ into itself, is {\it spawned out} of a global time reversal symmetry $I$,
as a parameter varies and changes the dimension of $\AA$ making it a
surface of dimension smaller than the dimension of the phase space $M$.

The latter property will seem at first sight quite special. However it is
enjoyed by a class of systems of interest in applications and at the same
time is a {\it structurally stable property} (\ie it remains valid under
small perturbations of the dynamics). The property was named ``Axiom C''
because it is a modification of the ``Axiom B'' property introduced in
\cite{Sm967}.

To visualize the geometry of the Axiom C property consider the simpler case
of a  time reversible map $S$ and imagine that the attracting
surface $\AA$ becomes disjoint from its time reversal image $I\AA$, because
a parameter controlling the evolution is raised above a critical value, see
Fig.2.

Then the stable manifolds of the points in $\AA$ are not entirely contained
in $\AA$ but extend out of $\AA$ and intersect $I\AA$ on manifolds which
are {\it unstable manifolds} for the points of $I\AA$. Likewise the
evolution $S^{-1}$ will have $I\AA$ as a attracting set out of which the
$S^{-1}$-stable manifolds of the points of $I\AA$ emerge and extend until
they intersect the surface $\AA$ on its unstable manifolds.

So out of each point $x$ of $\AA$ emerge two manifolds intersecting $\AA$
respectively on the contracting and expanding manifolds at $x$ restricted to
$\AA$ for $S_t$ and {\it at the same time} the two manifolds intersect also
$I\AA$ and the intersections are the stable and unstable manifolds for
$S$ at some point $x'=Ix$ (linked by a $1$-dimensional curve).

The correspondence $x'=Px$, thus established between $\AA$ and $I\AA$, {\it
  commutes} with the time evolution, because the manifolds whose
intersection defines the correspondence $P:I \AA\otto\AA$ are covariant
under the action of $S$: hence $PSx=S Px$ for all $x\in\AA$ or $x\in
I\AA$.  \*

\eqfig{220}{80}{
\ins{26}{9}{$x$}
\ins{83}{69}{$x'$}
\ins{145}{11}{$x$}
\ins{190}{15}{$\AA$}
\ins{190}{70}{$I\,\AA$}
\ins{144}{63}{$\wt I\, x$}}{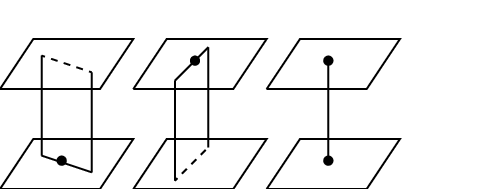}{}
\*
\0{\small Fig.2: \it Case of a map $S$. The first figure in Fig.2
  illustrates a point $x\in \AA$ and its attracting manifold, and a local
  part of its stable manifold that extends until $I\AA$ intersecting it in
  the hatched line (stable manifold on $\AA$ and unstable on
  $I\AA$). Likewise the second figure describes a point $x'$ on $I\AA$ with
  a local part of its stable manifold for $S^{-1}$ (extending to intersect
  $\AA$ on a unstable manifold, hatched). The third figure shows the
  ($1$-dimensional) intersection between the stable manifold of a point
  $x\in \AA$ and the unstable manifold of the point $x'\in I\AA$: in the
  figure such intersection is a unidimensional curve that connects $x$ with
  $x'$ (uniquely determined by $x$) {\it establishing the correspondence
    $P:\AA\to I \AA$ defining $P$}, with $x'=P x$. See caption to Fig.2.}  \*

The picture requires a few assumptions of technical nature to avoid
occurrence of some more complex possibilities (for instance it is necessary
to {\it exclude} that the contracting manifold emerging from $\AA$ wraps
around $I\AA$ rather than meeting it transversally): the mathematical
definition of the ``axiom C'' property can be found in \cite{BG997} and is
a modification of the notion of ``axiom B'', \cite{Sm967}.

A consequence is that the map $\wt I=P I$ maps $\AA$ into itself (as well
as $I\AA$ into itself) and is a time reversal symmetry for the restriction
of $S$ to $\AA$ (and to $I\AA$).

The above analysis exhibits a structurally stable mechanism, \cite{BG997},
which, if holding, implies that although time reversal is {\it lost as a
  symmetry} on an attracting set $\AA$, it might be accompanied by a new
map $\wt I$ on $\AA$ which can be regarded as a {\it new} time reversal
symmetry for motions evolving on $\AA$. Therefore it is interesting to see
whether a FR can also be established: heuristic ideas about such question,
with attention to a few possible applications will now be presented in the
rest of this section.

In presence of a pairing symmetry to the level $-\frac12\n$, Eq.\equ{e8.1},
suppose that the pairs of negative exponents describe the approach to the
attracting set and call $n_+$ the maximum number of {\it non negative}
Lyapunov exponents. Then the local exponents with labels $j=0,\ldots,n_+-1$
and the corresponding negative ones contribute
$\s_\AA(x)=\sum_{j=0}^{n_+}{\l_j(x)+\l_{2n-1-j}(x)}=n_+\n$ to the phase
space contraction and $n-n_+$ pairs of negative exponents should be
discarded in computing $\a_\AA$; so that the total average phase space
contraction on the attracting set $\s_{\AA,+}$ will be proportional to the
total average phase space contraction (\ie average $\s_+$ of minus the
divergence of the equation of motion) $\s_{\AA,+}=n_+\n=\frac{n_+}{n} n\n=
\frac{n_+}{n} \s_+$.

Remark that the number $n-n_+$ is defined in terms of the Lyapunov exponents:
hence it does not depend on the point $x$.  The conclusion is that in
systems with {\it time reversal and pairing symmetry} satisfying the CH and
axiom C, a fluctuation relation for the surface contraction of $\AA$,
$\s_\AA(x)=\frac{n_+}n\s(x)$, holds.%
\footnote{\small It would seem that smoothness
  of $\wt I$ should also be required because the axiom C implies only
  H\"older continuity for $\wt I$, see \cite{Po010}. However a careful
  examination of the FT proof shows that it is sufficient that the
  restrictions to $\AA$ of the stable and unstable manifolds of the points
  $a\in\AA$ are smooth manifolds and this is implied by the assumed
  smoothness of $\AA$ itself (by CH) and by the smoothness of the global
  manifolds.}

Set $\frac{n_+}n\equiv \frac{\NN_{attr}}{\NN}$ with $\NN_{attr}=$
dimension of the attracting surface and $\NN=$ dimension of the phase
space. Then the CH combined with Axiom C and a parity property will give,
for the probability of  $\t^{-1}\int_0^\t \frac{\s(S_tx)}{\s_+} d t=
\t^{-1}\int_0^\t \frac{\s_\AA(S_tx)}{\s_{\AA,+}} d t$, the relation
\be \frac{P_{srb}(p)}{P_{srb}(-p)}=e^{\t \frac{\NN_{attr}}{\NN}\s_+p + o(\t)}
\Eq{e9.1}\ee
in the notation of Eq.\equ{e6.4}, \cite{BGG997}, because
$\s_{\AA,+}=\frac{\NN_{attr}}{\NN}\s_+$: \ie the universal constant $1$ in
\equ{e6.4} is replaced by $\frac{\NN_{attr}}{\NN}$.

This covers the FR in systems verifying the pairing rule: {\it but, admittedly,
such systems are not really common in the applications}.

A much larger class of systems can be imagined if the Lyapunov exponents,
arranged as in Eq.\equ{e8.1}, satisfy $\frac12(\l_j+\l_{2n-1-j})=c_j$
with $c_j$ close to a constant (\ie $c_j\sim C(\frac{j}{2n})$ with $C(\x)$ a
smooth function.

This property arises in a few important cases. For instance in simulations
of reversible models for the 2D incompressible Navier-Stokes equation in
periodic geometry, \cite{GRS004,Ga019c}.

Suppose that the local Lyapunov exponents are such that
$\frac12(\l_j(x)+\l_{2n-1-j}(x))=c_j(x)$ and let $\lis \l_j,\lis c_j$ be
the respective time averages. Then it
can be still imagined that corresponding pairs $\lis \l_j,\lis\l_{2n-1-j}$
of exponents of {\it opposite sign} are exponents concerning the motion on
the attracting set; particularly if the system is close to one for which
the pairing rule holds and $\lis c_j$ is close to a constant.

By an argument similar to the one presented in the pairing symmetrical
cases above, a large deviations relation might be obtained, for the total
contraction $\s(x)$, similar to Eq.\equ{e9.1} with $\frac{\NN_{attr}}{\NN}
  \s$ replaced by $\s_\AA(x)=P\s(x)$ with $P\defi 1-\frac{\sum_{j\in L_-}
    \lis\l_j}{\sum_{j\in L} \lis\l_j}$, where $L$ is the set of Lyapunov
  exponents and $L_-$ the subset formed by the pairs of negative exponents.
  Leading to a FR {\it with a controlled modification of the slope} in $p$,
  at least if the pairing functions $c_j(x)$ can be found, see
  Sec.\ref{sec18} for a non trivial example.

The above may apply to time reversible systems with Lyapunov spectrum
obeying a pairing rule at least approximately; and could be extended,
possibly, to irreversible ones if the latter fall under the equivalence
properties mentioned in Sec.\ref{sec7}.
\*

\0{\it Remarks} (i) In this respect it is worth coming back to the issue
mentioned in the remark concluding Sec.\ref{sec6}. Which pointed out that
when considering many particle systems, like for instance the steel cube
brought up as an example, the phase space contraction might have an average
too large, due to the size of the Boltzmann's constant. And consequently
the FR would fail to be of any relevance for the fluctuations statistics.
\\
(ii) However in such cases, as in any macroscopic system, the phase space
contraction is very large: but the FR should be applied to the contraction
of the surface of the attracting set. To do so the above axiom C and a
pairing property may be of help: for instance in presence of exact pairing
the FR holds with the $\s_+$ replaced by $\frac{\NN_{attr}}{\NN} \s_+$.
The dimension $\NN_{attr}$ in large systems will usually be $\ll \NN$:
and in order to apply the FR the average phase space contraction must
be corrected by the factor $\frac{\NN_{attr}}{\NN}$.
\\
(iii) Since $\NN$ is, in the case (i), a number of the order of a multiple
of Avogadro's number this is sufficient to turn $k_B T_a$ into $R_0=\NN
k_B$, with $R_0$ being of the order of the gas constant: and this converts
Eq.\equ{e4.5} to $\NN_{attr}\sum_a \frac{\dot Q_a}{R_0 T_a}$
and $R_0T_a$ is no longer very small.  \\
(iv) Also $\NN_{attr}$, in the systems like the ones in the example in
Sec.\ref{sec4},\,\ref{sec5}, is typically {\it not of the order of
  Avogadro's number}: macroscopic systems often can be described by
macroscopic equations and the number of positive exponents can be
identified with the number of positive exponents of the ``equivalent''
macroscopic equation; the latter very often has a rather small number of
positive exponents so $\NN_{attr}$ can be small: in conclusion the FR could
be applied to classes of system which admit microscopic and macroscopic
equivalent representations.
\\
(iv) A non trivial example is analyzed in
Sec.\ref{sec17}\,,\,\ref{sec18}.

\def\SEC{\ Fluctuation Dissipation Theorem} 
\section{\SEC}
\label{sec10}
\iniz

Suppose that a dynamical system equations
\*
\0(a) depend on parameters $\V E=(E_1,E_2,\ldots)$ and satisfy a $\V
E$-independent, smooth, time reversal symmetry $I$,\footnote{\small To fix
  ideas think of a Hamiltonian system constrained to keep the total kinetic
  energy constant, for instance via a Gaussian constraint, as considered in
  many applications, \cite{EM990}: in absence of external forcing, and
  assuming CH, the SRB distribution is quite generally explicitly known and
  equivalent to the canonical distribution, \cite{EM990}.\label{example}}
\\
(b) for $\V E=\V 0$ the equations are supposed to satisfy the CH with
attracting set coinciding with the full phase space (as in the cases in the
footnote \footref{example}).\footnote{\small In other words at $\V E=\V0$ the
evolution is a Anosov system.}
\*

For $\V E\ne\V0$ the equations continue to be time reversal
symmetric with the same symmetry map $I$.\footnote{\small The symmetry
  could also depend on $\V E$, becoming $I_{\V E}$, however further
  assumptions would be needed, like differentiability, \cite{Po010b}.}

The dynamics is a Anosov system and it remains such at small $\V E$: the
attracting set coincides with the full phase space (by the structural
stability of CH) and the FT holds for the SRB
distributions.\footnote{\small Remark that this is an important case whose
  occurrence has been considered ``{\it rare to evanescent}'' in
  \cite[p.220]{Ho999}.}


It is therefore interesting to find whether the average phase space
contraction $\s_{\V E,+}$ is a function of $\V E$ with interpretation that
goes beyond its being a quantity associated with universal large
fluctuations of the dissipation.  In particular it is interesting to find
an interpretation of the multiple derivatives
$\BDpr^{\V n}_{\V E}\s_{\V E,+}|_{\V E=\V0}$.

The phase space contraction $\s_{\V E}(x)$, briefly $\s(x)$, will be supposed to have the
Taylor expansion:
\be
\s(x)=\sum_{i=1}^s E_i J^0_i(x)+ O(\V E^2)\Eq{e10.1}\ee
and, having assumed CH, the large deviation rate $\z(p)$ exists (model
dependent) and is analytic in $p$ for $p$ in the interval $(-p^*,p^*)$,
$p^*\ge1$, within which it can vary, \cite{Ga995b,Ga996a}.

On general grounds the function $\z(p)$ is the Laplace transform of
$\l(\b)=\lim_{\t\to\infty} \frac1\t\log\int e^{\b \t (p-1)} P_\t(d p)$
where $P_\t(d p)$ is the PDF of the variable $p=\frac1{\t \s_{\V
    E,+}}\int_0^t\s(S_t x)dt$ in the SRB distribution. Once $\l(\b)$ is
"known" then $\z(p)$ is recovered via a Legendre transform; $\z(p)=
\max_\beta\big(\beta\langle\s\rangle_+(p-1)-\lambda(\beta)\big) $,
\cite{Ga995b,Ge998}.

By using the cumulant expansion for $\lambda(\b)$ we find that
$\lambda(\b)={1\over 2!}\b^2 C_2+{1\over3!}\b^3 C_3+\ldots$ where
the coefficients $C_j$ are
$\int_{-\infty}^\infty\langle\s(S_{t_1}\cdot)\s(S_{t_2}\cdot)\ldots
\s(S_{t_{j-1}}\cdot)\s(\cdot)\rangle^T_+\,dt_1\ldots$ if
$\langle\ldots\rangle^T_+$ denote the cumulants of the variables
$\s(x)$.

In our case the cumulants of order $j$ have size $O(G^j)$ with $G\defi|\V
E|$, by Eq.\equ{e10.1}, so that:
\be
{\zeta(p)={\langle\s\rangle_+^2\over2 C_2}(p-1)^2+ O((p-1)^3
G^3)}\Eq{e10.2}\ee
\noindent{}(remark that the first term in r.h.s.  gives the central limit
theorem).  Eq.\equ{e10.2}, together with the FR Eq.\equ{e6.3}, yields at
fixed $p$ the key relations:
\be\langle\s\rangle_+={1\over 2}C_2+ O(G^3)\Eq{e10.3}\ee
Define, \cite{Ga995b,Ge998}: $J_i(x)=\partial_{E_i}\sigma(x)$ = {\it
  current}, $L_{ij}=\partial_{E_j}\langle J_i(x)\rangle_+|_{\V E=\V 0}$ =
{\it transport coefficients}; and study $L_{ij}$.

In the \rhs of the first of Eq.\equ{e10.3} discard $O(G^3)$: it becomes
quadratic in $\V E$ with coefficient $\frac12C_2$,
making use of the exponential decay of
SRB-correlations in Anosov systems, given by:
\be {1\over 2}\int_{-\infty}^\infty dt\, \big(\langle J^0_i(S_t\cdot)
J^0_j(\cdot)\rangle_+- \langle J^0_i\rangle_+\langle J^0_j\rangle_+
\big)\big|_{E=0}\Eq{e10.4}\ee
where convergence is implied by the strong mixing properties of the SRB
distribution due to the CH.

On the other hand the expansion of $\langle\sigma\rangle_+$ in the
\lhs of Eq.\equ{e10.3} to second order in $E$ gives:

\be\langle\sigma\rangle_+={1\over2}
\sum_{ij}\big(\partial_{E_i}\partial_{E_j}\langle\sigma\rangle_+
\big)\big|_{E=0} E_i E_j \Eq{e10.5}\ee
because the first order term vanishes, see Eq.\equ{e10.1}.

If $\m_+(dx)$ denotes the SRB distribution, the \rhs of Eq.\equ{e10.5} is the
sum of ${1\over2}E_iE_j$ times $\partial_{E_i}\partial_{E_j} \int \sigma(x)
\mu_+(dx)$ which equals the sum of the following three terms:\\
(i) 
$\int\partial_{E_i} \partial_{E_j}\sigma(x) \mu_+(dx)$,\\
(ii) 
$\int\partial_{E_i}\sigma(x)\partial_{E_j}\mu_+(dx)+ (i\leftrightarrow j)$
\\
(iii) $\int\sigma(x)\partial_{E_i}\partial_{E_j}\mu_+(dx))$, all
evaluated at $\V E=\V 0$.

The first addend is $0$ (by time reversal), the third
addend is also $0$ (as $\sigma=0$ at $\V E=\V 0$).  Hence:

\be\partial_{E_i}\partial_{E_j} \langle\sigma\rangle_+|_{E=0}=
\Big(\partial_{E_j}\langle{J^0_i}\rangle_++
\partial_{E_i}\langle{J^0_j}\rangle_+\Big)|_{E=0}
\Eq{e10.6}\ee

\noindent{}and it is easy to check, again by using time reversal, 
that:

\be
\partial_{E_j}\langle{J^0_i}\rangle_+|_{E=0}\,=
\,\partial_{E_j}\langle{J_i}\rangle_+|_{E=0}
= \,L_{ij}\Eq{e10.7}\ee
Thus equating r.h.s and l.h.s.  of Eq.\equ{e10.3}, as expressed
respectively by Eq.\equ{e10.4} and Eq.\equ{e10.6}  the matrix
${L_{ij}+L_{ji}\over2}$ is obtained, \cite{Ga996}.

At least if $i=j$ this is a ``Green-Kubo formula'', a relation sometimes
called "fluctuation dissipation theorem". It is however {\it very different}
from ``Onsager's reciprocity'' which would be $L_{ij}=L_{ji}$. The latter
will be discusses in the next section.

\def\SEC{\ Onsager's Reciprocity}
\section{\SEC}
\label{sec11}
\iniz

A far reaching extension is necessary to obtain $L_{ij}=L_{ji}$ which will
lead to reciprocity, \cite{Ga996a}, and to further extensions,
\cite{GR997}.

The main remark is that FT theorem can be {\it extended} to give properties
of {\it joint} SRB distribution of $\sigma(x)$ and of the observable
$q(x)=E_j\partial_{E_j}\sigma$.  Defining {\it dimensionless }
$j$-current $q=q_j(x)$ (at fixed $j$) as:

\be
{1\over\t}
\int_{-\t/2}^{\t/2}
E_j\dpr_{E_j} \s(S_tx)dt\defi
q
\Eq{e11.1}\ee %
where the factor $E_j$ is there only to keep $\s$ and $E_j\dpr_{E_j}\s$
with the same dimensions, the really essential property of $q_j(x)$ is
its odd {\it symmetry} under time reversal, as $\s(x)$.

Then if $P_\t(dp,dq)$ is the joint PDF of $p,q$ the {\it same} 
proof of the FT in \cite{Si977,GC995,Ga995b} yields also the existence of a
rate function $\z(p,q)$ for $P_\t$ with the symmetry:

\be \z(p,q)=\z(-p,-q)\,+\, p\,\s_{\V E,+}, \qquad
    {\rm for\ all} \ p,q \Eq{e11.2}\ee
for the {\it joint large fluctuations} of the variables $\s_{\V
  E}(x)$, $E_j\dpr_{E_j}\s(x)$.

The $\zeta(p,q)$ can be computed, in the same way as $\zeta(p)$ in
Sec.\ref{sec10}, by considering first the transform $\lambda(\b_1,\b_2)$:

\be
\lim_{\t\to\infty} \textstyle
{1\over\t}\log\kern-1pt\int e^{\t(\b_1\, (p-1)
\langle\s\rangle_++
\b_2\,(q-1)\langle E_j\dpr_{E_j}\s\rangle_+)}P_\t(dp,dq)
\Eq{e11.3}\ee 
and then the Legendre transform, abridging the SRB average
$\media{\cdot}_{\V E,+}$ with $\media{\cdot}_+$,

\be\eqalign{
  \max_{\b_1,\b_2}\big(&
\b_1\, (p-1)\langle\s\rangle_++
\b_2\,(q-1)
\langle E_j\dpr_{E_j}\s\rangle_+\cr
&-\lambda(\b_1,\b_2)\big)\,=\,\zeta(p,q)\cr}\Eq{e11.4} 
\ee

The function $\lambda(\Bb)$, $\Bb=(\b_1,\b_2)$, is 
evaluated by the cumulant expansion, as above, and one finds:

\be
\lambda(\Bb)={1\over2}\,\big(\Bb, 
C\,\Bb)+O(E^3)\Eq{e11.5}\ee  
where $C$ is the $2\times2$ matrix of the second order cumulants. The
coefficient $C_{11}$ is given by $C_2$ appearing in
Eq.\equ{e10.3},\equ{e10.4}; $C_{22}$ is given by the same expression with
$\s$ replaced by $E_j\dpr_{E_j}\s$ while $C_{12}$ is the mixed cumulant:

\be
\int_{-\infty}^\infty
\big(\langle\s(S_t\cdot)\,E_j\dpr_{E_j}\s(\cdot)\rangle_+
-\langle\s(S_t\cdot)\rangle_+\,\langle
E_j\dpr_{E_j}\s(\cdot)\rangle_+\big)dt\Eq{e11.6}\ee 
and convergence is again implied by the mixing properties of the SRB
distributions due to the CH.

Hence if $\V w=\pmatrix{(p-1)\langle\s\rangle_+\cr
(q-1)\langle E_j\dpr_{E_j}\s\rangle_+\cr}$ we get:

\be
\zeta(p,q)={1\over2}\,\big(C^{-1}\V w,\V w\big)+O(E^3)
\Eq{e11.7}\ee   
completely analogous to Eq.\equ{e10.2}. But the FT in Eq.\equ{e11.2}, implies
that $\zeta(p,q)-\zeta(-p,-q)$ is {\it $q$ independent}: this immediately
means:

\be -(C^{-1})_{22}
\langle E_j\dpr_{E_j}\s\rangle_+-(C^{-1})_{21}
\langle \s\rangle_+=0+O(E^3)\Eq{e11.8} \ee  
which, from $(C^{-1})_{22}=C_{11}/\det C$, becomes the analogue of
Eq.\equ{e10.3}:
\be
\langle E_j\dpr_{E_j}\s\rangle_+={1\over2}C_{12}+O(E^3)
\Eq{e11.9}   
\ee

\noindent{}Then, proceeding as in the derivation of Eq.\equ{e10.4} through
\Eq.\equ{e10.7} ({\it i.e.} expanding both sides of Eq.\equ{e11.9} {\it to
  first order} in the $E_i$'s and using Eq.\equ{e11.8} we get that
$\dpr_{E_i}\langle\dpr_{E_j}\s\rangle_+$ is given by the integral in
Eq.\equ{e10.4}.  This means that $L_{ij}=L_{ji}$ and the general Green-Kubo
formulae follow together with  Onsager's reciprocity.

Thus GK, and OR, are in the cases considered here, a consequence of FT and
of its extension, Eq.\equ{e11.2}, in the limit $E\to0$, when combined with
the expansion Eq.\equ{e10.2} for entropy fluctuations.  Those theorems and
the fast decay of the $\s\-\s$ correlations, \cite{Si977}, are all natural
consequences of (CH) for reversible systems (which are the
starting point of our considerations).  Reversibility is here assumed {\it
  both in equilibrium and in non equilibrium}: this is a feature of
Gaussian thermostat models but by no means of all models; the $\V
E$-independence of the reversibility map is also essential but in most
reversible models it is just the velocity reversal map, which is
independent of $\V E$.

Of course the OR and GK only hold around equilibrium, {\it i.e.}  they are
properties of $\V E$--derivatives evaluated at $\V E=0$; on the other hand
the expansion for $\lambda(\b)$ is a general consequence of the correlation
decay and the FT also holds for {\it non equilibrium} stationary states,
{\it i.e.}  for $\V E\ne\V0$ small\footnote{\small \ie as long as structural
stability maintains the system an Anosov system.}, and {\it can be
considered a generalization of the OR and GK}.

Evidence for the relation between $L_{ii}$, Green-Kubo formulae, and FT was
pointed out by P.Garrido in \cite{BGG997} in an effort to interpret results
of various numerical experiments and an apparent incompatibility between the
{\it a priori} known non Gaussian nature of the distribution $\pi_\t(p)$
and the "Gaussian looking" empirical distributions; the extension to the
reciprocity followed naturally (see also \cite{Ga996,Ga996a}).  In
\cite{BGG997} the situation arising at really large fields, when the
attractor is strictly smaller than the whole phase space, is also
discussed (eventually leading to the analysis in Sec.\ref{sec9} above).

\def\SEC{\ Fluctuation Patterns}
\section{\SEC}
\label{sec12}
\iniz

The derivation of Onsager's reciprocity for reversible Anosov systems
with a time reversal map $I$ smooth and parameters independent  (usually
just a ``velocity reversal''), and therefore for systems verifying the Chaotic
Hypothesis, suggests that the fluctuation relations might be extended to
fluctuations of more general observables. At least for small perturbations
of Anosov systems and for smooth Axiom C systems, see caption to Fig.2.

Consider first the fluctuations of the phase space contraction $\s(x)$ and
those of a second observable $\f(x)$ with {\it definite parity} under time
reversal: so $\s(Ix)=-\s(x)$ and $\f(x)=-\f(Ix)$ (or $\f(x)=\f(Ix)$).

Consider a SRB distribution $\m_{srb}$ for the system: let
$\media{\s}_+>0,\media{\f}_+$ be the SRB time averages  of $\s,\f$.
Call  ``fluctuation pattern'' $\p$ a function on $[0,\t]$:
$t\to \p(t)=(s(t),f(t))$.

The evolution of a point $x$ in phase space such that

\be \eqalign{|&s(t)-\s(S_t x)|<\e,\cr
  &|f(t)-\f(S_tx))|<\h\cr} \qquad {\rm for} \ t\in[0,\t]\ 
\Eq{e12.1}\ee
will be called a motion which {\it shadows the pattern $\p$ in the time
interval $[0,\t]$} and it will be
written
$x{\buildrel\t,\e,\h\over \sim}\p$.

The ``time reversal'' of the pattern $\p$ will be the pattern
$I\p=(-s(\t-t),-f(\t-t))$ (or if $\f(x)$ is even under time reversal
$I\p=(-s(\t-t),f(\t-t))$).

The SRB probability of a trajectory $x\to S_tx$ to follow a pattern $\p$
will be denoted
$P_\t(\{ x{\buildrel\t,\e,\h\over \sim}\p\})$; the argument at the basis of
the Fluctuation Theorem can be applied to study the ratio:

\be \frac1\t \log \frac{P_\t(\{ x{\buildrel\t,\e,\h\over \sim}\p\})}
{P_\t(\{ x{\buildrel\t,\e\over \sim}I\p\})}\Eq{e12.2}\ee
and for reversible Anosov systems leads, at first
surprisingly,  immediately to the result:

\be \frac{P_\t(\{ x{\buildrel\t,\e,\h\over \sim}\p\})}
    {P_\t(\{ x{\buildrel\t,\e,\h\over \sim}I\p\})}
    =e^{\t\, \media{\s}_+\, p + o(\t)}
\Eq{e12.3}\ee
asymptotically as $\t\to\infty$ and to lowest order in the precision $\e,\h$.

More generally several observables can be considered $\s,\f_1,\f_2,\ldots$
and the notion of pattern can be accordingly extended; with the same result
that the ratio of the probability of a fluctuation pattern to that of the
time reversed pattern is $e^{\t\, \media{\s}_+\, p + o(\t)}$, to leading
order as $\t\to\infty$ and in the precision, {\it independent on the
  specification} of the fluctuations of $\f_1,\f_2,\ldots$.

Also the $\f$ independence of Eq.\equ{e12.2} implies, given two fluctuation
patterns $\p$ for the observables $\s,\f$ and $\p'$ for the observables
$\s,\ps$, to leading order in $\e,\d,\t^{-1}$:

\be \frac{P_\t(\{ x{\buildrel\t,\e,\d\over \sim}\p\})}
    {P_\t(\{ x{\buildrel\t,\e,\d\over \sim}\p'\})}
    =\frac{P_\t(\{ x{\buildrel\t,\e,\d\over \sim}I\p\})}
    {P_\t(\{ x{\buildrel\t,\e,\d\over \sim}I\p'\})}
    \Eq{e12.4}\ee
    The above relations show that {\it once the rare event of a sign
      change of the entropy production } is realized then the time reversed
    patterns have the {\it same relative probability} that they have when
    the entropy production has the opposite sign.

In other words to see that time reversed patterns occur it is
``sufficient'' to just {\it change the sign of entropy production} (!):
``no further efforts'' are needed.

In Sec.\ref{sec11} the Eq.\equ{e11.2} has been shown to be essentially
equivalent to Onsager's reciprocity and it is a special case of the
general Eq.\equ{e12.3}: therefore the above Eq.\equ{e12.3} can be
considered an extension of Onsager's reciprocity to stationary
states of time reversible Anosov systems or more generally (if $\s_+$ is
intended as the average area contraction of the attracting surface $\AA$)
to systems verifying the CH and the ``axiom C'', see Sec.\ref{sec9}
and caption to Fig.2. 

\def\SEC{\ Irreversibility time scale}
\section{\SEC}
\label{sec13}
\iniz

The notion of ``reversible transformation'' between equilibrium states is
defined (often) to be an {\it infinitely slow} transformation through a
sequence of equilibrium states. The latter {\it oxymoron} is not really
satisfactory:\footnote{\small ``\`A la v\'erit\'e, les choses ne peuvent pas se
passer rigoureusement comme nons l'avons suppos\'e ...,
\cite[p.13-14]{Ca824}.}  it would be instead desirable to have a
quantitative definition or, better, a way to associate with a
transformation, defined by a given ``protocol'' leading from an equilibrium
state to another, a time scale $\Th$ whose size indicates how long it takes
to realize that the process is irreversible.

Then a reversible transformation should be one with $\Th=\infty$ (to be
interpreted that irreversibility is {\it impossible} to detect). If
$\Th<\infty$ then it should be said that the evolution irreversible nature
is revealed after time $\Th$ which could be taken as the {\it
  irreversibility time scale}, \cite[Ch.5-11]{Ga013b}.
 
So let $\m_0$ be the PDF of an equilibrium state and suppose that the
protocol of action on the system is enforced by a change on the parameters on
which the Hamiltonian depends: like the temperature of an external
thermostat, or the volume available to the molecules, in the case of a gas
enclosed in a container, or like the intensity of a volume force acting on
an incompressible fluid.

The protocol has a duration $\t$ and remains constant afterwards: during
the time $\t$ the system is no longer in equilibrium: the latter is reached
after the time $\t$ elapsed and the system remains isolated or in contact
with thermostats at the same temperature reaching the new equilibrium on a
characteristic time scale $\t'$.\footnote{\small Strictly speaking
  equilibrium will be reached after infinite time; however it can be
  considered reached for practical purposes after $\t'$, which has the
  meaning of a time scale.}

In the following the general system introduced in Sec.\ref{sec4}, see
Fig.1, will be considered to fix ideas.  The entropy production is given
by Eq.\equ{e4.5}.  It is a quantity {\it with dimension of an
inverse time}, coinciding with the phase space total contraction rate.

In nonequilibrium situations the thermostats temperatures can be time
dependent and also the force $f$ as well as the volume of the container
$\CC_0$ can be time dependent. The thermostats temperatures are fixed
phenomenologically and the mechanism of variation of the stirring forces
and of the geometric variation of the container shape or volume are more
difficult to understand and to model physically.

For instance the variation of the force $f$ can be imagined due to the
varying speed of a paddle, rotating in the gas contained in $\CC_0$, which
in turn can be imagined to be controlled by a motor; but it is impossible
to take into account, without a {\it Daemon} helping, how to keep control
of the direction and intensity of the collisions on the paddle. Hence
assuming that the paddle has constant speed, or that it follows a given
protocol of variation, is a phenomenological assumption.

There are experimental setups in which a paddle, or varying forces, are
present and act on the particles in the container $\CC_0$. Or the external
thermostats temperatures and the volume of $\CC_0$ change following
prescribed paths, \eg in the case of volume variations due to a moving
piston. Invariably the entropy production is measured via the amount of
work that the motor and the forces perform maintaining (or trying to
maintain) the external force constant, or constraining it to follow a
prefixed protocol, and via the heat ceded to the thermostats.

Here few cases will be considered in which the protocol contemplates only
variations of the external thermostats temperatures or of the volume of the
containers.

Given the general interpretation of the entropy production rate in terms of
phase space volume variation, the case of volume variation in the system of
Fig.1 (Sec.\ref{sec4}) can be treated {\it phenomenologically} by simply
adding to Eq.\equ{e4.5} the quantity $N\frac{\dot V}V$, which is the rate
of variation of the phase space volume $V^N$ allowed to the $N$ particles
in $\CC_0$.

Consider the system in Fig.1, Sec.\ref{sec4}, and express the
total phase space contraction per unit time, \equ{e4.5}, as
\kern-3mm
\be\s(x)\equiv\s_{tot}(\dot{\V X},\V X)= \sum_a \frac{\dot Q_a}{k_B
  T_a}-N\frac{\dot V_t}{V_t}-\dot U\Eq{e13.1}\ee
\kern-3mm

Let $[0,\t]$ be the time during which a transformation protocol $\G:\,t\to
(T_a(t),V(t))\equiv P(t)$ acts on an initial {\it equilibrium state} with
SRB distribution $\m_0(dx)$ (\eg a canonical ``Gibbs distribution''). Then
it is possible to define \*
\0(1) $\m_t(dx)=\m_0(S_{-t}dx)$, \ie the distribution into which
$\m_0$ evolves in time $t$ under the flow generated in phase space by the
equations of motion (remark that $S_t$ is not a group in $t$ because the
evolution is now time dependent).  \\
\0(2) the  $SRB$ distribution $\m_{srb,t}$ corresponding to the stationary
distribution that corresponds to parameters $(T_a,V)$ fixed
(``frozen'') at their value at time $t$, $P(t)=(T_a(t),V(t))$.
\\
\0(3) the ``relative'' phase space contraction
\be r(t)\defi (\lis \s_t -\lis\s_{srb,t})\Eq{e13.2}\ee
where $\lis\s_{srb,t}$ is the time average of the entropy production rate
in the SRB distribution corresponding to the control parameters $P=(T_a,V)$
frozen at time $t$, while $\lis \s_t$ is the average
phase space contraction in the
non stationary distribution $\m_t(dx)$ evolved from $\m_0$.
\footnote{\small The two terms in Eq.\equ{e13.1} have in general different
  dependence on particles number $N$: as of $O(N)$ in the case of
  volume variations or $O(N^{\frac23})$ if $P(t)$ only involves boundary
  temperature variations (hence the heat exchange is a boundary effect).}

\* Assuming the chaotic hypothesis the approach to the SRB states will be
exponential: the state $\m_t$ would evolve under the ``frozen evolution''
exponentially fast, on some time scale\footnote{\small Supposed to be the
  same for all $t$, for simplicity.}  $\k^{-1}$ to $\m_{srb,t}$.  Therefore
the integral (with a inverse time dimension):

\vskip-5mm\be \Th(\G)^{-1}=\int_0^\infty r(t)^2 dt\Eq{e13.3} \ee
will converge, provided the final values of the control parameters are
reached fast enough (\eg in a finite time, as in actual protocols which
last a prefixed finite time).

The {\it time scale of irreversibility} of the protocol could be defined by
$\Th(\G)$: the larger $\Th$ is, the closer to a quasi static one the
transformation is, as suggested by the following remarks.

A physical definition of ``quasi static'' transformation protocol is a
transformation that is ``very slow'' during its duration time
$\t$. This can be translated mathematically into an evolution in which
$P(t)\defi (T_a(t),V(t))$ evolves like, if not exactly, as

\kern-3mm
\be P(t)=P(0)+ (1-e^{-\e t})(P(\infty) -P(0))\Eq{e13.4}\ee
with $\e>0$ small. 

An evolution $\G$ ``close to quasi static'', but simpler for computing
$\Th(\G)$, would proceed changing $P(0)$ into $P(\infty)=P(0)+\D$ by $\t/\d$
steps of size $\d$, each of which has a time duration $t_\d$ long enough so
that, at the $k$-th step, the evolving system closely settles onto its
stationary state $P(0)+k\d$.

The $t_\d$ can be defined\footnote{\small Remark that the variation of
  $\lis\s_{(k+1)\d,+}-\lis\s_{k\d,+}$ is, in general, of order $\k\d$ as a
  consequence of the differentiability, \cite{Ru997b}, of the SRB states
  with respect to the parameters.} by $e^{-\k t_\d}\ll \k\d$ then by
Eq.\equ{e13.3}:
\be \Th(\G)^{-1}\simeq\, const\, \k^{-1}\,(\lis\s'\d)^2\,\log(\k\d)^{-1}
\Eq{e13.5}\ee
where $\lis\s'$ is an estimate of $\dpr_t \lis\s_{srb,t}$.  Therefore the
``slower'' is the protocol $\G$ (\ie the larger the time scale $\d^{-1}$ is)
the closer to $\infty$ is the irresversibility scale $\Th(\G)$.

Another way of reading the above: {\it the closer the actual entropy
  production $\lis\s_t$ is to the ``ideal'' $\lis\s_{srb,t}$ the longer is
  the irreversibility time scale, \ie the time beyond which the process
  cannot be considered reversible.}

\*
\0{\it Remark:} particularly interesting are adiabatic processes in which
external forces vary remaining conservative:
\\
(a) an example is an adiabatic 
expansion of a gas in a piston.  The irreversibility time scale can be
evaluated from the piston velocity, see 
\cite[Ch.5]{Ga013b}.
\\
(b) a second example is a rarefied gas, with mass $m$ molecules in a {\it
  fixed} adiabatic container, subject to a force of potential $mgz$. At
time $0$ the gas is in equilibrium at temperature $\b^{-1}$
and the process $G$ simply raises the acceleration $g$ to $g'>g$ at time
$0$ and then decreases it back to $g$ after a time $\t>0$ (or just stays
$g'$ forever).\footnote{\small This means that the initial potential energy
  is $m g N h_0$, where $h_0$ is the height of the center of mass, and
  varies at time $0$ to $m g' N h_0$.}  Since $\s_t\equiv0$ (by Liouville's
theorem) and $\s_{srb,t}\equiv0$ it is $\Th=\infty$: \ie the transformation
is {\it reversible} according to the above proposal of reversibility time
scale, as also suggested by Gibbs' entropy constancy in Hamiltonian
evolutions (even when the Hamiltonian is time dependent).  Nevertheless the
cycle leads to an intermediate temperature {\it variation} $\d T$ (with
$\frac{\d T}T\simeq\frac{\d g}{g}$, up to finite volume corrections): an
apparent disagreement with the independence on the rapidity of the process,
see Appendix \ref{appC} for details.  \*

\def\SEC{\ Chaos. Structure of Anosov systems. Their digital codes.}
\section{\SEC}
\label{sec14}
\iniz 

Since the early works on Statistical Mechanics the concept of {\it coarse
  graining} played a major role in relating macroscopic and microscopic
descriptions of mechanical systems, \cite{DGM984}.

Anosov systems, through the chaotic hypothesis, offer {\it new}
perspectives. For simplicity here will be considered the case of a discrete
time evolution via a map $S$ on a phase space $M$, which could be a
Poincar\'e's section of either a macroscopic model of evolution (possibly
infinite dimensional, like Navier-Stokes equation) or a microscopic one
(like Newton's equations for $10^{19}$ molecules) or a phenomenological
model (like Lorenz96 or Lorenz63 models or GOY shell model).

What follows can be extended to the case of Anosov flows,
\cite{BR975,Ge998}, essentially by reduction of the problem to the Anosov
maps case by replacing the flow with a Poincar\'e's map between timed
events (\ie by fixing a surface in phase space and studying the return map
to it).  Extension to axiom A maps or flows is also possible,
\cite{Bo970a,Bo975,BR975}.

The discussion below is necessarily somewhat technical as it tries to
convey the reason why Anosov maps lead to stationary states which can be
identified with equilibria of {\it one dimensional spin chains with short range
interactions}: the extreme simplicity of Anosov maps will be manifest after
understanding the formalism. It will reward the necessary time, thus
providing strong support to the statement (see Sec.\ref{sec1}) that Anosov
maps play, for chaotic systems, a role parallel to that of the harmonic
oscillators for ordered dynamics.

A main feature of Anosov maps is that the stable and unstable manifolds of
each point $x$ are smooth manifolds which depend ``almost'' differentiably
on $x$ (they are H\"older continuous and the H\"older exponent can be taken
as close to $1$ as wished, paying the price of a larger H\"older
constant). The manifolds can be used to build ``cells'' in $M$ enclosed
within boundaries which are unions of subsets of stable or unstable
manifolds.

The key remark, \cite{Si968a,Si968b}, is that the phase space $M$ can be
paved with cells, $\PP=(P_1,\ldots,P_N)$, which are connected sets,
closures of their interiors, and are either pairwise disjoint or have only
common boundary points; furthermore are ``{\it covariant}'' if transformed
by the map $S$ in the following sense: \*

\0(1) the boundary $\dpr P_j$ of $P_j$ has the form $\dpr_u P_j\cup \dpr_s
P_j$ with $\dpr_u P_j$ consisting of surface elements which are unions of
portions of unstable manifolds and $\dpr_s P_j$ consisting of surface
elements unions of portions stable manifolds: call $\dpr_u \PP=\cup_j
\dpr_u P_j$ and $\dpr_s \PP=\cup_j \dpr_s P_j$.  \\
(2) the images $S^{\pm1} P_j$ of the cell $P_j$ will have boundary still
consisting of stable or unstable surface elements (because images of stable
or unstable manifolds are still stable or unstable manifolds) and,
furthermore, will have the {\it covariance property}, see Fig.3:
\be S \dpr_s P_j \subset \dpr_s\PP,\quad
S^{-1} \dpr_u P_j \subset \dpr_u\PP\Eq{e14.1}\ee
This means that the $P_j$ are so deformed by $S$ (resp. $S^{-1}$) that no
new stable (resp. unstable) boundaries are created. Furthermore the points
$x$ in their evolution will never end up on any of the cells boundaries
with the exception of a set of zero volume (\ie the set
$\cup_{i=-\infty}^\infty S^i\dpr\PP$).\label{cup}
\\
(3) the $\dpr_u \PP,\dpr_s\PP$ have $0$ volume
\*

\eqfig{150}{50}{}{fig0}{}

%
\0{\small Fig.3: \it Very symbolically, as $2$-dimensional squares, a few
  elements of $\PP$ are shown as an array of squares.  An element $P_i$
  (shaded, left) of $\PP$ is transformed by $S$ into $S P_i$ (shaded,
  right) in such a way that the part of the boundary that contracts ends up
  exactly on a boundary of some element among $P_1,P_2,\ldots,P_n$. A
  similar figure with horizontal and vertical lines exchanged would
  illustate the action of $S^{-1}$.}  \*

Obviously if $\PP$ is a pavement of $M$ with the above properties (1),(2)
then also the pavement whose elements are $P_{i,j}=S P_i\cap P_j$ has the
  same properties: hence the hyperbolicity of the map yields that there
  exist pavements with elements with diameter smaller than a prefixed
  $\e>0$.
  
Such pavements $\PP=(P_0,\ldots,P_n)$ of $M$ are called {\it Markovian
  partitions} if the maximum diameter of the $P_i$'s is so small that
the intersection between $S^{\pm} P_i$ and $P_j$ is a connected set:
as inthe figure.%
\footnote{\small Disconnected intersections may happen if the maximum
  diameter of the $P_i$ can be dilated by the action of $S$ or $S^{-1}$ to
  become larger than the diameter of $M$.}

The hyperbolicity of $S$ implies existence of
Markovian partitions and they can be constructed iteratively,
\cite{Si968b,Bo970a,FR981,GBG004}.


The elements of $\PP$ are called ``rectangles'' as they have boundaries
formed by portions of stable and unstable manifolds which in the case of
the simplest Anosov maps, \ie algebraic hyperbolic maps of the
$2$-dimensional torus, are really quadrilaterals with opposite sides
parallel and equal: algrebraic means that the maps are defined by a
constant matrix with integer entries, no eigenvalue with
modulus $1$ and determinant $\pm1$ (like $\pmatrix{0&1\cr1&1\cr}$).

In dimension $2$, in general, they have the aspect of deformed rectangles
(as the manifolds constituting their boundaries are neither parallel nor
flat) with smooth boundaries. If the map $S$ is an algbraic map of the
$2$-torus (\ie $S$ is a $2\times2$ matrix with integer entries)
they are rectangles, in the literal sense.

In $\ge 3$ dimensions the intersections between the stable manifolds and
the unstable manifolds meeting at the edges of the rectangle are not
smooth: in general a portion of unstable manifold of dimension $u>1$,
contained in $\dpr P_i$, may have a boundary\footnote{\small \ie the
  intersection with the stable manifolds in $\dpr P_i$.} which does not
contain a smooth surface of dimension $u-1$ (\eg in dimension $3$ and if
the unstable manifold had dimension $2$ it does not contain a
differentiable arc, as one might naively imagine, \cite{Bo978}: \ie the
edge is not a smooth line).

Likewise a portion of stable manifold, of dimension $s>1$, contained in
$\dpr P_i$, may have a boundary which does not contain a smooth surface of
dimension $s-1$. So the rectangles edges may be {\it quite
  rugged}. Nevertheless the boundaries of the sets $P_i$ can be shown to
have {\it zero volume}.

\*

Given a point $x\in M$ its {\it history}
$\Bs=(\s_k)_{k=-\infty}^\infty$, $\s_k=1,2,\ldots,n$, on a Markovian
partition $\PP$ is 
defined by
\be S^k x \in P_{\s_k}, \quad \forall k\in(-\infty,\infty)\Eq{e14.2}\ee
{\it uniquely} with the exception of the set, with zero volume in $M$, of
the points $x\in\cup_{-\infty}^\infty S^k\dpr \PP$, \ie except the set of
points which in their evolution fall on the boundary of some of the
$P_i\in\PP$. The history is a {\it digital code} for the points of M and the
labels $k$ can, naturally, be called {\it ``times''}.

The history is very convenient as it transforms the evolution $x\to Sx$
into the simple ``translation'': if $x$ has history
$\Bs=\{\s_i\}_{i=-\infty}^{\infty}$ and $x$ evolves into $Sx$ then its history
$\Bs$ evolves into $\t \Bs=\{\s_{i+1}\}_{i=-\infty}^\infty$.

Define the $n\times n$ ``transitivity matrix'' $T_{\s,\s'}=1$ if there
is an interior point $x\in P_\s$ whose image $Sx$ is an interior point of
$P_{\s^\prime}$ and $T_{\s,\s^\prime}=0$ otherwise.  Then only sequences
$\Bs$ with $T_{\s_k,\s_{k+1}}\equiv1$, that will be called ``$\PP$-compatible'',
can arise as histories of points.

Viceversa given any $\PP$-compatible history $\Bs$ there is at least one
$x\in M$ whose history is $\Bs$ and the correspondence $x\otto\Bs$ is
one-to-one with the exception of points $x$ in the above mentioned zero
volume set $\cup_i S^i\dpr\PP$. This geometric property follows from
hyperbolicity and the covariance of the boundaries of $\PP$, \cite{Ga013b}.
The history $\Bs$ determines the corresponding point $x$ ``exponentially
fast'', meaning that there is a constant $\k>0$ such that the
$\{\s_i\}_{i=-n}^n$ determines $x$ within $const\,e^{-\k n}$ (and $\k$ can
be taken any smaller than the minimum of the expansion rates for $S$ and
$S^{-1}$).

In Anosov maps there are points with a dense trajectory (see
\footref{hyperbolic} in Sec.\ref{sec1}): hence the compatibility matrix $T$
is ``transitive'', \ie there is $K>0$ such that $T^K_{\s,\s^\prime}\ge1$
for all pairs $\s,\s^\prime$: this means that among compatible histories it
is possible that any symbol $\s$ is followed by any other symbol
$\s^\prime$ after at most $K$ steps.

The symbolic history can, therefore be used to code the distributions
$\m_0(dx)=\r(x)dx$, with density $\r(x)$ with respect to the volume element
$dx$ in $M$, into {\it stochastic processes}, \ie into probability distributions
on the space of the compatible histories.\footnote{\small The code that
  associates with $x\in M$ the history $\Bs$ of $x$ is closely related to
  the ``structural stability'' of Anosov maps. Structural stability of
  Anosov maps means that if an Anosov map $S$ is smoothly perturbed to
  become a map $S_\e$, as a function of a parameter $\e$, with $S_0=S$
  then, if $\e$ is small enough, also $S_\e$ is an Anosov map and $S_\e$ is
  conjugated to $S$ in the sense that there is a H\"older continuous
  homeomorphism $\Th$ of $M$ such that $S_\e=\Th\,S_0\,\Th^{-1}$.  \\
An essential step to prove this property is to show that a Markovian
partition $\PP$ for $S$ can be deformed ``by continuity'' into a Markovian
partition $\PP_\e$ for $S_\e$ {\it with the same transition matrix}: so
that the conjugation is the map $x\otto x'$ associating pairs with the same
histories: under $S_0$ on $\PP$ and under $S_\e$ on $\PP_\e$.%
\label{struct-stab}}

\def\SEC{\ Volume as stochastic process.
  SRB as Ising spin chain equilibrium}
\section{\SEC}
\label{sec15}
\iniz

The key to the theory of Anosov maps is the {\it representation of the volume
measure} as a probability distribution on the set of compatible sequences,
\ie as a stochastic process, which in the above case has been proved to be
a ``Gibbs process'' with a short range potential, which {\it however}, in
general, {\it is not translation invariant}, \cite{Si968a,Si968b,Si972a},
see below. A connection with the Gibbs processes emerges naturally also
when attempting to interpret results of simulations,
\cite[Sec.3]{BGG997},\cite{LS999,Ma999}.

Given a Anosov map $S$, its phase space $M$ can be thought as the space of
states of a spin system on a $1$-dimensional lattice: evolution of $x\in M$
being just the shift of the history $\Bs$ on a Markovian partition
$\PP=(P_1,P_2,\ldots,P_m)$, see Sec.\ref{sec14}, into which $x$ is
coded. Therefore points of $M$ are still digitally represented although {\it
the usual digital sequences} for their Cartesian coordinates) are abandoned.

\* \0{\it Remark:} Representation via histories on Markovian partitions in
not universal, like the the one via the digits of the cartesian
coordinates, but is specifically adapted to the particular dynamical system
$(M,S)$.  \*

The normalized volume is then coded into a probability distribution
$\m_{vol}(d\Bs)$ on the space $C({\bf Z})$ of compatible strings. In the
language of Statistical Mechanics, it would be an ``Ising model'', in which
the $\Bs$'s can be regarded as sequences of {\it spins},\footnote{\small
  Here a {\it spin} is a variable that can assume a finite number of
  values, \eg $\s=\pm1$ or $\s=1,2,..,m$.} so that the time label $i$ of
$\s_i\in\Bs$ becomes the location of the spin on a ({\it one dimensional})
lattice.

The $\m_{vol}$ can be contructed via a function $\BF$, called ``{\it
  potential}'', defined for all integers $a\le b$
on the finite strings $\Bs=\{\s_a,\ldots,\s_b\}\in
C([a,b])$  that are compatible (\ie that are restrictions to $[a,b]$ of
a string in $C({\bf Z})$. The $\BF$ has the ``short range'' property, \ie
$\BF_{[a,b]}(\Bs)$ tends to $0$ {\it exponentially} if $b-a\to\infty$ and {\it
  uniformly} in $a$ as

\be||\BF||=\sup_{a}\sum_{b\ge a}\sum_{{\Bs}\in C([a,b])} |\BF_{[a,b]}(\Bs)|
e^{\k|b-a|}<\infty\Eq{e15.1}\ee
for some $\k>0$: at fixed time $a$ the potential $\BF$ is {\it
  exponentially localized} at time $a$.

The potential $\BF$ will attribute to spin
configurations $\Bs\in C([-\t,\t])$, an ``energy'':
\be U(\Bs,\t)=\sum_{B\subset [-\t,\t]}\BF_B(\Bs_B)\Eq{e15.2}\ee
where $\Bs_B$ is the part of $\Bs$ with time labels in the interval $B$ and
the summation is over the intervals $B$ in $[-\t,\t]$.

The basic property concerns the set of $x$'s whose history $\Bs$ restricted
to $i\in\L=[-\ell,\ell]$ coincides with a given $\Bs_\L\in C(\L)$: by the
definition of history of $x$ this set is simply $P_{\Bs_\L}=\cap_{k\in\L}
P_{\s_k}$. Fixed $\Bs_\L\in C(\L)$ the normalized volume
$\m_{vol}(P_{\Bs_\L})$ is expressed in terms of the potential $\BF$ as:
\be \m_{vol}(P_{\Bs_\L})=\lim_{\t\to\infty}
\frac{\sum_{\Bs\in C([-\t,\t])}^\L e^{U(\Bs,\t)}}
     {\sum_{\Bs\in C([-\t,\t])} e^{U(\Bs,\t)}}\Eq{e15.3}\ee 
where the superscript $\L$ restricts the sum in the numerator to the
configurations $\Bs$ coinciding with $\Bs_\L$ in the sites of $\L$. 

The $\BF$ is a suitable potential, expressible in terms of the
representation of the expansion and contraction rates at the point $x$ coded
into $\Bs$, \cite{Si968a,Si968b,Ga013b}.

Eq.\equ{e15.3} can be fairly easily checked in systems of dimension $2$
(particularly if $S$ is an algebraic map of the torus, see also
\cite{FR981}) because the description, see Sec.\ref{sec14}, of the
Markovian partition can be well visualized via geometric drawings, see
Fig.3 Sec.\ref{sec14}, but requires some effort in higher dimension,
\cite{Bo970a}.

Furthermore the potential $\BF$ tends asymptotically to become $\BF^+$ to
the right of the origin but it becomes asymptotically $\BF^-$ to the left
and $\BF^\pm$ are translationally invariant.\footnote{\small
  If $B=[a,b]$ and $B+t=[a+t,b+t]$ then $\BF^\pm_{B+t}(\Bs_{B})\equiv
  \BF^\pm_{B}(\Bs_{B})$ for all $t$.}
This means:

\be \sum_{B}^\th\sum_{\Bs_B}|\BF_B(\Bs_B)-\BF_B^\pm (\Bs_B)| e^{\k
  (|B|+d(0,B))}, \ \th=\pm
\Eq{e15.4}
\ee
where $B$ are intervals $[a,b]$ to the right 
of the origin if $\th=+$ or to the left if $\th=-$ (respectively) and
$d(0,B)$ is the distance of $B$ to the origin.

Hence, if
$\Bs=(\s_{-\t},\ldots,\s_{\t})=(\Bs_-,\Bs_+)$, with
$\Bs_-=\{\s_k\}_{k=-\t}^0$ and $\Bs_+=\{\s_k\}_{k=1}^\t$,
$U(\Bs)$ can be split as

\be U(\Bs,\t)= U_-(\Bs_-,\t)+ U_+(\Bs_+,\t) +\BPs(\Bs,\t)\Eq{e15.5}\ee
where $U_\pm(\Bs^\pm,\t)=\sum_{B\subset [0,\pm\t]} \BF^\pm_B(\Bs_B)$ and 
  $\BPs(\Bs,\t)$ can be expressed in terms of a potential
$\BPs$ which satisfies a bound like Eq.\equ{e15.1} with $\BPs_B=0$
unless $B$ contains at least one of the three sites $\pm\t,0$: in other
words $\BPs$ is a {\it suitable interpolation} between $\BF^\pm$ and $\BF$.

The limit Eq.\equ{e15.3} exists as a consequence of the $1$-dimensionality
of the $\Bs$'s, of the short range of $\BF^\pm,\BF$ and of the absence of
phase transitions in stochastic processes with such potentials: the usual
SM analysis is presented only for the case of translation invariant
potentials, \cite[Sec.5.8]{Ga000}, but it works, essentially word-by-word,
also for non translation invariant potentials like the above $\BF$'s.  \*

The proof of the Eq.\equ{e15.3} is technical, \cite{Si968b},\cite{Si968a}:
in {\it heuristic} form can be found in Ch.3 of \cite{Ga013b} where it is
discussed together with several important corollaries which are summarized
in the following remarks, see also Ch.6 in \cite{GBG004}.  \*

\0{\it Remarks} 
(1) As a byproduct of the proof of Eq.\equ{e15.3} an interesting expression for
the phase space contraction emerges. Let $x$ be selected in $P_\Bs$ with
$\Bs=(\Bs^-,\Bs^+)$ (as above), then the logarithm of the total phase space
contraction in the interval $[-\t,0]$ at $y=S^{-\t}x$ can be expressed by
\be \eqalign{
  \t \s_{t[-\t,0]}(y)\defi&-\log |\det \dpr_i S^{\t}_j(y)|\cr
=&-U_-(\Bs^-,\t)+U_+(\Bs^-,\t)\cr} \Eq{e15.6}\ee
{\it up to a correction} $\BPs'(\Bs,\t)$ with $\BPs'$ a potential with the
same properties as $\BPs$ in Eq.\equ{e15.5}, hence up to a $\t$-independent
constant.
\\
(2) Eq.\equ{e15.6} is a function which has average towards the future equal
to $\t \s_+$ with $\s_+$ being the SRB average of the single step phase
space contraction $-\log |\det \dpr_i S_jx|$. Eq.\equ{e15.6} says that the
r.h.s. $-U_-(\Bs^-,\t)+U_+(\Bs^-,\t)$ and be used to replace $\sum_{i=-\t}^\t
\s(S^ix)$ up to a correction bounded by a $\t$ independent constant.
\\
(3) A second byproduct, see
Eq.3.8.5, Eq.3.11.2 in \cite{Ga013b},is
\be \frac{\m_{srb}(P_{\Bs})}{\m_{vol}(P_{\Bs})}=e^{-\t\s_{[-\t,0]}(S^{-\t}
    x) +\ldots}\Eq{e15.7}\ee
where the dots indicate a correction which is bounded by a $\t$-independent
constant: this gives details about the singularity of the SRB distribution
with respect to the volume.
\*

Existence of $\BF^+,\BF^-$ is  behind the theorem on
Anosov maps, \cite{Si968a,Si968b,Si972a}, stating that the SRB distribution
$\m_{srb}$ can be naturally represented as a PDF on the set of compatible
sequences associated with a Markovian pavement $\PP$ (any one, as there are
infinitely many of them to choose). From the general theory of the
one-dimensional Gibbs states, and from Eq.\equ{e15.3}, it can be read:
\*
\0(a) the SRB is given by
Eq.\equ{e15.3} with $\BF=\BF^+$,\\
(b) the volume distribution has the form in
Eq.\equ{e15.3} with $\BF$ in general $\ne\BF^\pm$, and
\\
(c) the SRB distribution for the {\it backward evolution}, $S^{-1}$, is
given by Eq.\equ{e15.3} with $\BF=\BF^-$: (a),(b),(c) together imply the
theorem of Sec.\ref{sec1}.
\\
(d) the phase space contraction $\s(x)$ is expressed in terms of the
symbolic history $\Bs$ of $x$ and of the potentials $\BF^\pm$ via
Eq.\equ{e15.6}. This is the key to derive the FT.
\*

With the above ``Ising model interpretation'' of the phase space volume, the
short range nature of the potentials $\BF,\BF^-,\BF^+$ and the
$1$-dimensionality of the time (\ie of the labels of the strings $\Bs$)
imply, from a SM viewpoint and as a {\it theorem}, that the
volume distribution is a simple stochastic process with very strong
ergodicity properties.

Therefore a randomly chosen point $x$ (except for a set of $x$ in a set
with zero volume) will have a well defined statistics, the SRB statistics,
such that $S^tx$ is coded into a string $\Bs_t=\{\s_{i+t}\}$ which, for
$t>0$ and large, is a {\it typical string} for the process with the ``future
potential'' $\BF^+$, while for $t<0$ and large is a typical string for the
process with the ``past potential'' $\BF^-$.

With probability $1$, with respect to the volume measure, or to any one
which has a density with respect to the volume, a point $x$ will generate a
well defined SRB statistics, {\it in general different} for the evolution
$S$ towards the future or for the evolution $S^{-1}$ towards the past. This
explains why in general the SRB distribution for $S$ and that for $S^{-1}$
are singular with respect to each other and to the volume.

The result can be suitably adapted to Anosov flows and also extended to
more general maps or flows, called Axiom A maps or flows,
\cite{BR975,Ge998}.

The structure of Anosov systems as a stochastic process with potential
$\BF$ is basic in the derivation of the fluctuation relation in
Sec.\ref{sec6}\,; it also indicates a urgent problem: namely that what said
so far might be simply insufficient to define a local phase entropy
production and to formulate a {\it local fluctuation relation} dealing with
some of the fluctuations taking place in a small region.

The problem is interesting as the fluctuations of the phase space
contraction, just because of its physical meaning, will be often
macroscopic quantities which, therefore, will be difficult to observe in
measurements.\footnote{\small Very large fluctuations can hint at
  ``violations'' of the second principle, \cite{ECM993}, hence cannot be
  observed in large systems.}

Nevertheless there is some relation that can be established between the
latter problem and the structure of the just described global SRB
distributions, and it indicates that a fluctuation relation valid for
locally observed fluctuations (\ie observed in small regions compared to
the system size) might be possible:\footnote{\small The importance of the
  problem is made obvious by a few recent experimental works, \eg
  \cite{STX005,VCBCDGGP019}}: more details are deferred to Appendix
\ref{appB}.

\def\SEC{\ Entropy\,? Stationarity \& Approach to it}
\section{\SEC}
\label{sec16}
\iniz

Boltzmann's $H$-theorem for {\it rarefied gases} led to the general
definition of equilibrium entropy as $S=k_B \log W$, as written by Planck,
where $W$ is the volume of phase space where the equilibrium distribution
is concentrated. In the $H$-theorem $S$ is the limit value of the more
general $H$-function, defined even for a nonequilibrium distribution of a
rarefied gas, which reaches its maximum $S$ on the equilibrium state.

Therefore $H$ can be regarded as an extension to nonequilibrium evolutions
(of rarefied gases {\it not in equilibrium}, but isolated and evolving
towards equilibrium) with the main feature that it is a ``Lyapunov
function'' varying with time and approaching (monotonically) a maximum
value, namely the equilibrium entropy.

Recently the Boltzmann's formula $S=k_B\log W$ has been extended to general
evolutions towards equilibrium, \cite{GGL004}, defining appropriately the
volume $W$ as the volume in phase space of the macrostate associated with
the initial microscopic state, determined by a local a coarse grained
empirical density and by the total energy (initial data consisting of
single (typical) phase space points and for a dense gas), and showing that
the new quantity appears to increase monotonically in time (towards an
equilibrium state).

This is different from a natural question arising here:
namely whether an entropy function can be associated with a
nonequilibrium stationary state, and if it even admits an extension to the
evolution towards stationary states which plays the role of a Lyapunov
function.

Going back to the origin of the ergodic hypothesis imagine the phase space
compatible with the constraints {\it as a discrete set of points} located
in the usual continuum phase space.

This is tempting as it would bring back the idea that a phase space point
wanders visiting successively all other points: it would explain the
existence of a unique stationary distribution, to be {\it therefore},
identified with the SRB distribution, which would be simply the
distribution giving {\it equal weight to all points}, whether in isolated
systems (\ie Hamiltonian evolutions) or in systems out of equilibrium (\ie
under the action of non conservative forces and thermostats): thus a
unification of the equilibrium and nonequilibrium phenomena would be
achieved.

To discuss the latter question consider a chaotic system {\it defined by a
map} on a manifold $M$ (and satisfying the CH).

Form a Markovian partition $\PP$ of the continuum phase space of a system
into finitely many ``cells'' $P_i$ and call $\m_{srb}(P_i)$ the SRB
probability of each set, \ie the frequency of visit to $P_i$ from a
randomly chosen initial data. $\m_{srb}(P_i)$ is well defined although
singular, {\it \ie not expressible} in general via an integral over $P_i$
of a density function: hence it is different from the volume $\m_{vol}(P_i)$ of
$P_i$. Then replace the continuum phase space by a {\it finite number} of
points $\NN_0$, with $\NN_0 \m_{srb}(P_i)$ of them in each $P_i\in\PP$.

The evolution should be a {\it one cycle permutation} of the phase space
points: in this way each cell $P_i$ is visited, in a very long time, with a
frequency which is, therefore, uniquely determined and is a
representation of the SRB distribution, at least for the computation of the
averages of observables whose variations in the cells $P_i$ can be
considered negligible. The time necessary might even be not be too long if
the cells $P_i$ are not too small and contain a order $1$ fraction of the
total number of (discretized) points.

But in the case of nonequilibrium the equations of motion are no longer
Hamiltonian, and are dissipative. This means that, in general, the
divergence $-\s$ of the equations of motion is not $0$ (as it is for the
isolated evolutions, \ie in the Hamiltonian cases) and must have a non
negative average $\media\s\ge0$.\footnote{\small The average $\media\s$
  cannot be $<0$, \ie phase space cannot keep expanding forever if a
  stationary state can be reached, \cite{Ru996}.}  If $\media{\s}>0$ this
means that motion evolves towards an attracting set which has zero volume:
it can be imagined (by CH) dense on a smooth surface $\AA$ of dimension
lower than that of phase space and initial data $x$ starting out of it
evolve in time with their distance to $\AA$ tending to $0$ exponentially
fast.

A discretization of phase space should therefore be a discrete
representation of the attracting set $\AA$. Under the chaotic hypothesis
heuristic arguments can be developed to {\it estimate the number $\NN$ of
  discrete points} necessary to give an accurate description of the motions
of data on the attracting set $\AA$, \cite[Ch.3.11]{Ga013b}.

The points on which the dynamics develops can be obtained by covering phase
space with a {\it uniform lattice} with meshes $\d p,\d q$ in momentum and
position and representing the dynamics as a map on the discrete set of
$\NN_0$ points so obtained: then {\it select $\NN\le \NN_0$ of them which
  are recurrent}; such points exist, having supposed that the motion can be
represented on a regular discrete lattice and the SRB is ergodic. Here one
should have in mind the {\it numerical simulations} of chaotic dynamical
systems: there the evolution is {\it literally} simulated as a map
(``code'') of a discrete set of points, digitally represented and regularly
spaced.

Then it is natural to try to define entropy of the SRB state the quantity
$S=k_B \log \NN$.  A {\it heuristic estimate} of $\NN$, under the CH, has
been proposed in \cite[Ch3.11]{Ga013b} as sketched below.

First refine the Markovian partition $\PP=\{P_\s\}_{\s=1}^n$ into $\wh
  \PP=\vee_{-\t}^{\t} S^k\PP$, \ie define the partition whose elements have
  the form $\cap_{i=-\t }^{\t } P_{\s_i}=P_{\Bs}$, choosing $\t$ so large
  that the size of each element is so small that the {\it few observables}
  of interest have a {\it constant value} in each $\wh P=P_{\Bs}$.

Therefore choose $\t$ so that $e^{-\l \t}\d=\d'$ where $\d$ is the maximal
linear dimension of the $P\in\PP$, $\d'$ is the maximal linear dimension of
$\wh P\in\wh \PP$ and $\l$ is the minimum Lyapunov exponent of $S$: thus
$\t$ depends on the precision
$\t=\l^{-1}\log\frac{\d}{\d'}$.\footnote{\small The CH implies that there
  is no vanishing exponent for the map.}

Let $\NN$ be the number of the points on the attractor and $\NN_0$ be the
number of points in the regular lattice over which the dynamics is
discretized: then in a cell $P_\Bs$ the numbers of points will be
respectively $\NN\m_{srb}(P_\Bs)$ and $\NN_0\m_{vol}((P_\Bs))$.

Therefore a simple estimate, \cite{Ga013b}, of the number points of the
uniform lattice that must be recurrent to guarantee a ``faithful'' discrete
representation of the dynamics over a time $\t$ is
\be \NN\le \NN_0 \min_\Bs
\frac{\m_{srb}(P_\Bs)}{\m_{vol}(P_\Bs)}\Eq{e16.1}\ee 
where the minimum is over the histories $\Bs\in C([-\t,\t])$.  The
Eq.\equ{e15.7} leads to: $\NN\le \NN_0 e^{-\media{\s}\t}$ with $\media{\s}$
equal to the SRB average of the phase space contraction $\s(x)$. Hence:
\be S=k_B\log\NN\le k_B(\log\NN_0
-\frac{\media{\s}}{\l}\log\frac{\d}{\d'})\Eq{e16.2}\ee
and changing the precision of the observations, \ie changing the
observables determining $\d'$, $S$ changes by a quantity which depends on
the SRB distribution, if $\media{\s}>0$, via $\media{\s}$ and the smallest
non zero Lyapunov exponent, except when {\it $\media{\s}=0$},
\cite{Ru996}). This provides some evidence that $S$ is not defined just up
to an additive constant.

This is in sharp contrast with the equilibrium result (in which
$\media{\s}=0$) where changing the precision changes $\log\NN$ by a
constant {\it independent of the particular equilibrium state} studied.
And, although the derivation of the estimate is heuristic (and is an
inequality), it seems to indicate that entropy, as a {\it function of
  state}, might not be definable for stationary states out of equilibrium,
\cite[Sec.3.10,3.11]{Ga013b}.

Nevertheless one of the main features of the {\it extension} of entropy, as
$S=k_B\log W$, to rarefied gases not in equilibrium but isolated and
evolving towards equilibrium, is that it is a ``Lyapunov function'' varying
with time and approaching (monotonically) a maximum value as a limit value,
namely the equilibrium entropy, \cite{GGL004}.

It is conceivable that {\it also} in the evolution to a stationary state it
could be possible to define a Lyapunov function with the same property of
evolving (possibly not monotonically) to a maximum which is reached at
stationarity, \cite[Ch3.11]{Ga013b}, as briefly discussed below.

Consider as an initial non stationary distribution {\it a delta function on
  a single point in phase space}, for simplicity.  Then the fraction
$P(\x,t)$ of times that the point $\x\in\AA$ is visited tends to
$\frac1\NN$, as prescribed by the SRB distribution in the above discrete
representation, where $\NN$ is the number of points in $\AA$. Therefore:
\be \kern-3mm\eqalign{&S(t)=k_B\sum_\x -\lis{P(\x,t)}\log\lis{P(\x,t)}\cr
  &\tende{t\to+\infty} S_\infty=k_B \sum_\x -\frac1\NN \log \frac1\NN=k_B
  \log\NN\cr}\Eq{e16.3}\ee
Hence $S_\infty$ is the maximum value that $S(t)$ can
reach: so that $S(t)$ can play the role of a Lyapunov function.

Although $S_{\infty}$ depends {\it non trivially} on the precision of the
discretisation used still, for all choices of the discretisation, the $S(t)$
will have the property of evolving to reach ({\it however not necessarily
  monotonically}) the maximum value on the SRB distribution, \ie on the
natural stationary state. Entropy might be not defined in general
stationary states, as a function of state, although in the approach to
stationarity it could be a Lyapunov function (not unique) extending the
equilibrium entropy function, \cite[Sec.3.12]{Ga013b}.
 
\def\SEC{\ Viscous Fluids and Reversibility}
\section{\SEC}
\label{sec17}
\iniz

The analysis of the previous sections deal essentially with systems of
particles and leaves out the important class of stationary distributions
that arise in systems normally described via PDE's, but often can be also
described by properties of assemblies of microscopic particles, via
suitable scaling limits, \cite{Pr009}.

This suggests that it should be possible to apply the same ideas to
macroscopic systems, like fluids. Of course the theory of
chaos was developed precisely for such systems,
\cite{RT971,Ru989,Ru989b,Ru995}: however, if systems like fluids are
considered, the reversibility is usually lost in the macroscopic
descriptions. 

Yet friction, responsible for the loss of reversibility, is a
phenomenological notion and it can be thought that the same systems could
admit equivalent descriptions via other equations, possibly even reversible.

A key might be the theory of ``ensembles'' for stationary non
equilibrium states, following the proposals considered in
Sec.\ref{sec7}\,-\,\ref{sec9}. An attempt in this direction is presented now
focusing attention on the incompressible Navier-Stokes fluids. A first
step is to propose, via the example of the NS equations, that the
stationary states of macroscopic systems that are scaling limits derived
from microscopic molecular evolutions can be described, in suitable
circumstances, by reversible equations, and equally well.

In the case of the NS equations the proposal goes back, in a related
context, to \cite{SJ993} and, in the form proposed below, to works
summarized in \cite{Ga013b}:
it appeared already in
\cite{GC995,BGG997,Ga996b,Ga997b,Ga000a,Ga000b,Ga001a,Ga006d}.

The classical NS equation in dimension $d=2,3$, for a velocity
$\uu(\xx)=\sum_{\kk\ne\V 0} e^{-i\kk\cdot\xx} \uu_\kk$, with periodic boundary
conditions in $[0,2\p]^d$, is
\be\dot \uU+(\uu\cdot\BDpr) \uU=\n\Delta \uU + {\bf f} - \BDpr p ~, \quad
  \BDpr \cdot \uu = 0
\Eq{e17.1}\ee
where the external forcing ${\bf f}$ is supposed to be {\it concentrated on the
large scale} Fourier components, actually it will be supposed to have only
one Fourier's component $f_{\pm\kk_0}$ with $|f_{\pm\kk_0}|=\frac1{\sqrt2}$ and
$\kk_0=(2,-1)$, to fix ideas.%
\footnote{\small More generally the forcing can be supposed to have
  $f_\kk\ne0$ only for $|\kk|<F$, with $F$ being a fixed cut-off. The cases
  $\kk=(0,\pm1)$ and $\kk=(\pm1,0)$ are somewhat trivial, see
  \cite{Ma986}.}  The equation is not reversible for the time reversal map
$I\uu=-\uu$ and will be called INS.\footnote{\small Viscosity plays the
  role of a model of thermostat: the fluid keeps a constant temperature in
  spite of the viscosity: therefore viscosity is a model for the
  undisclosed mechanism keeping the temperature constant.}

In the above dimensionless form the viscosity is written $\n=\frac1R$,
where $R$ is usually called ``Grashof's number''. The viscosity is a
phenomenological notion derived from reversible microscopic equations of
motion, \cite{Ma867-b}, and it is possible to think that the coefficient
$\frac1R$ could be replaced by a Lagrange multiplier designed to hold
constant a property characteristic of the flow.

The dissipation per unit time is $\n\DD(\uu)\equiv\frac1R\DD(\uu)$ with:
\be \DD(\uu)=\frac1{(2\p)^2}
\int (\BDpr\uU)^2d\xx=\sum_{\kk\ne\V0} \kk^2|u_\kk|^2\Eq{e17.2}\ee
which is called {\it enstrophy}, controls statistical properties
of the flow through its average. Therefore a first proposal is to replace
the viscosity $\frac1R$ with a multiplier such that
$\dpr_t\DD(\uu)\equiv0$. This leads immediately to
\be\eqalign{
&\dot \uU+(\uu\cdot\BDpr) \uU=\a(\uU)\Delta \uU + \V f - \Tdpr p ~, \quad
  \Tdpr \cdot \uU = 0\cr
&\a(\uu)= \frac{\sum_\kk \kk^2\lis {{\bf f}}_\kk\cdot \uu_\kk}
    {\sum_{\kk\ne\V0} \kk^4 |\uu_\kk|^2}\cr
}\Eq{e17.3}\ee 
in space dimension $d=2$. The equation is reversible for the time reversal
map $I\uu=-\uu$ and will be called RNS.%
\footnote{\small In dimension $d=3$: $\a=\a(\uu)$ has to be modified by
  adding to the numerator of Eq.\equ{e17.3} be quantity $
  \sum_{\kk_1,\kk_2} (\kk_1+\kk_2)^2$ $(\uU_{\raise 3mm
    \hbox{$\kk_1$}}\cdot i\, {\T\kk}_{\raise 3mm \hbox{$\scriptstyle2$}})
  (\lis\uu_{\kk_1+\kk_2}\cdot\uu_{\kk_2})$.}

Let $N$ be a cut-off and consider the evolutions for INS and RNS
above in $d=2$ for simplicity. Then the evolution equation for
$\uu_\kk=i\frac{\kk^\perp}{|\kk|} u_\kk$ are

\be \eqalign{
  \dot u_\kk=&-\sum_{\kk_1+\kk_2=\kk}
\frac{(\kk_1^\perp\cdot\kk_2)(\kk_2^2-\kk_1^2)}
     {2\,|\kk_1|\,|\kk_2|\,|\kk|}u_{\kk_1} u_{\kk_2}\cr
&-\b\, \kk^2u_\kk + f_\kk,\quad |\kk_2|,|\kk_2|,|\kk|\le N\cr}\Eq{e17.4}\ee
where $\b=\frac1R$ in the case of INS and $\b=\a(u)$ in the case of RNS and
in both cases $|\V u_{\V k}|=0$ for $|\V k|>N$ or $|\V k|=0$.

The size of the parameter $R$ controls the stability of the evolution: {\it for
  simplicity} it will be supposed that if $R$ is large enough then all
initial data, with the exception of a set of zero volume, evolve towards a
unique attracting set and define a unique stationary distribution at least
if $N$ is not too small.

The stationary distributions of the two equations will be parameterized by
$R$ for INS and by $D$, the constant value of the enstrophy. And the
question will be whether there is a correspondence $R\otto D$ which
associates distributions which are ``equivalent'', \ie assign equal
averages to suitable classes of observables.  \*

\0{\it Remarks:} (1) It is well known that at fixed $N$ it is not true that
there is a unique stationary distribution at given $R$ or $D$: at small
$R$ (\eg $R<60$) by direct calculations by and accurate simulations this is
shown. In 
\cite{BF980,FT979,FTZ984,FT985,FGN988,BF991} the phenomenon of
``hysteresis'', \ie coexistence of several attracting sets, is discussed in
detail.
//
(2) One of the reasons behind the phenomenon (but by no means the only
one) is the ``gauge'' symmetry of the NS equations: if there is $\V a$ such
that $\kk\cdot\V a=0$ for all $\kk$ for which $f_\kk\ne0$, then if $u_\kk$
is a solution also $e^{i \kk\cdot\V a}u_\kk$ is a solution. Hence if
$f_\kk=c\,\d_{\kk,\pm\kk_0} e^{\pm i\g}, \, c\in R$ also $u_\kk
e^{i\th_0\kk\cdot\kk_0^\perp}$ is a solution: for each $\th_0$ an invariant
set of data is then defined.  \\
(3) In the several cases some (or all) of the invariant sets may be stable and
several stationary states will coexist.\footnote{\small For instance if
  $f_\kk$ is real and $\kk\cdot \V a=0$ if $f_\kk\ne0$, there is a solution
  with $u_\kk$ real: and we have infinitely many invariant sets in which
  $u_\kk$ has the form $v_\kk e^{\th_0\,\kk\cdot\V a}$, parameterized by
  $\th_0$.}
\\
(4) Symmetry, or more generally existence of more than one attracting surface,
divides the stationary distributions into equivalence classes: and the
particular stationary distribution that is reached starting from a given
initial data $u$ may depend on $u$, see remark (iv) in Sec.\ref{sec6}. In
these instances equivalence may be difficult to check. Unless for all data
$u$, aside a set of zero volume, the stationary distribution is unique and
we are interested only in generic behavior in the space of velocity fields
$u$ {\it with complex components}.
\\
(5) The simplification of uniqueness of the attracting set (on which the CH
holds) that will be used below means that all invariant sets, except one,
become unstable at large $R$.
\\
(6) In the following we shall proceed under the above uniqueness
assumption. However this is not essential: if there are several possible
attracting sets then they will have to be distinguished by labels $\g$ and
the stationary distributions will be parameterized by the labels:
equivalence becomes in this case the existence for both equations of an
equal number of attracting sets and all parameters determining them can be
put in correspondence so that the corresponding stationary distributions
assign the same averages to the local observables.  \*

Call $S^{irr,N}_t, S^{rev,N}_t$ the evolutions generated on the phase space
(of dimension $4N(N+1)$ if $d=2$) by the two equations. The SRB
distributions will be parameterized respectively by $R,N$ or by $D,N$ where
$D=\DD(u)\defi\sum_{\kk}\kk^2 |u_\kk|^2$ is the (constant) enstrophy and
constitute elements of the ``ensembles'' $\EE^{irr,N}$ and $\EE^{rev,N}$
respectively, whose elements will be denoted $\m^{irr,N}_\n$ and
$\m^{rev,N}_D$, respectively.

The discussion in Sec.\ref{sec7}-\ref{sec9} suggests considering the two
collections of SRB distributions and estabilsh a correspondence $\sim$
between $\m^{irr,N}_R\in \EE^{irr,N}$ and $\m^{rev,N}_D\in \EE^{rev,N}$
by, see Eq.\equ{e17.2},
\be \eqalign{
  &\m^{irr,N}_R \sim \m^{rev,N}_D\quad {\rm if} \quad
  \m^{irr,N}_R(\DD)=D,\cr
}
\Eq{e17.5}\ee
\0{\it Conjecture:} {\it If $O(u)$ is a ``local'' observable, in the sense
  that $O$ depends only on the components $u_\kk$ with $|\kk|<K$:
\be \lim_{N\to\infty} \m^{irr,N}_R(O)=\lim_{N\to\infty} \m^{rev,N}_D(O), \
\forall K\ {\rm prefixed}\Eq{e17.6}\ee
provided the equality in Eq.\equ{e17.5} holds as a relation between
$R\equiv\frac1\n$ and $D$.}  \*

Multiplying both sides of Eq.\equ{e17.4} by $ u_{-\kk}$ yields that the
time derivative $\dot E$ of the energy $E=\frac12\sum_\kk |u_\kk|^2$ is given by
$-\frac1R \DD(u)+W(u)$ or $-\a(u)D+W(u)$ with $W(u)=\sum_\kk f_\kk
u_{-\kk}$: where $W$ is the work done per unit time by the
external force $\V f$.

Hence since $W$ is a local observable, as $f_\kk$ has been
supposed such, the average of $W$, which will be called $W^a$ for
$a={(rev,N)},{(irr,N)}$ respectively, has to be the same in equivalent
stationary states if $N\to\infty$, \ie:
\be W^{irr,N}\equiv\n\m^{irr,N}_R({\cal }D), \quad
  W^{rev,N}\equiv\m^{rev,N}_{D}(\a) En\Eq{e17.7}\ee
because the average of $\dot E$ has to vanish. Hence the equivalence condition
$\m^{irr,N}_R({\cal D})=D$ immediately implies:
\be R \m^{rev,N}_D(\a)\tende{N\to\infty}   1\Eq{e17.8}\ee
which becomes a key preliminary test of the conjecture when initial data are
randomly chosen and the evolution has a unique stationary state.

And the equivalence condition, if the conjecture
holds, receives the interpretation that the average work done by the
forcing and dissipated per unit time {\it is the same in the two
  evolutions}.  

In the cases in which there are several attracting sets, hence several SRB
distributions, the conjecture has to be modified (see remarks (iv) in
Sec.\ref{sec6} and (6) above) simply by saying that if $\g,\g'$ are labels
distinguishing the extremal distributions in $\EE^{irr,N},\EE^{rev,N}$,
with a given $R$ and the corresponding $D,$ then a correspondence between
$\g$ and $\g'$ is eventually, for $N$ large enough, possible so that
Eq.\equ{e17.8} holds.

If holding, the conjecture would establish a strong analogy between, on one
hand, the theory of the thermodynamic limit of the canonical and
microcanonical equilibrium ensembles and, on the other hand, the above
proposed equivalence of ensembles of SRB distributions for the INS and RNS
equations. The $\EE^{irr,N}$ is analogous to the canonical ensemble with
$\n=R^{-1}$ corresponding to $k_BT$ and $\EE^{rev,N}$ is analogous to the
microcanonical ensemble with $D$, the enstrophy, corresponding to the
energy. The observables $O$ play the role of the local observables and
their localization in momentum corresponds to the localization in space in
the thermodynamical equilibrium ensembles.

The above conjecture can be tested and some tests are being made in
simulations. It is also emerging that the conjecture could be strengthened
to cover also the Lyapunov spectra of equivalent elements of the two
nonequilibrium ensembles.

\def\SEC{\ Simulations on 2D-NS}
\section{\SEC}
\label{sec18}
\iniz

Consider the two equations Eq.\equ{e17.4} and fix $R=2048$: the 
conjecture stated in the previous section can be tested in simulations. The
cut-off will be set, in the tests that follow, at $960$ Fourier's modes \ie
$|k_i|\le15$. The first test is to check the Eq.\equ{e17.8}: in all cases
below the evolution is empirically chaotic.

The figure, as well as the subsequent ones, is obtained after running the
{\it irreversible evolution} at $R=2048$, with $960$ modes for a long time
to obtain the average value $D$ for the enstrophy: this realizes the
equivalence condition Eq.\equ{e17.5}. Then the conjecture would predict
that in {\it reversible evolution}, run from an initial data with enstrophy
$D$, the average of $\a(\V u)$ should be $\frac1R$ . The first simulations
yields Fig.4.

\eqfig{200}{120}{}{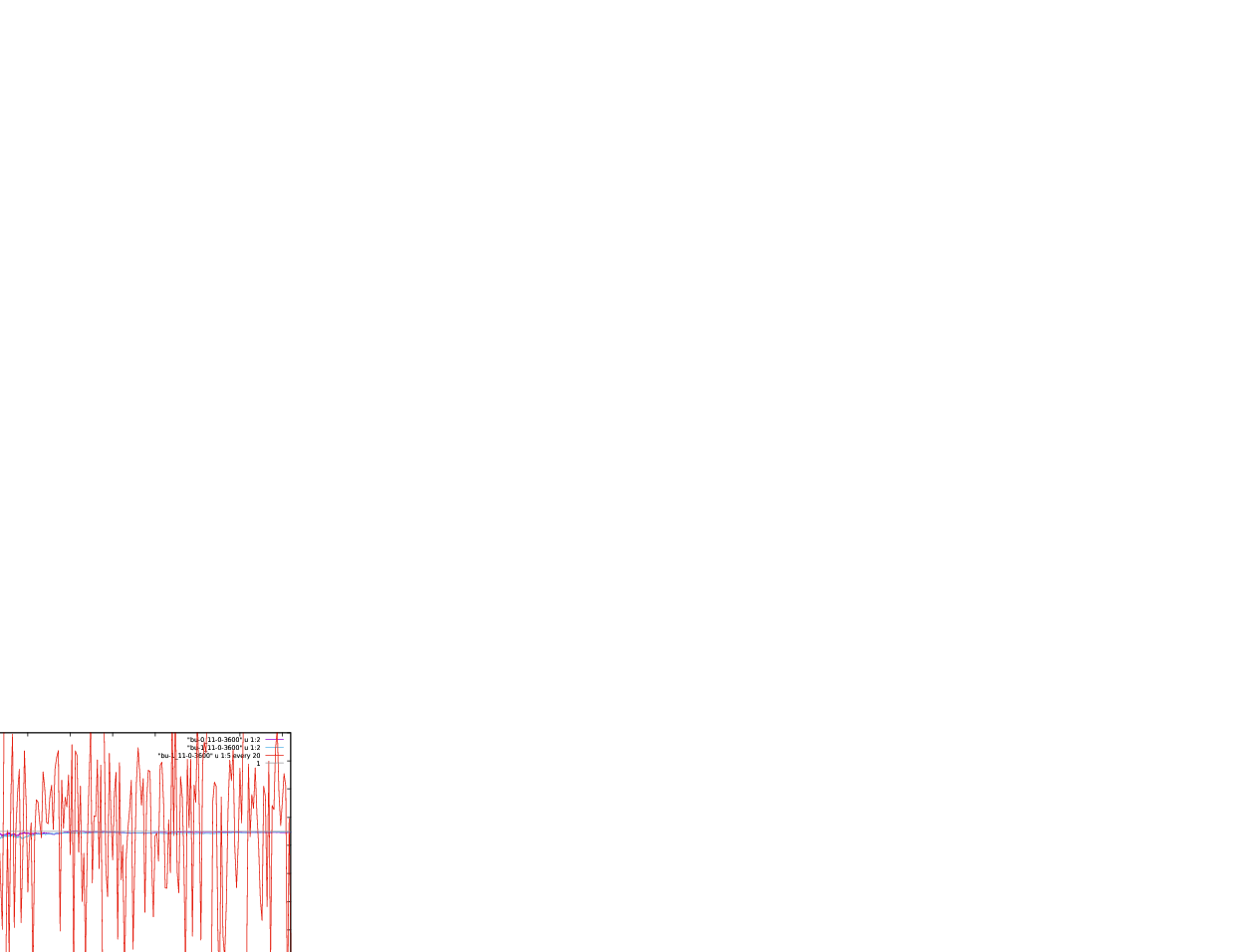}{}

\0{\small Fig.4:\it Reversible evolution \alert{$NS_{rev}$}: running average
  of the ``reversible friction'' $R \a(u)\equiv R\frac{2 Re(f_{-\kk_0}
    u_{\kk_0})\kk_0^2}{\sum_\kk \kk^4|u_\kk|^2}$, superposed to the
  conjectured value $1$ and to the fluctuating values $R \a(u)$:
  R=2048, 960 modes, $\l_{max}$= max. Lyapunov exp. $\simeq
  1.5$, integration step $h=2^{-17}$, $x$-axis time unit $4h$, 
  forcing $f_\kk=0$ except $f_{\pm(2,-1)}= {e^{\pm i\p/3}}/{\sqrt2}$; hence
  time unit in abscissa corresponds to $2^{19}$ integration steps: data are
  plotted by lines at such time intervals.
  Superposed also to the running average of $R\a(u)$ in the
  equivalent irreversible NS eq. The two running averages and the line 1
  are not easy to distinguish on the scale of the drawing.}  \*

A daring test, which goes beyond the conjecture, deals with the equivalence
of the exponents of the Jacobian matrix evaluated in the two equations
under the equivalence conditions: the result for the same truncation of the
equations ($960$ modes) are drawn, on the same frame which reports the
exponents for the Jacobian matrix over a time of the order of
$\l_{max}^{-1}$ where $\l_{max}$ is (an approximation of) the largest
Lyapunov exponent. Such exponents will be called ``local Lyapunov
exponents'',\footnote{\small They are defined in
  terms of the diagonal elements of the QR decomposition of the matrix
  $\dpr_u (S_\t u)$, linearizing the flow, with $\t$ fixed: in the picture
  $\t=h 2^{10}$ was chosen, which is a small fraction of the time unit
  fixed by the integration step (which is in the pictures
  $h=2^{-17}$). Approximating the matrix $\dpr_u (S_\t u)$ as $V=(1+h
  J(u))^{\t h^{-1}}$, where $J(u)=\dpr_u\dot u$ is the Jacobian matrix of
  the flow, the QR decomposition of $V$ gives a triangular matrix $R>0$ and
  the logarithms of its eigenvalues divided by $\t$, denoted $\L_k(u)$,
  $k=1,2,\ldots$, are sampled as $u(t)$ evolves in time every $4 h^{-1}$
  steps and their averages are the definition adopted here of the local
  Lyapunov exponents $\l_k$. A much more accurate definition would be
  replacing $V$ with the time ordered product $\prod_{k=0}^{\t h^{-1}} (1+h
  J(S_k u))$: but this greatly increases the computation time (by a factor
  $2^7$ here). Alternatively one could consider the eigenvalues of the
  symmetric part of $J(u)$, \cite{Ru982}: but this also requires a large
  computation time.\label{QR0}} and are different from the Lyapunov
exponents whose evaluation would require substantially larger computation
time), \cite{Ru982,Li984}.

\eqfig{180}{120}{}{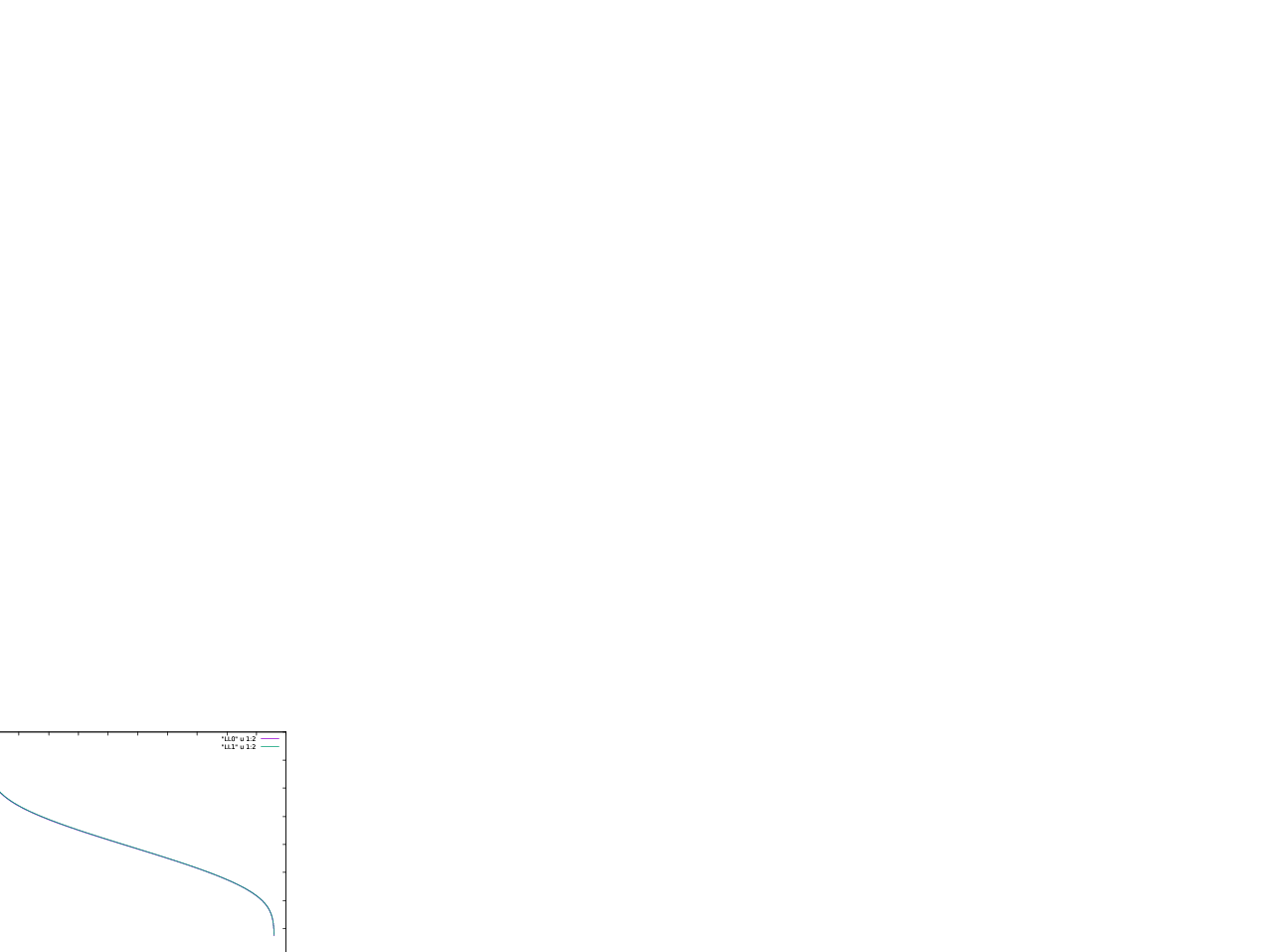}{}

\0{\small Fig.5:\it The {\it local}, over a time step $4h^{-1}$, Lyapunov
  spectra for $960$ modes truncation: {\it reversible and irreversible
    superposed}. The sum of the (local) exponents in Fig.4 is $<0$.  } \*

The two spectra look quite identical: and the relative difference of
corresponding exponents
($|\l_i^{rev}-\l_i^{irr}|/{\max(|\l_i^{rev}|,|\l_i^{irr}|)}$ is perhaps
more informative:

\eqfig{180}{120}{}{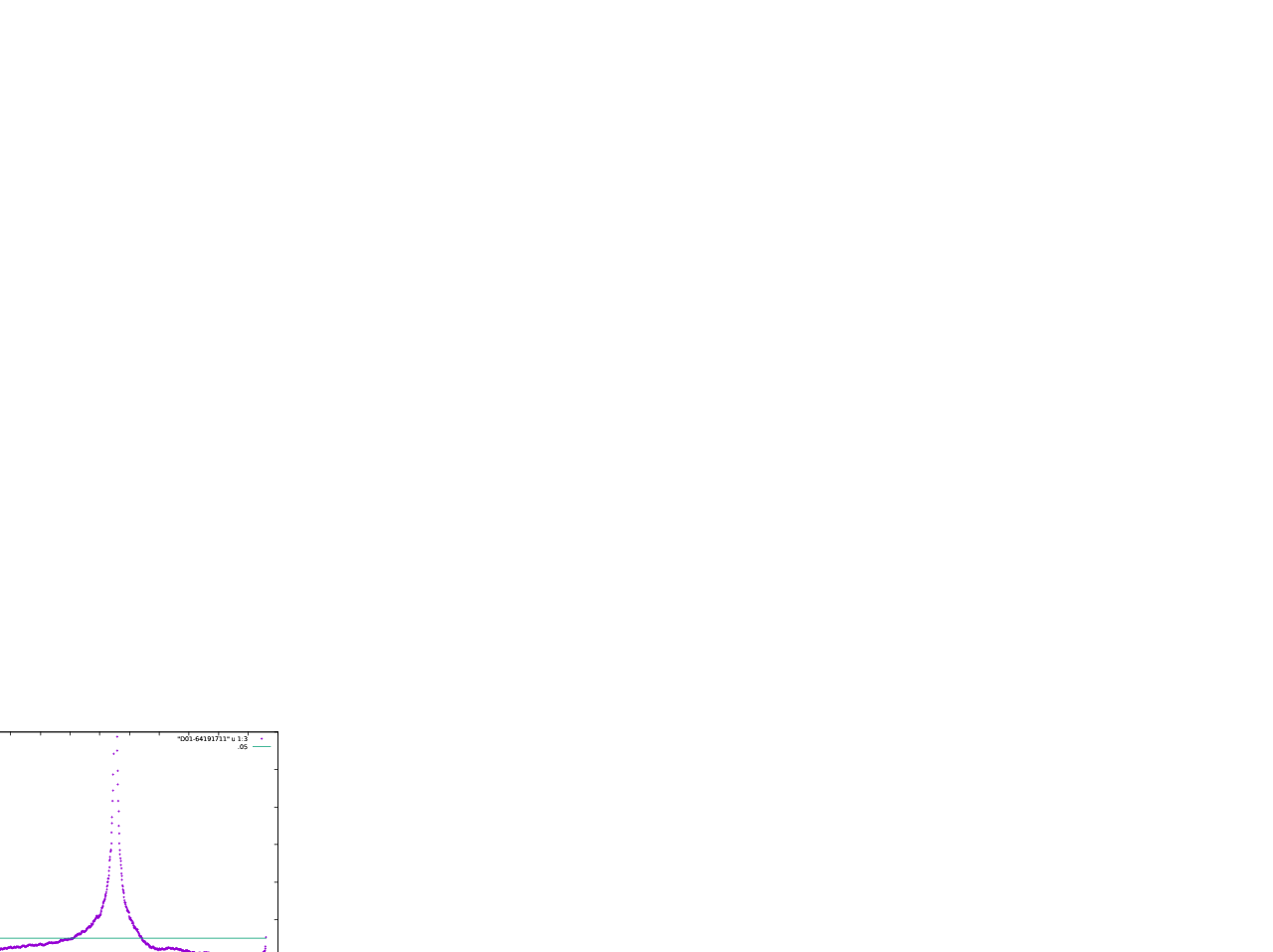}{}

\0{\small Fig.6: \it Relative difference betweeen (local) Lyapunov exponents
  in the previous Fig.5; \alert{R=2048}, $960$ modes. The bar marks the $5\%$
  discrepancy, and the lines are visual aids.
}
\*

It is remarkable that the (local) Lyapunov exponents may provide an example
of a pairing rule, see Sec.\ref{sec8}:

\eqfig{180}{120}{}{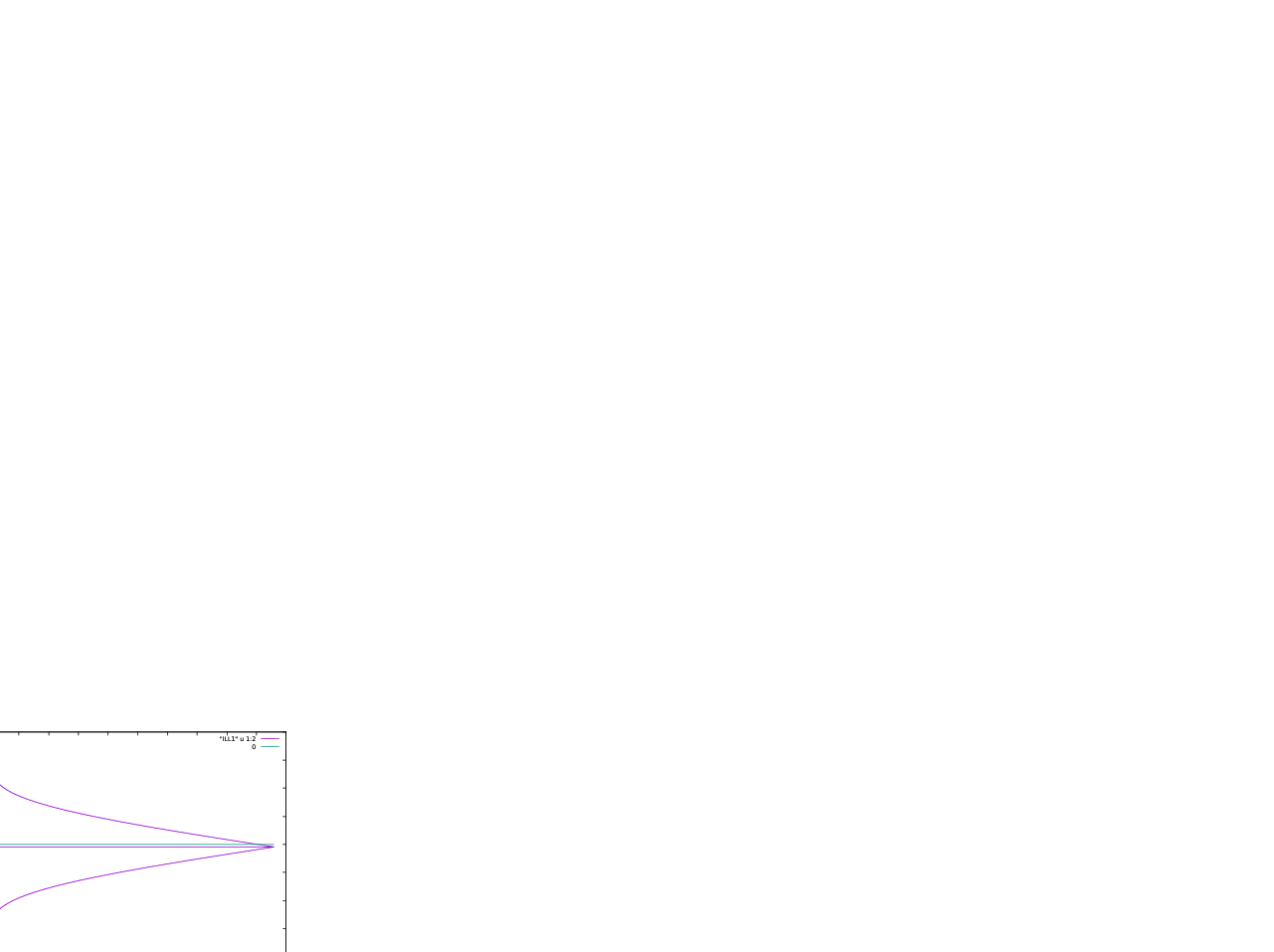}{}

\0{\small Fig.7: \it The approximate pairing rule: graph of
  $\frac12(\l_k+\l_{n-1-k})$, $n=960$, with the $\l_k$ the local exponents
  in the previous Fig.5; \alert{R=2048}, $960$ modes. 
}
\*

A pairing rule emerges from Fig.7. This remarkable fact possibly {\it
  suggests} that the pairs consisting of two negative exponents are
associated with the attraction by the attracting set and the dimension of
the latter is therefore twice the number of exponents $>0$, while the
fractal dimension of the attractor is the KY dimension computed using only
the pairs of exponents of opposite sign. In the case of the previous
picture the following Fig.8 provides a detail with a clearer
pairing illustration:

\eqfig{180}{120}{}{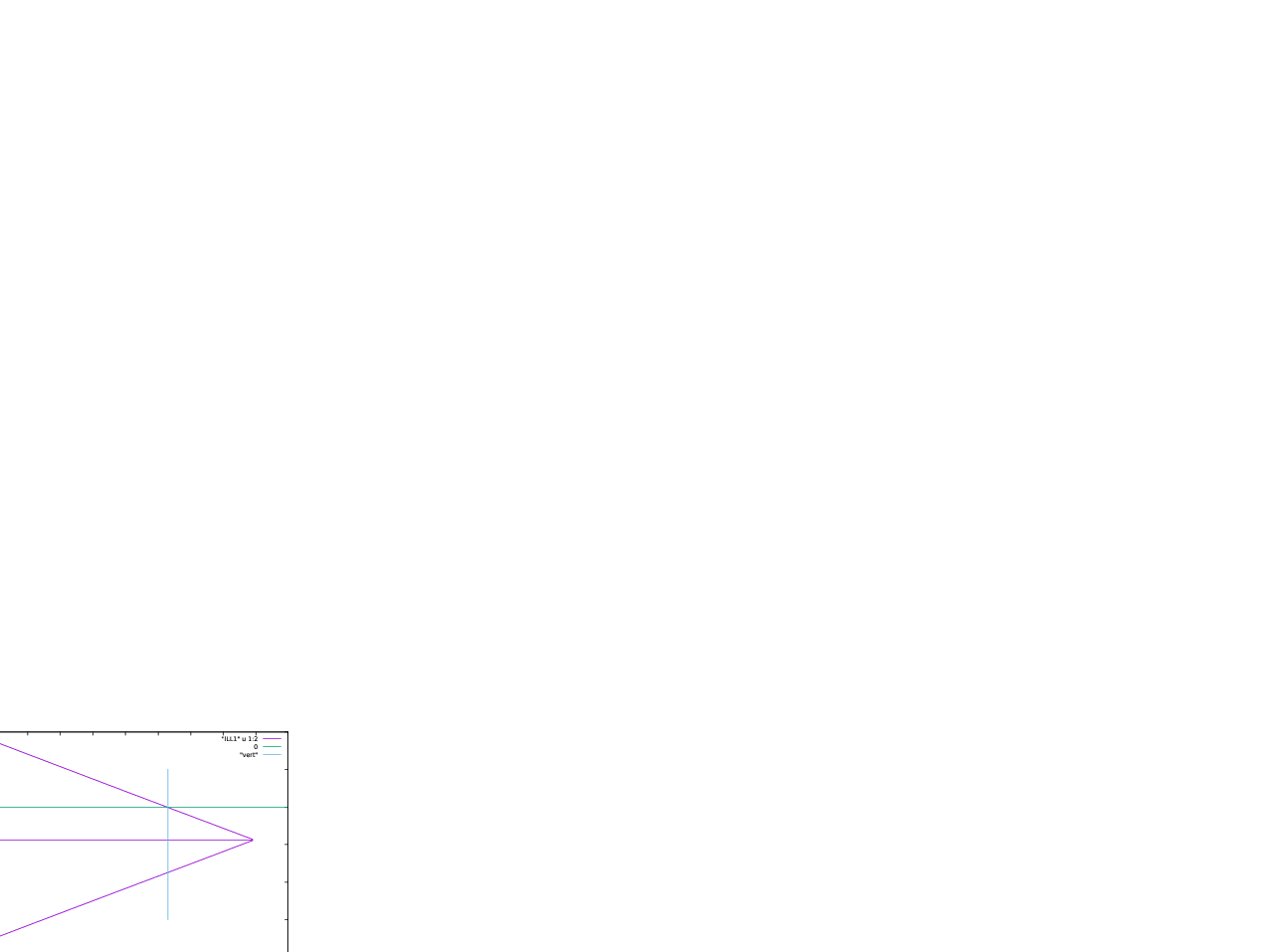}{}

\0{\small Fig.8: \it Detail of Fig.7 showing the pairs of opposite sign and the
  ones of equal (negative) sign. The vertical line marks the $k\simeq452$
  where the negative pairs begin to appear: hence suggest a dimension of
  the attracting set $~904$ out of $960$. Graph of
  $\frac12(\l_k+\l_{n-1-k})$, $n=960$, with the $\l_k$ the local exponents
  in the previous Fig.5.  } \*

The pairing appears exact, but is is not: as it could be seen by drawing
the pairing line on a larger scale. Still even on the scale of Fig.5 it is
not possible to distinguish the pairing line from an exactly horizontal
line.

A pairing property, quite manifest in Fig.5,6, was proposed in
\cite{Ga997b} as possible in NS fluids. It could be an approximate pairing
reflecting an exact one which should hold for the spectrum of the fluid
equations with ``Ekmann friction'' (\ie with viscosity force $-\n \uu$
instead of $\n\D\uu$).

For zero viscosity and forcing the equation can be considered a
Hamiltonian equation with conjugate variables $(\Bd,\uu)$, called
Arnold-Euler equation, where $\Bd(x)$ is the displacement (with
respect to an initial configuration of fluid particles) of the
``fluid particle'' that reaches the point $x$ at the instant in
which the fluid velocity at $x$ is $\uu(x)$. So $\uu$ is a
momentum variable while $\Bd$ is a position variable and
$\frac{d}{dt}\Bd(x)=\uu(x)$ (\eg see \cite{Ga997b}).


Formally the Ekmann's equation, aside the infinite
dimensionality, is covered by a pairing theorem, being
Hamiltonian, if viscosity and forcing vanish: hence its Lyapunov
spectrum should have the exponents paired to $0$.  The NS
equation viscosity is not proportional to $\uu$ and forcing is
not a gradient: the argument in \cite{Dr988} does not apply, not
even formally. Still in \cite{Ga997b} the possible pairing in the
NS spectrum is discussed (called ``barometric formula'') and for
large cut-off is proposed to pair $\l_k,\l_{n-1-k}$ to a suitable
curve $c_k$ (which would be close to a constant in large
intervals of $k$).

A few more simulations have been performed to test the conjecture, all in
2D, because the 3D case is too demanding. 
For a few further tests in systems with $48,224,960,3968$ modes (\ie
increasing the cut-off $N$) and for $R$ up to $8192$, see \cite{Ga019}
where particular attention is dedicated to the approximate pairing, see
Sec.\ref{sec8}, of the Lyapunov exponents. Very few tests have been done
for $R$ small: but the conjecture should hold even in the laminar regimes;
\ie when at given forcing the attractors can be coexisting stable periodic
motions.

Furthermore changing the forcing to allow a $\V f$ with more than a single
mode, but still keeping it acting only on the large
scale $\V k$'s and of size $||\V f||_2=1$, the average enstrophy can change
substantially but the results on the equivalence remain encouraging.
Also the precision, \ie the integration time step $h$, can have strong
influence: even hysteresis may appear if $h$ is not small enough even
though it disappears for smaller $h$.

The results are still preliminary and hopefully will be continued not only
to check those so far obtained but also to study further tests and
refinements.
\*

\0{\it Remarks:} 
(1) Since a real force $f_\kk$ transforms real data $u_\kk$ into real ones,
there will be an invariant distribution concentrated on real velocity
fields $u_\kk$: it may, as $N$ (or $R$ or both) grows, become unstable to
perturbations of $u$ which break the symmetry (\ie reality of
$u_\kk$). Nevertheless such distribution may be unique among those which
are generated by a real initial $u_\kk$: hence, as mentioned in
Sec.\ref{sec17},  to check equivalence it
becomes necessary, in general, to identify other invariant conditions on
$u$ on which to base the selections of pairs of equivalent distributions
besides the corresponding $R$ and $E$; for instance compare only
distributions concentrated on real $u_\kk$'s.
\\
(2) The simplest checks of the equivalence concern ``gauge invariant''
observables: at least possible different stationary states related by the
symmetry (\ie that can be transformed into each other by application of the
symmetry) will attribute the same averages to such observable. The
$R\media{\a}_{rev}=1$ is an example and the average work
$\media{W}_{rev}=\media{W}_{irr}$ or also
$\media{|u_\kk|^2}_{rev}=\media{|u_\kk|^2}_{irr}$ are examples. 

\def\SEC{\ Other relations. Comments.}
\section{\SEC}
\label{sec19}
\iniz

Several {\it universal} relations have been proposed in the recent
literature. I select below two among them.

\subsection{Transient fluctuation theorem}
Deals, \cite{ES994}, with reversible evolutions starting from random
initial data {\it chosen from an equilibrium distribution} of particles
(hence, in the nontrivial cases, not stationary), of {\it Boltzmann-Gibbs kind},
or more generally from a distribution {\it symmetric under time reversal}
(which, in most cases, is velocity reversal) and with density with
respect to the volume.

In this case the statement is that the probability density that a phase space
volume contracts by a factor $e^A$ compared to the probability that it
contracts by $e^{-A}$ in a time interval $\t$ is such that:

\vglue-6mm
\be \frac{P(A)}{P(-A)}=e^{A}, \quad \forall \t< \infty\Eq{e19.1}\ee
which is an immediate consequence of the definition (\ie of the above few
lines preceding it): {\it no further assumption is necessary}.

Since Eq.\equ{e19.1} is sometimes compared to Eq.\equ{e6.3} (or
Eq.\equ{e6.4}) then, for the purpose of comparison, the $\s_+$ should be
defined as the average as $\t\to\infty$ of $\frac{A}\t$ and $p$ should be
set $p\defi \frac{A}{\t \s_+}$. In terms of $p,\s_+,\t$ Eq.\equ{e19.1}
becomes formally identical to Eq.\equ{e6.4}.

It is claimed that Eq.\equ{e19.1}, being valid for all $\t$, will imply
the fluctuation relation for the stationary state reached by the evolution
at infinite time and for the variable $\frac1{\t\s_+} A$. However the
stationary state in nonequilibrium cases is singular with respect to the
initial state and typical fluctuations observed in time $\t$ have the form
$p\s_+\t$ so, if $\s_+>0$ as it is in nonequilibrium cases, the quantity
$A$ in Eq.\equ{e19.1} has an unclear meaning when the system has reached a
stationary state and a time $\t=+\infty$ has already elapsed.%
\footnote{\small Simple examples of the meaning of Eq.\equ{e19.1} compared
  to FR can be constructed: which exhibit systems, {\it as chaotic as
    wished}, evolving towards a stationary state with average phase space
  contraction $\s_+>0$ and which for every finite time satisfy
  Eq.\equ{e19.1} but at infinite time do not satisfy the FR. An example of
  such a map follows: let $S_0$ be a map on the unit
  circle $\TT$ defined by the evolution at time $t=1$ (say) of
  $\dot\f=-\sin\f$: it has $\f=\p$ as an unstable fixed point and $\f=0$ as
  a stable fixed point (with Lyapunov exponents $\l_0=\pm1$,
  respectively). Let $I$ be the reflection of the point $\f$ at the circle
  center. Then the evolution is $I$-reversible and the distribution
  $\m_0(d\f)=\frac{d\f}{2\p}$ is $I$-symmetric. Hence Eq.\equ{e19.1} holds
  for all finite $\t$: at $\t=\infty$ the distribution of $p=\frac1\t
  \sum_{k=0}^\t \cos(S_0^k\f)=\frac1\t A$ evolves to $\d(p)$ which does not
  satisfy the FR for any $p>0$ although $\s_+=1$. The example can be easily
  adapted to deal with a chaotic evolution: it is enough to consider the
  dynamical system acting on pairs $(\f,x)$ which evolve in $(S_0\f,S_\X
  x)$, where $\X$ is any Anosov map reversible under a map $J$. This is
  reversible under the time reversal $(\f,x)\to (I\f,J x)$. Then
  Eq.\equ{e19.1} holds but leads to a relation with slope $\s_+=\s(\X)+1$,
  where $\s(\X)$ is the phase space contraction of the map $\X$), while FR
  predicts the correct slope $\s(\X)$, because the example is a simple
  example of a system with a smooth hyperbolic attracting set (\ie the
  pairs $(0,x)$), hence it satisfies the Chaotic Hypothesis: a case in
  which the FR is a theorem. Likewise a flow example can be easily
  constructed. The example is due to F.Bonetto.\label{TrFl}}

A proof of any relation between Eq.\equ{e19.1} and the FR discussed in the
present review, in any event,
has never been published, in spite of several announcements.

\subsection{The Jarzinsky relation}
The Eq.\equ{e19.1} is an identity but nevertheless it can be useful, as
shown by its applications in various domains and this might be the
explanation of the lack of interest on the FR and the Chaotic
hypothesis.

In this respect there are other relations which are exact and useful
identities with several interdisciplinary applications in nonequilibrium
phenomena.

An example is provided by an implementation of the simplest ``Monte Carlo
method'': here the general purpose of the Monte Carlo methods is intended
as the use of a controlled random number generator to produce random events
with a prescribed distribution.

For instance suppose that it is necessary to produce spin configurations on
a $N$-points lattice $\LL$ with a distribution proportional to
$e^{-\b U(\Bs)},\,\b>0$, with $U(\Bs)=\sum_R J_R \s_R$, where $J_R\ge0$ are
are given ``couplings'' for the spins $\s_r=\pm1$ with $r$ in a subset
$R=(r_1,\ldots,r_N)\subset \LL$. Suppose available a random number
generator $G_0$ able to generate a known distribution of $\Bs$, for
instance a Bernoulli shift $(\frac12,\frac12)$ distribution; then follow
the algorithm, also called a ``protocol'':
\*
\0(1) generate a spin configuration $\Bs$ for the Bernoulli distribution
using a deterministic random number generator $G_0$ (initialized beforehand
once and for all with a fixed number). This plays the role of selection of
initial data from a known initial state (here a sample Bernoulli path). And
compute the weight of the $\Bs$ in the Bernoulli shift (which in the case
under consideration would be $e^{-\b U_0(\Bs)}\equiv 1$, \ie probability
$Z_0^{-1}=2^{-N}$).  \\
(2) compute
$U(\Bs)$ and the ratio $e^{-\b (U(\Bs)-U_0(\Bs))}$
\\
(3) attribute to $\Bs$ the weight $e^{W(\Bs)}\equiv e^{-\b (U(\Bs)-U_0(\Bs))}$.
\*

Repeat the protocol many times: the statistic determined on the outputs
$\Bs_1,\Bs_2,\ldots$ by assigning them the weights $e^{-\b
  (U(\Bs_i)-U_0(\Bs_i))}$ acquires eventually the PDF $P(\Bs)=Z^{-1} \exp
-\b U(\Bs)$ and the relation
\be \media{e^W}=\frac{Z}{Z_0}\Eq{e19.2}\ee
holds with the average being
taken with respect to the initial distribution (\ie the Bernoulli shift in
the present case).

The procedure can be used to generate the Gibbs distribution at temperature
$\b$ and Hamiltonian $H_1(x)$ from a Gibbs distribution with Hamiltonian
$H_0(x)$, \cite{Ja997}. Imagine to have at hand a system in equilibrium
with Hamiltonian $H_0(x)$ and a way (``protocol'') to force the evolution
of a configuration via equations of motion following a time dependent
Hamiltonian $H_t(x)$ which evolves from $H_0$ at time $0$ to $H_1$ at time
$1$: \*

\0(1) generate an initial state $x$ by picking a sample out
of the initial distribution, and
\\
(2) act, by changing the parameters of the Hamiltonian, so that $x$ evolves
with the time dependent Hamiltonian $H_t$ as $x\to S_{0,t}x$ and keep track
of the energy $W_t(x)=H(S_{0,t}x)-H_0(x)$\\
(3) weigh the output at time $t=1$ with $e^{-\b
  W_1(x)}$: eventually the statistics of the weighted outputs will be the
distribution $Z^{-1}e^{-\b H_1(x)}$, as it is immediately checked using the
Liouville theorem $S_{0,t} dx=dx$ (where $dx=dpdq$ in canonical
coordinates).
\*

The above two protocols are realizations of a (naive) ``Monte Carlo''
method: the second can be particularly useful, aside numerical simulations,
even in applications to bio-systems where it has been possible to find a
way to measure $W_1$ at each run of the protocol.

The quantity $W_1$ has been identified, in several cases,
with the work performed on the system during one iteration of the protocol:
it has then been used particularly to measure the free energy variation
between two different equilibria at the same temperature: $\b\D F=-\log
\media{e^{-\b W_1}}$ (with the average being over the statistics of the
initial data), \cite{Ja997}.

Notice that access to $W_1$ is the only requirement necessary: the random
generator being the {\it initial equilibrium state} and the evolution $H_t$
only needs to be always the same, each time the protocol is run. Of course
it is necessary to be able to justify that the measurements of $W_1$ really
evaluate the work $W$ done on the system: in concrete cases it may be not easy
to be sure that all forces are taken into account, in particular the ones
that change $H_0$ to $H_t$. It may also be difficult to make sure that the
protocol used is always exactly the same.

\subsection{Ruelle-Lieb bounds}

There are remarkable {\it rigorous} bounds on the averages, with respect to
the stationary distributions, of the eigenvalues of the symmetric part of
the Jacobian matrix $J$ for the NS equations, symbolically given by
$J_{i,j}=\n\d_{ji}\D-\frac12(\dpr_i u_j+\dpr_j u_i)$, acting on the
incompressible velocity fields.\footref{QR0} The averages over time of such
eigenvalues are a kind of ``local exponents''. The estimates give an upper
estimate $\wt N_J$ to the maximum number of exponents which add up, ordered
by decreasing size, to a non negative value, hence also an upper bound on
the number of non negative Lyapunov exponents.

The numbers $\wt N_J$ are, \cite{Ru982,Li984}, bounded in dimension
$2$ and $3$ (and more); and {\it in dimension $2$} the bounds can be expressed,
\cite[Eq(34)]{Li984}, in terms of the average
$D=\media{\sum_\kk \kk^2
  |u_\kk|^2}_{irr}$ of the enstrophy $D(u)$:
\be\eqalign{
    \wt N_J\le&A (2\p)^2 \sqrt{R^2D},\qquad A=0.55...
    \cr}\Eq{e19.3}\ee
As seen in Sec.\ref{sec17} $\frac1R D=\media{W}_{irr}$ with $W$ the 
power spent by the external force, see comment on Eq.\equ{e17.7}.

The $2$-dimensional estimates are in \cite[Eq/(43)]{Li984}: there are also
found similar estimates, in higher dimension, extending earlier ones in
\cite{Ru982}. In $\ge3$ dimensions the $\wt N_J$ are not bounded in terms
of $ \m^{a}_\cdot(\int dx(\BDpr\uu(x))^2),\ a =rev,irr$, but involve powers
of $\BDpr\uu$ higher than $2$.

The estimates apply to the irreversible NS
equations, truncated at arbitrary ultraviolet cut-off, and involve only the
eigenvalues of $\frac12(J+J^*)$ averaged in time: which can be evaluated in
simulations. Being rigorous they can be important in checks of the accuracy
of simulations. In the reversible equations $R^2 D$ should be replaced
by $\media{\a(u)^{-2} D(u) \ch(\a(u))}_{rev}$ where $\ch(z)=1$ if $z\ge0$
and $0$ otherwise. 

\subsection{Wishes}

Several tests of the FR have been performed in the literature.
Unfortunately the FR is confused with the similar relation called above
the ``transient fluctuation theorem'' (possibly even omitting the
qualification of transient).

It is quite interesting that in most {\it simulations} the tests performed
really deal with the FR; hence it would be very interesting to mount
experiments to test the FR (in nonequilibrium situations).

Many experimental works, very delicate and difficult, that claim to have
tested the FR have, unfortunately, instead only tested the above transient
relation; and could be, {\it perhaps even easily}, devoted to a real test
of FR. Or the tests have been devoted to check the linearity in the
symmetry relation Eq.\equ{e6.3} for the fluctuations of a quantity identified 
with $p$ but neither attempting to check the relation with the phase space
contraction nor examining the validity of the assumption that the attracting
set has dimension equal to the of phase space.

Beautiful laboratory experiments on nano materials,
proteins, granular materials, ... have been performed and are, very often,
remarkable and innovative from the technical view point but in all cases,
that I am aware of, at best they test the transient theorem.
The study of the FR has not yet attracted sufficient interest, with the
notable exception of the many numerical simulations.

The check of the fluctuation relation in stationary states of systems in
nonequilibrium is a test of the Chaotic Hypothesis, which is a {\it
  physical assumption}, unlike the checks of the transient fluctuation
relation (which {\it per se} does not test any physical assumption, because
testing an identity or a theorem does not provide new information). 

Also some experimental works limit the analysis to studying the PDF of the
work done or of the heat arising in the experiment but, in my view, not
always enough attention is dedicated to check that all forces acting are taken
into account.  This leads, sometimes, to claim (more or less openly) that
the PDF of the work or heat generated at temperature $T$ in a process
cannot be normalized with $k_B T$, as in Eq.\equ{e4.3}: therefore, it is
concluded, the stationary FR is false.

The problem seems to be a certain {\it resilience} to invest time to follow
the ideas behind the chaotic hypothesis (\ie Anosov systems) and the
general Axiom A attractors: an example of the attitude towards these ideas
is in the quoted statements in \cite{Ho999}; see also Appendix \ref{appA} below.
In spite of all the above remarks I still hope that the FR relation will be
tested in ``real'' experimental contexts.

\APPENDICE{1}
\def\SEC{\ About certain comments on CH}
\section{\SEC}
\label{appA}

In the above sections quotes from \cite{Ho999} have been reproduced
without really commenting them. The reason is that the quotes were written
when the Chaotic Hypothesis had been just developed and many had not
yet had the time to really study the subject.

But the same comments have appeared in a second edition of the just quoted
book, \cite{HG012}, which I have seen only very recently (after completion
of the present text). Since the comments have had some resonance the next
few lines try to clarify some of the issues.

The Author of \cite{HG012} criticizes the use of the Anosov systems as
paradigm of chaotic motions. The full section from p.344 to p.347 discusses
the merits and demerits of Anosov systems. On p. 344 begins
\*
\0{\it ... has discussed the possibility that the useful
properties exhibited by certain oversimplified and quite rare dynamical
systems, termed "Anosov systems", have counterparts in the more usual
thermostatted systems studied with nonequilibrium simulation
methods. Anosov systems are oversimplifications, like square clouds or
spherical chickens...}
\*

This seems to refer to the proposal that the ``Axiom A'' systems should be
the right paradigm for generic chaotic systems, \cite{Ru989,Ru999}: a
proposal which however is not centered on Anosov systems. The Axiom A
systems are systems which have an the attracting set $\AA$ on which motion
has strong chaotic properties (is essentially hyperbolic).

And the CH just proposes, in its final formulation, (1996), that for many
purposes the axiom A paradigm can be strengthened and simplified by
requiring {\it in addition} that $\AA$ is an attracting surface, possibly
of dimension lower than that of phase space, on which the motion is an
Anosov system. Even in time reversible cases $\AA$ {\it can be different
from its time reversal image}. This is explicitly stated with related
problems and examples in \cite{BGG997,BG997} and in several successive
publications.

The underlying idea being that it is not possible to distinguish, in a
system of physical interest, a fractal of Hausdorff dimension
$=10^6+3.1415...$ from a surface of exactly $10^6$ dimensions.

In summary the Chaotic Hypothesis only assumes that the dynamics under
consideration behaves (in some respects) like Anosov dynamics.  This is
after all not too astonishing if the most relevant
degrees of freedom are chaotic like those of Anosov systems.

Most of the subsequent criticism in \cite{HG012} is anchored on keeping the
identification between the CH and the proposal that the whole dynamical
system is an Anosov system.

On p.346 the fluctuation theorem is called a ``retrospective result'' and
identified with the true Fluctuation Theorem, Sec.\ref{sec6} above,
claiming: \*
``{\it These same "results" were actually given earlier by Denis Evans and
several of his coworkers, for more general circumstances and through more
elementary arguments.}''
\*
\0but no reference is made here to the applicability of the ``earlier
retrospective result'' to stationary nonequilibria to which the Fluctuation
Theorem applies, see\footref{TrFl}.

Then on p. 347 the view is found that: 
\*
``{\it Theoretical constructs such as ``measures'', should be viewed with a
healthy suspicion until algorithms for evaluating them are supplied. The
chaos inherent in {\it interesting} differential equations guarantees that
our only access to the "strange sets" which constitute attractors and
repellers will be representative time series from dynamical simulations. In
no way can we construct, or even conceive of constructing, a
Sinai-Ruelle-Bowen measure for an interesting system.}''
 \*

 \0However for most purposes {\it by the CH Hamiltonian systems should be
   considered Anosov systems} (literally, except of course the integrable
 ones). Hence the assumption that the attracting set is the full phase
 space is not always unreasonable.

Furthermore it is useful to stress that there are easy examples of systems
satisfying {\it the CH, with equal or disjoint attracting and repelling
  surfaces, time reversible, with as many degrees of freedom and negative
  Lyapunov exponents as wished (unrelated to the number of positive ones)
  and whose SRB measure is explicitly and completely constructed},
\cite[Sec.10.2]{GBG004}.

\APPENDICE{2}
\def\SEC{\ Local Fluctuations. An example.}
\section{\SEC}
\label{appB}
\iniz

The phase space contraction in the evolution of a macroscopic system
is typically a macroscopic quantity: whether it is the  amount of heat
ceded to the thermostats or the amount of work performed by the systems.

Therefore the average phase space contraction $\s_+$ which controls the
large fluctuations, Sec.\ref{sec6} and the occurrence of ``anomalous''
patterns, Sec.\ref{sec12}, cannot be really observed in measurements on
macroscopic systems.

Avoiding comments on the many experimental fluctuations observations which
claim to check the FR,\footnote{\small Sometimes claiming to have checked
  it and sometimes claiming the opposite, while very often dealing with
  unrelated transient phenomena.} the question asked here is whether a kind
of fluctuation relation could be defined, and constrain quantities
depending on events that can be observed in very small parts of the system.

In other words is it possible to give a meaning
to a {\it local fluctuation relation}? \cite[Ch4.9]{Ga013b}.

The following relies on Sec.\ref{sec15}: it is inserted as it provides a
quite interesting example on how to make use of the symbolic dynamics
representation of the Anosov systems.

A simple example, in a system with time reversal symmetry, will be 
discussed in which a {\it local entropy production rate} can be defined and
checked to satisfy a local version of FR. A general view on the matter can
be found in \cite{Ga998b,Ga997c}.

The analysis deals again with maps rather than flows.\footnote{\small This
  time {\it the reason is not ``for simplicity''}, but because in the case
  of {\it coupled flows}, even if the coupling has short range and is weak,
  there seems to be no detailed and constructive general theory of the SRB
  distributions, because no simple conditions are known that, via
  perturbation techniques, yield hyperbolicity of the flow and allow
  studying its properties. For a glimpse on the kind of
    complications which arise when studying flows consult,
    \cite{RW001,GGG006}.  Instead, at least in the case of coupled maps,
  the theory is quite well understood, \cite{PS991,BK995,BK997,GBG004}, as
  in the example in Eq.\equ{eB.1} below, at small coupling $\e$.}

Consider a system with a translation invariant spatial structure, \eg a
periodic chain, or a $d$-dimensional square lattice $[-L,L]^d$ with
periodic boundary, of $(2L)^d$ weakly interacting Anosov maps.

The phase space of the system is $M=\{\V x=(\ldots
x_1,x_0,x_1,x_2,\ldots)\}=\MM_0^{(2L)^d}$, where $x_i\in \MM_0$ are points in
a manifold $\MM_0$: to fix ideas we {\it take $\MM_0$ to be a torus}, on which
an Anosov map $\lis S_0$ acts; then define the ``coupled map'':

\be \lis S_\e(\V x)_i=\lis S_0 x_i+\e g(x_{i-1},x_i,x_{i+1}), \ i=...,0,1,\ldots
\Eq{eB.1}\ee
where $g$ is a smooth perturbation, \ie a smooth periodic function on
$\MM_0^3$.

If $\e$ is small and the perturbation has short range it is proved in
\cite{PS991,BK995,BK997} that, defining the map $\lis S_\e$ as in
Eq.\equ{eB.1} with periodic boundary condition (\ie identifying the site
$-L $ with $L $), the map $\lis S_\e$ remains, if $\e$ is
small enough, still Anosov. It is conjugated to $\lis S_0$, via a H\"older
continuous correspondence $\BTh_\e$, see\footref{struct-stab}, by
associating points $x$ and $x'$ with the same history under $\lis S_0$ and
$\lis S_\e$. Furthermore there is $\e_0>0$ such that the above holds for
$|\e|<\e_0$ {\it uniformly} in the system size $L$.
\footnote{\small
The SRB distribution for the evolution $\lis S_\e$ with $L=\infty$ could
also
be
defined via the SRB distribution $\m_{srb,\e}^L$ for the system $\lis S_\e$
consisting of the sites labeled from $-L $ and $L $ and then taking the
limit $\m_{srb}=\lim_{L\to\infty}\m_{srb,L}$, \cite{PS991}.
\\
This is possible because of the uniformity in $\e<\e_0$: below, however,
$L<\infty$ will be fixed, keeping in mind that the results will hold for
all $L$ if $|\e|<\e_0$.}

Here the purpose is to study whether a local version of the FR can hold at
least in an example derived from $\lis S_\e$: but the $\lis S_\e$ is, in
general, not reversible. A related reversible map $S^{rev}_\e$ can be
easily constructed on the ``doubled'' phase space $\MM=\MM_0\times\MM_0$ by
setting:
\be S_\e^{rev}(\V x,\V y)= (\lis S_\e(\V x),
(\lis S_\e)^{-1}(\V y))\Eq{eB.2}\ee
which is reversible for the time reversal map $I:(\V x,\V y)=(\V y,\V
x)$. In the rest of this section this system will be considered in more
detail.

A Markovian partition $\PP^L_0$ for $\lis S^L_0\times (\lis
S^L_0)^{-1}$ will be chosen to be the product of partitions $\PP_{-\frac
  L2},\ldots,\PP_{\frac L2-1}$ for the single site maps $\lis S_0$ and
$\lis S_0^{-1}$; and $\PP^L_\e$ will be the partition $\BTh_\e\PP^L_0$
existing and defined by the structural stability map $\BTh_\e$, conjugating
$\lis S_0^{rev}$ to $\lis S_\e^{rev}$, \cite[Sec.10.2]{GBG004}.

Hence the history of a point $x$ will be a sequence of labels $\s_{i;j}$
with $i\in M$ and $j\in (-\infty,\infty)$: naturally
$i$ can be called a ``space label'' while $j$ a ``time label''.  The
superscript $rev$ will be omitted in what follows, to simplify notations.

The analysis in Sec.\ref{sec15} applied to the Anosov map $S_\e$, will give
a representation of the volume distribution $\m_0$ and of the SRB
distribution $\m_{srb,\e}$ for $S_\e$ via, respectively, suitable
potentials $\BF_\e,\BF_{\e}^\pm$.

Let $[0,\t]$ be a time interval and $\L=[-\frac12L,\frac12L]^d=M$,
$|\L]=L^d$. Via the Jacobian matrix $J_\L(x)=\dpr_x (S_\e x)$ define the
  {\it phase space contraction} and the {\it time averaged contraction} per
  site as, respectively:
\be \eqalign{
\h_{\L,\e}(x)&=-\frac1{|\L|}\log |\det (\dpr_x
(S_\e x))|\cr
\h_{\L,\e,\t}&=\frac1{|\L|}\lim_{\t\to\infty}
\frac1\t\sum_{j=0}^\t\h_{\L,\e}(S_{\e}^j x)\cr
}\Eq{eB.3}\ee
The limit of $\h_{L,\e,+}$ in Eq.\equ{eB.3} as $\t\to\infty$ exists with
probability $1$ with respect to the volume $\m_{vol}$, as well to the SRB
distribution $\m_{srb,\e}$, and is $x$-independent aside $x$'s in a set of
$0$ volume: because the statistical properties of the volume distribution
are those of the Gibbs distribution with potential $\BF^+_\e$, hence enjoy
strong ergodicity properties, as any SRB distribution, with respect to time
translations.
\footnote{\small Likewise the space-time limit $\h_{\e,+}$ exists because
  of the space-time ergodicity of the short range Gibbs processes
  describing the volume as well as the SRB distributions.}

The phase space contraction $\sum_{t=0}^\t \h_{L,\e}(S_\e^t x)$ can be
expressed, see Sec.\ref{sec15}, via the potentials
$\BF^+_{\e},\BF^-_{\e},\BF_{\e}^\f$, where $\BF^\f_{\e}$ is a potential
that describes the interpolation between $\BF_{\e}^-$ to $\BF_{\e}^+$ and
which is therefore ``localized'' (see comment to Eq.(\equ{e15.1})) in the
sense that $\BF^\f_{\e}(\Bs_I)\ne0$ only if $I$ contains the sites $0$ or
$L$ and $|\BF_{\e,I}^\f(\Bs_{I})|\le C e^{-\k |I|}$ for some $C,\k>0$).
Given the symbolic history $\Bs$ of $x$, the Eq.\equ{e15.6} can be
expressed as:
\be\eqalign{
\frac1{\t L^d}
\sum_{K\subset M\times [0,\t]}
  (\BF^+_{\e,K}(\Bs_K)-\BF^-_{\e,K}(\Bs_K))+\ldots\cr}
\Eq{eB.4}\ee
where $K=I\times [a,b]$ is a {\it parallelepiped} in $\L\times[0,\t]$,
and $\BF^z_{\e,K}\defi\sum_{t\in[a,b]}\BF^z_{I+t}$ for $z=\pm,\f$, 
and the $\ldots$
indicate a correction $\sum_K \BF^\f_{\e,K}(\Bs_K)$,\footnote{\small
Relatively vanishing as $L^{-1}$ uniformly in $x$.}.
t
A natural mathematical definition of the ``local average phase space
contraction'' could be the $-\frac1{\t\L_0}\sum_{t=0}^\t \log J_{\L_0}(S^x)$
where $J_{\L_0}(x)=|\det (\dpr_i (S_\e x)_{i'})|$. But this is a quantity
difficult to express in a useful way.

However it is also possible to propose a {\it different} definition of
local average phase space contraction based on the representation
Eq.\equ{eB.4} of the average of the logarithms of the full Jacobian. The
latter can be expressed as Eq.\equ{eB.4} up to a quantity uniformly bounded
in $L$: and the contribution to Eq.\equ{eB.4} from the parallelepipeds
$K$'s {\it entirely contained} in $\L_0\times[0,\t]$ is:

\be \h_{\L_0,\e,\t}^{loc}
  \defi\frac1{\t L_0^d}
   \sum_{K\subset\L_0\times[0,\t]}
  \Big(\BF_{+,K}(\Bs)- 
  \BF_{-,K}(\Bs)\Big)
\Eq{eB.5}\ee
see Eq.\equ{e15.1},\equ{e15.4}; the $ \h_{\L_0,\e,\t}^{loc}$ can be,
heuristically, called the ``local contraction rate''. It can be uniformly
bounded (in $\t,L$).

Given $\L_0$ let $\h^{loc}_{\L_0,\e,+}$ be the time average
$\h_{\L_0,\e,\t}^{loc}$ define:
\be p'=\frac1\t
\frac{\h_{\L,\e,\t}(x)}{\h_{\L,\e,+}},\quad p=\frac1\t
\frac{\h_{\L_0,\e,\t}^{loc}(x)}{\h_{\L_0,\e,+}^{loc}}\Eq{eB.6}\ee
and remark that $\h_{\L_0,\e,+}^{loc}=\h_{\L,\e,+}+o(L_0^{-1})$
(because of the SRB distribution representation of a Gibbs process).

It can also be shown that {\it to leading order} as $L_0,L,\t\to\infty$
the large deviation rates for $p',p$ in Eq.\equ{eB.6}
have the form $\t L^d\z_\infty(p'),\ \t L_0^d\z^0_\infty(p)$, with
$\z_\infty=\z^0_\infty$ because
$\z_\infty$ is obtained as a thermodynamic limit of a kind of partition
function: for a proof
see \cite[(5.14)]{Ga999b}.

Therefore by the FT applied to $S_\e$ it is
$L^d\z_\infty(p')-L^d\z_\infty(-p')= L^d p' \h_{\L,\e,+}$ and, since
$\h_{\L_0,\e,+}^{loc}= \h_{\L,\e,+}+o(L_0^d)$, the large deviations rate
for $p$ in Eq.\equ{eB.6} satisfies a FR of the form
\be \eqalign{
  &L^d_0(\z_\infty(p)-\z_\infty(-p))=p \,L_0^d\h_{\L_0,\e,+}^{loc}\cr&=
p\, r\, (L^d\h_{\L,\e,+})\cr}\Eq{eB.7}\ee
with $r=\frac{L_0^d}{L^d}$ and $|p|\le p^*, p^*\ge1$, up to corrections of
$O(L_0^{d-1})$: which means that the global and local large fluctuations
rates are proportional and trivially related by a rescaling which equals
$r=(\frac{L_0}{L})^d$ up to a correction bounded $\k^{-1} L_0^{d-1}$ with
$\k$ bounding the range of the SRB potential, as in Eq.\equ{e15.1}.

The universal slope $1$ in the global FR is modified into
$r=(\frac{L_0}{L})^d$ in the local FR. The Eq.\equ{eB.7} {\it can be proved
  for the system in} Eq.\equ{eB.2}.

However $p$ in Eq.\equ{eB.6} {\it is not related to a measurable quantity},
as it cannot be hoped to be able to measure directly the 
local phase space contraction .defined as in Eq.\equ{eB.5}. 

Still the phase space contraction is often related to the amount of heat
ceded or the work done on the surroundings by a system in a stationary
state, as exemplified in the case of Eq.\equ{e4.4}: hence it is tempting to
test, in cases in which the latter quantities are accessible to local
measurements, whether Eq.\equ{eB.7} holds. This is attempted in some
simulations, \cite{GL014}.

The interest of the above special example lies in the statements
independence on the total size of the systems: they also mean that the
fluctuation theorems {\it may lead to observable consequences} if one looks
at the far more probable microscopic f\/luctuations of the local entropy
production rate,\cite{PS991,BK995,BK997}. For more details see
\cite{Ga995b}.

\APPENDICE{3}
\def\SEC{\ Reversible heating}
\section{\SEC}
\label{appC}

Imagine a rarefied gas enclosed in a cubic container of side $L$ described
by a canonical distribution at inverse temperature $\b^{-1}$.  The potential
energy is $\sum_{i=1}^N mg z_i +\sum_{i,j} v(x_i-x_j)\defi M\,g\,H +V$,
with $M$=total mass and $H$ the height of the center of mass. The
initial free energy if $F(\b,g)=-\b^{-1} \log \int e^{-\b(V+gP)}
d^{3N}pd^{3N}q$. The entropy can be computed via Gibbs' formula $S_0=-\int
\r(p,q)\log\r(p,q)d^{3N}pd^{3N}q$.

The gas is set out of equilibrium by changing the gravity $g$ to a new value
$g'$ for instance suddenly at time $t=0$ or following a given prescription
$t\to g(t), t\in [0,\t]$ with $g(\t)=g', \t<\infty$. Then it is let to
evolve.

Since the evolution is Hamiltonian (although not autonomous) $\r(p,q)$
evolves in $\r(p,q;t)$ and the latter tends, as $t\to\infty$, to a new
equilibrium state in the gravity potential $m g'z$; but $-\int
\r(p,q;t)\log\r(p,q;t)d^{3N}pd^{3N}q$ remains equal to $S_0$.  Therefore at
the end of the evolution the new distribution $\r(p,q,\infty)$ will be an
equilibrium state of the system in the modified gravity field.

It will not be, however, any more a canonical Gibbs state at temperature
$\b^{-1}$ in a gravity field with acceleration $g'$; if the system is
ergodic on the energy surface then the final distribution reached at
infinite time after suddenly increasing the gravity $g$ to a new value $g'$
will be (integration over $p',q'$ only)''
\be \m^\infty(dpdq)=\int \m_\b(dp'dq') \m^{mc}_{E(p',q')}(dpdq)\Eq{eC.1}\ee
where $\m^{mc}_{E}(dpdq)$ is the microcanonical  distribution with energy
$E$ and $E'(p',q')= K(p')+V(q')+Mg'H(q')$ is the sum of the kinetic energy,
internal potential energy and energy of the center of mass in a gravity
acceleration $g'$.

The distribution Eq.\equ{eC.1} will be equivalent to a canonical Gibbs
distribution (with temperature different from $\b^{-1}$) only in the
thermodynamic limit: in the finite system that we are considering it will
be different by corrections vanishing in the thermodynamic limit. Yet the
new state will be a stationary state close to a canonical (or any other
equivalent) equilibrium state,

To estimate, actually to define, the temperature $\b^{'-1}$ of the new
state imagine to identify the above $\m^\infty$ with a canonical
distribution $\m_{\b'}$, \ie {\it neglect the finite volume
  corrections}. Computing the Gibbs entropy $S_\infty$ of the new equilibrium
(reached after infinite time) and make use of the identity between the
Gibbs entropies of the initial and final states:
\be\kern-3mm\eqalign{
  &F=-\frac1\b\log \kern-1mm\int e^{-\b(V+gP)} d^{3N}pd^{3N}q=-T \log Z_0\cr
&S=-\dpr_T F=-\log Z_0-\b\media{V+ gP}\cr
&\dpr_g S|_\b=\b^{2}(\media{P V}-\media{P}\media{V})
+g(\media{P^2}-\media{P}^2)\cr}\Eq{eC.2}\ee
with $T=\b^{-1}$, $P=MgH$ and $\dpr_g S|_\b\ne0$ at $g=0$: thus the new
equilibrium {\it cannot} have the same entropy as the initial state if the
temperature remained the same unless $\b'\ne\b$ (because in general $\dpr_g
S|_\b\ne0$, \eg if $V\simeq0$ it is $\media{P^2}> \media{P}^2$). If the final
state has to become a canonical distribution at some temperature (\eg the
above estimated $\b^{'-1}$) then the system will have to be attached to a
thermostat and some heat exchange will take place and the entire
transformation will be irreversible: in any event, if the system container
was really adiabatic and at any finite (or infinite) time the gravity
acceleration was dropped back to the initial value, then the system should
in the same time return to the initial canonical state.  See also
\cite{SSHT002} for the analysis of equally interesting cases.  \*

\0{\bf Acknowledgments:} I profited from discussions with L.Biferale,
M.Cencini, M.De Pietro, V.Lucarini in Sec.\ref{sec18}\,, and L.S.Young
in Sec.\ref{appB}\,, whom I thank warmly.
\*

\0{{\bf Erratum:} In versions 1,2 at the last page of Sec.18 the
  statement: ``Formally the Ekmann equation, aside the infinite
  dimensionality, is covered by the pairing theorem...'' is
  incorrectly supported in appendix D (mainly because the forcing
  $\bf f({\bf x})$ cannot be a gradient). The statement {\it has
    been weakened in the present version 3, and the Appendix D
    removed,} without affecting the rest of Sec.18 and of the
  paper.}

\bibliographystyle{unsrt}



\end{document}